\documentclass[11pt,a4paper, epsfig]{article}
\usepackage{amsfonts}
\usepackage{amsmath}
\usepackage{amsthm}
\usepackage{graphicx}
\usepackage{subfigure}
\usepackage{amssymb}
\usepackage{bm}
\usepackage[dvips]{color}
\usepackage{longtable}
\usepackage{hyperref}
\usepackage{makeidx}
\usepackage{rotating}
\usepackage[toc]{appendix}
\usepackage{setspace}

 \usepackage{anysize}
 \marginsize{2.5cm}{2.5cm}{1.5cm}{1.5cm}

\newcommand{\ra}{\rangle}
\newcommand{\la}{\langle}
\newcommand{\ua}{\uparrow}
\newcommand{\da}{\downarrow}

\newcommand{\bfx}{{\bf x}}
\newcommand{\bfa}{{\bf a}}
\newcommand{\bfb}{{\bf b}}
\newcommand{\bfc}{{\bf c}}
\newcommand{\bff}{{\bf f}}

\newcommand{\tboxed}[1]{
\setlength\fboxrule{0.5pt}\setlength\fboxsep{2mm}
  \[\fbox{%
      \addtolength{\linewidth}{-2\fboxsep}%
      \addtolength{\linewidth}{-2\fboxrule}%
      \begin{minipage}{\linewidth}%
    #1
      \end{minipage}%
    }\]%
}

\newcommand{\tbox}[1]{
\setlength\fboxrule{2pt}\setlength\fboxsep{2mm}
  \[\fbox{%
      \addtolength{\linewidth}{-2\fboxsep}%
      \addtolength{\linewidth}{-2\fboxrule}%
      \begin{minipage}{\linewidth}%
    #1
      \end{minipage}%
    }\]%
}

\theoremstyle{remark}

\makeindex

\begin{document}
\begin{center}
{ \LARGE Some Topics in Quantum Games} 
\ \\
\ \\
Yshai Avishai \\
\ \\
\ \\
Department of of Physics and Ilse Katz Institute for Nano-Technology\\
\ \\
Ben Gurion University of the Negev, Beer Sheva, Israel.
\ \\
\ \\
\ \\
\ \\
Based on a Thesis Submitted on August 2012 to the \\
\ \\
Faculty of Humanities and Social Sciences, Department of Economics \\ 
\ \\
 Ben Gurion University of the Negev, Beer Sheva, Israel\\
 \ \\
in Partial Fulfillment of the Requirements for the Master of Arts Degree\\

\end{center}


\newpage
\noindent
{\bf abstract}
\begin{spacing}{1.5}
This work concentrates on simultaneous move quantum games of two players. Quantum game theory models the behavior of strategic agents (players) with access to quantum tools for controlling their strategies. The simplest example is to envision a classical (ordinary) two-player two-strategies  game $G_C$ given in its normal form (a table of payoff functions, think of the prisoner dilemma) in which players communicate
with a referee via a specific quantum protocol, and, motivated by this vision, construct a new game $G_Q$ with
greatly enlarged strategy spaces and a properly designed payoff system. The novel elements in this scheme consist of three axes. First, instead of the four possible positions (CC), (CD), (DC) and (DD) there is an infinitely continuous number of positions represented as different quantum mechanical states. Second, instead of the two-point strategy space of each player, there is an infinitely continuous number of new strategies (this should not be confused with mixed strategies).  Third, the payoff system is entirely different, since it is based on extracting real numbers from a quantum states that is generically a vector of complex number. The fourth difference is apparently the most difficult to grasp, since it is a conceptually different structure that is peculiar to quantum mechanics and has no 
analog in standard (classical) game theory. This very subtle notion is called {\it quantum entanglement}. Its significance in game theory requires a non-trivial modification of one's mind and attitude toward game theory and choice of strategies. 
Quantum entanglement is not always easy to define and estimate, but in this work where the classical game $G_C$ is simple enough, it can be (and will) be explicitly defined. Moreover, it is possible to define a certain 
continuous real parameter $0 \le \gamma \le \pi/2$ such that for $\gamma=0$ there is no entanglement, while for $\gamma=\pi/2$ entanglement is maximal.  

Naturally, a substantial part of this work is devoted to settling of the mathematical and physical grounds for the topic of quantum games, including the definition of the four axes mentioned above, and the way in which a standard (classical) game can be modified to be a quantum game (I call it {\it a quantization of a classical game}). 
The connection between game theory and information science is briefly explained. While the four positions of the classical game are formulated in terms of {\it bits}, the myriad of positions of the quantum game are formulated in terms of {\it quantum bits}. While the two strategies of the classical game are represented by a couple of simple $2 \times 2$ matrices, the strategies of a player in the quantum game are represented by an infinite number of complex unitary $2 \times 2$matrices with unit determinant. The notion of entanglement is explained and exemplified and the parameter controlling it is introduced. The quantum game is formally defined and the notion of pure strategy Nash equilibrium is defined. \\
With these tools at, it is possible to investigate some important issues like existence of pure strategy Nash equilibrium and its relation with the degree of entanglement. The main achievement of this work are as follows:
\begin{enumerate} 
\item{} Construction of a numerical algorithm based on the method of best response functions, 
designed to search for pure strategy Nash equilibrium in quantum games. The formalism is based on the discretization of a continuous variable into a mesh of points, and can be applied to quantum games 
that are built upon two-players two-decisions classical games.  based on the method of best response functions
\item{} Application of this algorithm to study the question of how the existence of pure strategy Nash equilibrium is related to the degree of entanglement (specified by the parameter $\gamma$ mentioned above). It has been proved (and I prove it here directly) that when the classical game $G_C$ has a pure strategy Nash equilibrium that is not Pareto efficient, then the quantum game $G_Q$ with maximal entanglement ($\gamma= \pi/2$) has no pure strategy Nash equilibrium. By studying a non-symmetric prisoner dilemma game, I find  that there is a critical value $0 < \gamma_c < \pi/2$ such that for $\gamma<\gamma_c$ there is a pure strategy Nash equilibrium and for $\gamma \ge \gamma_c$ the is not. The behavior of the two payoffs as function of $\gamma$ start at that of the classical ones at $(D,D)$ and approach the cooperative classical ones at $(C,C)$. 
\item{}  Bayesian quantum games are defined, and it is shown that under certain conditions, there {\it is} a pure strategy Nash equilibrium in such games even when entanglement is maximal. 
\item{} The basic ingredients of a quantum game based on a two-players {\it three} decisions classical games. This requires the definition of trits (instead of bits) and quantum trits (instead of quantum bits). It is shown that in this quantum game, there is no classical commensurability in the sense that the classical strategies are not obtained as a special case of the quantum strategies. 
\end{enumerate} 
\end{spacing}
\begin{spacing}{1.5}
\newpage
\begin{center}
\section{\small Intoduction} \label{Section1}
\end{center}
This introductory Section contains the following parts: 1) A prolog that specifies the arena of the thesis and sets the relevant scientific framework in which the research it is carried out. 2) An acknowledgment expressing my gratitudes to my supervisor and for all those who helped me 
passing an enjoyable period in the department of Economics at BGU. 3) An abstract with a list of novel results 
achieved in this work. 4) A background that surveys the history and prospects of the topics 
discussed in this work. 5) Content of the following Sections of the thesis.  

\subsection{Prolog}
This manuscript is based on the MA thesis written by the author under the supervision of professor Oscar Volij, as partial fulfillment of academic duties toward achieving second degree in Economics in the Department of Economics at Ben Gurion University. The subject matter is focused on the topic of quantum games, an emergent sub-discipline of physics and mathematics.  It has been developed rapidly during the last fifteen years, together with other similar fields, in particular quantum information to which it is intimately related. Even before being acquainted with the topic of quantum games the reader might wonder (and justly so) what is the relation between quantum games  and Economics . This research will not touch upon this interface, but numerous references relating quantum games and Economics will be mentioned. 
Similar questions arose in relation to the amalgamation of quantum mechanics and information science. If information is stored in our hard disk in bits, what has quantum mechanics to do with that? But in 1997 it was shown by Shor that by 
using quantum bits instead of bits, some problems that require a huge amount of time to be solved on ordinary computers could be solved in much shorter time using quantum computers. It was also shown that quantum computers can break secret codes in a shorter time than ordinary computers do, and that might affect our everyday life as for example, breaking our credit card security codes or 
affecting the crime of counterfeit money. Game theory is closely related with information science because taking a decision (like confess or don't confess in the prisoner dilemma game) is exactly like determining the state of a bit, 0 or 1. Following the crucial role of game theory in Economics, and the intimate relation between game theory and information science, it is then reasonable to speculate that the 
dramatic impetus achieved in information science due to its combination with quantum mechanics might repeat itself in the application of quantum game theory in Economics. 

As I stressed at the onset, the present work focuses on some aspects of quantum game theory, especially, quantum games based on simultaneous games with two players and two or three point strategic space for each player. The main effort is directed on the elucidation of pure strategy Nash equilibria in quantum games with full information and in games with incomplete information (Bayesian games).  I do not touch the topic of the interface between quantum games and economics, since 
this aspect is still in a very preliminary stage.  

Understanding  the topics covered in this work requires 
a modest knowledge of mathematics and the basic ingredients of quantum mechanics. Yet, the writing style is not mathematically oriented. Bearing in mind that 
the target audience is mathematically oriented economists,  I tried my best to explain and clarify every topic that appears to be unfamiliar to non-experts.  It seems to me that mathematically oriented economists will encounter no problem in handling this material. The new themes required beyond the central topics of mathematics used in economic science include complex numbers, vector fields, matrix algebra, 
group theory, finite dimensional Hilbert space and a tip of the iceberg of quantum mechanics. But all these topics are required on an elementary level, and are covered in the pertinent appendices.

\subsection{Background}
There are four scientific disciplines that seem to be intimately related. Economics, Quantum Mechanics, 
Information Science and Game Theory. The order of appearance in the above list  is chronological. The birth of Economics as an established scientific discipline  is about two hundred years old. Quantum mechanics has been initiated more than hundred years ago by Erwin Schr\"odinger, Werner Heisenberg, Niels Bohr, Max Born, Wolfgang Pauli, Paul Dirac and others. It has been
established as the ultimate physical  theory of Nature. 
The Theory of Information has been developed by Claude Elwood Shanon in 1949
\cite{Shannon}, and Game Theory has been developed by John Nash in 1951\cite{Nash}.  

The first connection between two of these four disciplines has been discovered in 1953 when  the science of game theory and its role in Economics has been established by von Newmann and Morgenstern \cite{von} (Incidentally, von Newmann laid the mathematical foundations of quantum mechanics in the early fifties). Almost half a century later, in 1997, the relevance of quantum mechanics for information  was established\cite{Shor} and that marked the birth of a new science, called {\it quantum information}.

These facts invite two fundamental questions: 1) Is quantum mechanics relevant for game theory? That is, can one speak of {\it quantum games} where the players use the concepts of quantum mechanics in order to design their strategies and payoff schemes? 2) If the answer is positive, is the 
  concept of quantum game relevant for Economics? 
   
The answer to the first question is evidently positive. In the last two and a half decades, the theory of quantum games has emerged as a new discipline in mathematics and physics and attracts the attention of many scientists. Pioneering works before the turn of the century 
include Refs. \cite{Wiesner, Ekert, Lev, Meyer}. The present work is inspired by some works published after the turn of the century that developed the concept of  quantum games that are based on standard (classical) games albeit with quantum strategies and a referee that imposes an entanglement\cite{EWL, Hayden, Flitney,  Piotrowski1, Flitney1,
 Landsburg1} and others. Quantum game theory combines game theory, that is, the mathematical formulation of competitions and conflicts, with the physical nature of quantum information. 
 
 The question why game theory can be interesting and what it adds to classical game theory was addressed in some of the references listed above.  Some of the reasons are: 
 \begin{enumerate} 
 \item{} The role of probability in quantum mechanics is rather fundamental. Since classical games also use the concept of probability, the interface between classical and quantum game theory promises to be conceptually rich. 
 \item{}  Since quantum mechanics is {\it the theory of Nature}, it must show up also in people mind when they communicate with each other. 
 \item{} Searching for quantum strategies in quantum game may lead to new quantum algorithms designed to solve complicated problems in polynomial time. 
 \end{enumerate} 
 
 The answer to the second question, the relevance of quantum game to economics  is less deterministic. Numerous works were published on this interface\cite{Piotrowski2}  and they give stimulus for further investigations. I feel however that this topics is still at a very early stage and requires a lot of new ideas and breakthroughs before it can be established as a sound scientific discipline. 
 
As I have already indicated,  the present thesis rests within the arena of quantum games and does not touch the interface between quantum games and economics. Its main achievement is the suggestion and the testing of a numerical method based on best response functions in the quantum game for searching pure strategy Nash equilibria. 
 \subsection{Content of Sections} 
 \begin{itemize} 
 \item{} In Section \ref{Section2} we cast the classical 2-player 2-strategies game in the language of classical information. Using the prisoner dilemma game as a guiding example we present the four positions on the game table 
 (C,C),(C,D),(D,C) and (D,D) as two bit states  (0,0),(0,1),(1,0) and (1,1) and define the classical strategies as operations on bits, that is known in the theory of information as classical gates. At the end of this Section we briefly discuss the information theory representation of 2-player three strategies classical games. 
 \item{} In Section \ref{Section3} All the quantum mechanical tools necessary for the conduction of a quantum game are introduced. These include a very short introduction to the concept of Hilbert space (discussed in more details in Section \ref{Section7}), followed by the definition of quantum bits, that is the fundamental unit of quantum information. Then the quantum strategies of the players are defined as unitary $2 \times 2$ complex matrices with unit determinant. The quantum states of a two players in a quantum game are then defined, and their relation to 
 the two qubit states is clarified. This leads us to the basic concept of entanglement and entanglement operators $J$ that play a crucial role in the protocol of the quantum game. In addition, the concept of partial entanglement is explained (as it will be used in Section \ref{Section5}). 
 \item{} Section \ref{Section4} is devoted to the definition of the quantum game, and its planning and conduction culminated in Fig.~\ref{Fig1}. The concept of pure strategy Nash equilibrium or a quantum game is defined and its relation to the degree of entanglement is explained. 
 \item{} In Section \ref{Section5} we introduce our numerical formalism to construct the best response functions and to search for pure strategy Nash equilibrium by identifying the intersections of the best response functions. 
 The method is then used on a specific game and the relation between the payoffs and the degree of entanglement is clarified. 
 \item{} In Section \ref{Section6} we briefly discuss more advanced topics such as Bayesian quantum games, 
 mixed strategies, quaternionic formulation of quantum games and quantum games based on two-players three decision classical games. These requires the introduction of quantum trits (qutrits) and the definition of strategies as $3 \times 3$ complex unitary matrices with unit determinant. 
 \item{} Finally, in Section \ref{Section7} we collect the minimum necessary mathematical apparatus in 
 a few appendices, including complex numbers, linear vector spaces, matrices, elements of group theory, introduction to Hilbert soace and, eventually, the basic concepts of quantum mechanics. 
 \end{itemize} 
 
\end{spacing}

\begin{spacing}{1.5}
\begin{center}
\section{\small Information Theoretic Language for Classical Games} \label{Section2}
\end{center}
The standard notion of games as appears in the literature will be referred to as 
a \underline {\bf classical games}, to distinguish it from the notion of \underline {\bf quantum games} 
that is the subject of this work. In the present Section we will use the language of 
information theory in the description of simultaneous classical games. Usually these games 
will be represented in their normal form (a payoff table). Except for the language used, nothing is new here. 
\subsection{Two Players - Two Decisions Games: Bits}
Consider a two player game with pure strategy such as the prisoner dilemma, given below in Eq.~(\ref{PD}). The formal 
definition is,  
\begin{equation} \label{0}
\Gamma=\la N=\{1,2\}, A_i=\{C,D \}, u_i: A_1 \times A_2 \to \mathbb{R} \ra~.
\end{equation}
Each player can choose between two strategies $C$ and $D$ for Confess or Don't Confess.  
Let us modify the presentation of the game just a little bit in order to adapt it to the nomenclature of quantum games. 
When the two prisoners appear before the judge, he tells them that he assumes that they both confess 
and let them decide whether to change their position or leave it at C. This modification does not 
affect the conduction of the game. The only change is that instead of choosing C or D as strategy, 
the strategy to be chosen by each player is either to {\it replace C by D} or {\it leave it C as it is}. Of course, if the judge would tell the prisoner that he assumes that prisoner 1 confesses and prisoner 2 does not, then the strategies will be different, but again, each one's strategy space has the two points \{ Don't replace, Replace \}. 

Now let us use different notations than C and D say 0 and 1. This has nothing to do with the numbers 0 and 1, they just 
stand for the two different symbols. We can equally consider two colors, red and blue. 
Such two symbols form a \underline {bit}. We thus have:\\
\underline{Definition:} A bit is an object that can have two different states. \\
%

%
A bit is the 
basic ingredient of information science and is used ubiquitously in numerous 
information devices such as hard disks, transmission lines and other information storage devices. 
There are several  notations used in information theory to denote the two states of a bit. The simplest one is just to 
say that the bit state is 0 or 1. But this notation is inconvenient when it is required to perform some operation on bits 
like {\it replace} or {\it don't replace}.  A more informative description is to consider bit states as two dimensional 
 vectors (see below). Yet a third notation that anticipates the formulation of quantum games is to denote the two states of a bit as $|0 \ra$ and $|1 \ra$. This {\em ket notation} might look strange at first glance but it proves very useful in analyzing quantum games. In summary we have, 
 \begin{spacing}{0.8}
\begin{equation} \label{1}
\mbox{bit state 0} = \begin{pmatrix} 1\\0 \end{pmatrix}=|0\ra, \ \  \mbox{ bit state 1} = \begin{pmatrix} 0 \\ 1 \end{pmatrix}=|1 \ra. 
\end{equation}
\end{spacing}
\subsubsection{Two Bit States} 
\vspace{-0.1in}
Looking at the game table in Eq.~(\ref{PD}), 
the prisoner dilemma game table has four squares marked by (C,C), (C,D),(D,C), and (D,D). In our modified language, 
any square in the game table is called a {\it two-bit state}, because each player knows what is his bit value 
in this square. The corresponding four two-bit states are denoted as (0,0),(0,1),(1,0), (1,1). 
In this notation (exactly as in the former notation with C and D) it is understood that the first symbol (from the left) belongs to player 1 and the second belongs to player 2. 

Thus, in our language, when the prisoners appear before the judge 
he tells them "your two-bit state at the moment is (0,0) 
and now I ask anyone to decide whether to replace his bit value from 0 to 1 or leave it as it is".  
As for the single bit states that have several equivalent
 notations specified in Eq.~(\ref{1}), two bit states have also several
different notations. 
  In the vector notation of Eq.~(\ref{1}) the four two-bit states listed above are 
obtained as \underline{outer products} of the two bits
\begin{spacing}{0.8}
\begin{equation} \label{2}
 \begin{pmatrix} 1\\0 \end{pmatrix} \otimes  \begin{pmatrix} 1\\0 \end{pmatrix}= \begin{pmatrix} 1\\0 \\0\\0 \end{pmatrix} , \ \  \begin{pmatrix} 1\\0 \end{pmatrix} \otimes  \begin{pmatrix} 0 \\1 \end{pmatrix}= \begin{pmatrix} 0\\1 \\0\\0 \end{pmatrix}, \ \ \begin{pmatrix} 0\\1 \end{pmatrix} \otimes  \begin{pmatrix} 1\\0 \end{pmatrix}= \begin{pmatrix} 0\\0 \\1\\0 \end{pmatrix} , \ \  \begin{pmatrix} 0\\1 \end{pmatrix} \otimes  \begin{pmatrix} 0 \\1 \end{pmatrix}= \begin{pmatrix} 0\\0 \\0\\1 \end{pmatrix}~.
 \end{equation}
 \end{spacing}
 \ \\
Again, it is understood that the bit composing the left factor in the outer product belongs to player 1 (the column player) and the the right factor in the outer product belongs to player 2 (the row player).  
 Generalization to $n$ players two-decision games is straightforward. A set of $n$ bits can exist in one of $2^n$ different configurations and 
 described by a vector of length $2^n$ where only one component is 1, all the others being 0. \\
 {\bf Ket notation for two bit states:} The vector notation of Eq.~(\ref{2}) requires a great deal of page space, 
 a problem that can be avoided by using the ket notation. In this framework, the four two-bit states are respectively denoted as 
 (see the comment after after Eq.~(\ref{2})),
 \begin{equation} \label{2a}
 |0\ra \otimes |0 \ra=|00 \ra, \ \  |0\ra \otimes |1 \ra=|01 \ra, \ \  |1 \ra \otimes |0 \ra=|10 \ra, \ \  |1 \ra \otimes |1 \ra=|11 \ra.
 \end{equation} 
 For example, in the prisoner dilemma game, these four states correspond respectively to
 \\
  $(C,C),(C,D),(D,C),(D,D)$. 
 \subsubsection{Classical Strategy as an Operation on Bits}
 Now we come to the description of the classical strategies (replace or do not replace) using our 
 information theoretic language. 
Since we have agreed to represent bits as two components vectors, execution of operation of each player on his own bit (replace or do not replace) is represented by a $2 \times 2$ real matrix. 
In classical information theory, operations on bits are referred to as {\it gates}. Here we will be concerned with the two simplest operations  performed on bits changing them from one configuration to another. An operation on a bit state that results in the same bit state is accomplished by the unit $2 \times 2$ matrices ${\bf 1}=\binom{1 \ 0 }{0 \ 1}$. An operation on a bit state that results in the other bit state is accomplished by a $2 \times 2$ matrix denoted as 
$Y \equiv\binom {~~0 \ 1}{-1 \  0}$. 
\tboxed{\underline{An important notational comment:} The -1 in the matrix $Y$ is designed to guarantee that det$[Y]$=1, in analogy with the strategies of the quantum game to be defined in the following Sections. As far as the classical game is concerned, this sign has no meaning, because a bit state $|0 \ra$ or $|1 \ra$ is not a number, it is just a symbol. 
 So that we can agree that for classical games, the vectors $\binom {1}{0}$ and $\binom {-1}{0}$ represent the {\it same} bit, $|0 \ra$ and the vectors $\binom {0}{1}$ and $\binom {0}{-1}$ represent the {\it same} bit, $|1 \ra$
 }
\begin{spacing}{0.9}
\begin{equation} \label{3} 
\begin{pmatrix} 1 & 0 \\ 0 & 1\end{pmatrix} \binom {1} {0} = \binom {1} {0}, \ \ \begin{pmatrix} 1 & 0 \\ 0 & 1\end{pmatrix} \binom {0} {1} = \binom {0} {1}, \ \
\begin{pmatrix} 0 & 1 \\ -1 & 0\end{pmatrix}   \binom {1} {0} = \binom {0} {-1}, \ \ \begin{pmatrix} 0 & 1 \\ -1 & 0\end{pmatrix} \binom {0} {1} = \binom {1} {0}~. 
\end{equation}
\end{spacing}
\ \\
Written in ket notation we have,
\begin{equation} \label{3a}
{\bf 1}|0\ra=|0 \ra, \ \ {\bf 1}|1\ra=|1 \ra, \ \ Y |0\ra=|1 \ra, \ \ Y |1 \ra=|0 \ra~.
\end{equation}
\tboxed{
In the present language, the two strategies of each player are the two $2 \times 2$ matrices 
${\bf 1}$ and ${Y}$ and the four elements of $A_1 \times A_2$ are 
the four $4 \times 4$ matrices, 
\begin{spacing}{0.5}
\begin{equation} \label{3b}
{\bf 1} \otimes {\bf 1}, \ \ {\bf 1} \otimes Y, \ \ Y \otimes {\bf 1}, \ \ Y \otimes Y~.
\end{equation}
\end{spacing}
}
In this notation, following the comment after Eq.~(\ref{2}), the left factor in the outer product  is executed by player 1 (the column player) on his bit, while the right factor in the outer product  is executed by player 2 (the row player).  In matrix notation each operator listed in Eq.~(\ref{3b}) acts on a four component vector as listed in Eq.~(\ref{2}). \\
\underline{Example:} Consider the classical prisoner dilemma with the normal form,
\begin{center}  $~~~~~~~~~~~~~~~~~~$Prisoner 1
\end{center}
\begin{equation} \label{PD}
 \mbox{
  $~~~~~~$ Prisoner 2$~~$   \\
\begin{tabular}{|l| |l| |l| |l| |l| |l|} \hline  &~ {\bf 1} (C) ~  & ~Y (D) \\ \hline ~{\bf 1} (C)~ & -4,-4 & -6,-2 \\  \hline ~Y (D)~ & -2,-6 & -5,-5 \\  \hline \end{tabular} 
}
\end{equation}
The entries stand for the number of years in prison. 
\subsubsection{Formal Definition of a Classical Game  in the Language of Bits}
\begin{figure}[!h]
\centering
\includegraphics[width=8truecm, angle=0]{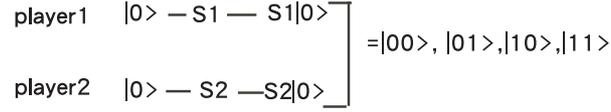}
\caption{\footnotesize{ A general protocol for a two players two strategies classical game showing the flow of information. To be followed on the figure from left to right. 
Here $S_1=I, Y$ and similarly $S_2=I, Y$. There are only four possible finite states of the system. }}
\label{Fig0}
\end{figure}

The formal 
definition is,  
\begin{equation} \label{0B}
G_C=\la N=\{1,2\}, |ij\ra, A_i=\{{\bf 1},Y \}, u_i: A_1 \times A_2 \to \mathbb{R} \ra~.
\end{equation}
The two differences between this definition and the standard definition of Eq.~(\ref{0}) is that 
the players face an initial two-bit state $|ij\ra \ \ i,j=0,1$ presumed by the judge (usually $|00\ra=(C,C)$ 
and the two-point strategy space of each players contains the two gates $({\bf 1},Y)$ instead of $(C,D)$.  
  The conduction of a pure strategy 
classical two-players-two strategies simultaneous game 
given in its normal form (a $2 \times 2$ payoff matrix) follows the following steps:
\begin{enumerate}
\item{}  A referee declares that the initial configuration is some fixed 2 bit state. This initial state is one of the four 2-bit states listed in Eq.~(\ref{2a}). The referee's choice does not, in any way, affect the final outcome of the game, it just serves as a starting point. For definiteness assume that the referee suggests the state $|00 \ra$ as the initial state of the game. We already gave an example:
In the story of the prisoner dilemma it is like the judge telling them that he assumes that they both confess.  
\item{} In the next step, each player decides upon his strategy
(${\bf 1}$ or $Y$) to be applied on his respective bit. 
For example, 
 if each player choses the strategy $Y$ we note from Eq.~(\ref{3}) that 
 \begin{equation} \label{twobit}
Y \otimes Y |00 \ra=Y|0 \ra \otimes Y|0 \ra=|1 \ra \otimes |1 \ra=|11 \ra=|DD \ra.
\end{equation} 
Thus, a player can choose either to leave his bit as suggested by the referee or to change it to the second possible state.  
As a result of the two operations, the two bit state assumes it final form. 
\item{} The referee then ``rewards" each players according to sums appearing in the corresponding payoff matrix. Explicitly, \\
$~~~~~~~~~~~~~~~~~~~~~u_1({\bf 1},{\bf 1})=u_2({\bf 1},{\bf 1})=-4, \ u_1({\bf 1},Y)=u_2(Y ,{\bf 1})=-6, \\ 
~~~~~~~~~~~~~~~~~~~u_1(Y,{\bf 1})=u_2({\bf 1},Y )=-2, \ 
u_1(Y,Y)=u_2(Y,Y)=-4. $
\end{enumerate}
The  procedure described above is schematically shown in Fig.~\ref{Fig0}. 
\ \\
 A pure strategy Nash equilibrium (PSNE) 
 is a pair of strategies $S^*_1,S^*_2 \in \{ {\bf 1}, Y \}^2$ such that 
 \begin{eqnarray}
&& u_1(S_1,S_2^*) \le u_1(S^*_1,S_2^*) \ \forall S_1 \ne S_1^* \nonumber \\
&& u_2(S^*_1,S_2) \le u_2(S^*_1,S^*_2) \ \forall S_2 \ne S_2^*.
\end{eqnarray}
 In the present example, it is easy to check that, given the initial state $|00 \ra$ from the referee,  the pair of strategies leading to NE is $(S_1^*,S_2^*)=Y \otimes Y$. However, this equilibrium is not {\it Pareto efficient}, namely there is a strategy set $S_1,S_2$ such that $u_i(S_1,S_2) \ge u_(S_1^*,S_2^*)$ for $i=1,2$. In the present example 
 the strategy set $I \otimes I$ leaves the system in the state $|00\ra$ and 
 $u_i({\bf 1},{\bf 1})$=$-4>u_i(Y,Y)$=$-5$.\\

\subsubsection{Mixed Strategy in the Language of Bits} 
This technique of operation on bits is naturally extended to treat, mixed strategy games. 
 Then by operating 
on the bit state $\binom{1}{0}$ by the matrix $p {\bf 1}+(1-p) Y$ with $p \in [0,1]$, 
we get the vector, 
\begin{spacing}{0.8}
\begin{equation} \label{4}
\left [p \begin{pmatrix} 1 & 0 \\ 0 & 1 \end{pmatrix}+(1-p) \begin{pmatrix} 0 & 1 \\ 1 & 0 \end{pmatrix} \right ]\begin{pmatrix} 1\\0 \end{pmatrix} =  \begin{pmatrix} p \\ 1-p \end{pmatrix},
\end{equation}
\end{spacing}
that can be interpreted as a mixed strategy of choosing pure strategy $|1 \ra$  with probability $p$ and pure strategy $|0 \ra$ with probability $1-p$. Following our example, assuming 
player 1 choses ${\bf 1}$ with probability $p$ and $Y$ with probability $1-p$ 
and player 2 choses ${\bf 1}$ with probability $q$ and $Y$ with probability $1-q$ the 
combined operation on the initial state $|00 \ra$ is, \\
$[p {\bf1}+(1-p) Y] \otimes [q {\bf1}+(1-q) Y] |00\ra=pq|00\ra+p(1-q)|01 \ra+(1-p)q|10\ra+(1-p)(1-q)|11 \ra$. 
\end{spacing}
\begin{spacing}{1.5}
\begin{center}
\section{\small The Quantum Structure: Qubits} \label{Section3}
\end{center}
In quantum mechanics, the analog of a bit is a {\it quantum bit}, briefly referred to as {\it qubit}.  Physically, this is a {\it two level system}. The most simple example is the two spin states of an electron. In order to explain this concept we need to carry out some preparatory work.  
\footnote{For understanding this section, the reader is assumed to have gone through the Appendix on Quantum Mechanics.}
\subsection{Two Dimensional Hilbert Space}
As discussed in the Appendix \ref{HS}, a Hilbert space ${\cal H}$ is a 
linear vector space above the field of complex numbers, see Appendix \ref{CN}. 
The dimension of a Hilbert space is the maximal number 
of linearly independent vectors  belonging to ${\cal H}$. 
A Hilbert space might have any dimension, 
including infinite. In quantum information we mainly encounter finite 
dimensional Hilbert spaces. In quantum games the dimension of Hilbert space pertaining to a given player $i$ is equal to the number of his {\it classical} strategies. One of the simplest cases relevant to game theory is a classical game with two players-two decisions game. Therefore, for the time being 
we will be concerned with two-dimensional 
Hilbert space, denoted as ${\cal H}_2$. 
As we learn from Appendix \ref{HS},  we can define a set of two linearly independent orthogonal vectors (kets) in ${\cal H}_2$ denoted as $|0 \ra , |1 \ra \in {\cal H}_2$. 
The fact that the notation of basis states is the same as that used for bits is 
of course not accidental. 

An arbitrary state (or vector) $|v \ra \in {\cal H}_2$ 
is written as  $|v \ra = a|0 \ra+b|1 \ra$. As we also recall from 
Appendix \ref{HS} the Hilbert space ${\cal H}_2$ is endowed with an 
inner product, that is, a mapping $F: {\cal H}_2 \times {\cal H}_2 \to \mathbb{C}$ 
written as $f(|u \ra,| v \ra)=\la u | v \ra =\la v|u \ra^* \in \mathbb{C}$. 
The basis states have the following properties, 
\begin{enumerate} 
\item{} \underline{Orthogonality and normalization:} $ \la 1|0\ra=\la 0|1 \ra=0,  \ \ \la 0|0 \ra=\la 1|1 \ra=1$. 
\item{} \underline{Linear independence} If $a,b \in \mathbb{C}$ (namely, they are complex numbers, see Appendix \ref{CN}), then $a|0\ra+b|1 \ra={\bf 0} \ \mbox{(the zero vector)} \ \Leftrightarrow \ a=b=0$. 
\item{} \underline{Expanding vectors:} Every vector (state) $|v \ra \in {\cal H}_2$ can be expressed as a linear combination \\
 $|v \ra=a|0\ra+b|1 \ra$, with $a=\la 0|v \ra, \ \ b=\la 1|v \ra \in \mathbb{C}$. \\
 The last equality is obtained by performing the inner products $\la 0|v \ra$ and 
 $\la 1|v \ra$ and using the orthogonality of the bases states discussed in item 1. A more concrete way to say it is that we ``multiply" the two sides of the expression 
 $|v \ra=a|0\ra+b|1 \ra$ on the left once by $\la 0|$ and once by $\la 1|$. This show the power of the Dirac notation. 
\end{enumerate} 
\subsection{Qubits}
The quantum bit (shortly qubit) is the basic unit of quantum information, in the same token that bit is the basic unit of classical information. While the notion of bit is familiar to anyone who has a basic knowledge in information storage (on a hard disk for example) and information transfer, the notion of qubit is much less familiar. Until a few years ago it could be argued that qubit are simple quantum system that
cannot be used in such discipline as information science, 
economics, computational resources and cryptography. This is definitely not the case nowadays as the fields of quantum information and quantum computation become 
closer and closer to reality. For economists, in general, and for game theorists in particular, the concept of qubit 
requires some change of mind in the sense that a decision 
(a strategy) is not simple {\it yes or no} (for pure strategy) or simple 
{\it yes with probability $p$ and no with probability $(1-p)$}. 
Similar to the classical game, where a decision is an operation on bits (see Eq.~(\ref{3a}) a strategy is an operation on qubit. However, since a qubit has a much richer structure than a bit, a quantum strategy is much richer than a classical one. But before speaking of quantum games and quantum strategy we need to define the basic unit (like the hydrogen atom in chemistry).
\subsection{Definition and Manipulation of Qubits}
Now we come to the central definition:
\tboxed{
\underline {Definition} A qubit is a vector $|\psi \ra = a|0\ra+b|1 \ra \in {\cal H}_2, \ \ a,b \in \mathbb{C}$  
 such that $|a|^2+|b|^2=1$. \\
 \underline{ The collection $\Sigma \equiv \{ |\psi \ra \}$ of all qubits is a set and not a space} (the vector sum of two qubits is, in general, not a qubit, and hence it has no meaning in what follows). 
 The cardinality of the set of qubits is hence $\aleph$ (recall that there are 
 only two bits). Two qubits $| \psi \ra$ and $e^{i \phi}| \psi \ra, \ \ \phi \in \mathbb{R}$ that differ by a unimodular 
factor $e^{i \phi}$ (see appendix \ref{CN}) are considered identical. This is called {\em phase freedom}. 
 }
 A convenient way to underline the difference between  bits and qubits  is to write them as vectors,  
 \begin{spacing}{0.8}
\begin{equation} \label{5}
\mbox{bit state 0} = \begin{pmatrix} 1\\0 \end{pmatrix}, \ \  \mbox{ bit state 1} = \begin{pmatrix} 0 \\ 1 \end{pmatrix}, \ \ \mbox{ qubit=} a\begin{pmatrix} 1\\0 \end{pmatrix}+b\begin{pmatrix} 0 \\ 1 \end{pmatrix}=\begin{pmatrix} a \\ b \end{pmatrix}, \ \ |a|^2+|b|^2=1~.
\end{equation}
\end{spacing}
\ \\
Another standard notation is to write the basis states in terms of arrows. The three notations 
\begin{spacing}{0.8}
\begin{equation} \label{5a}
|0 \ra=\binom{1}{0}= |\ua \ra, \ \ |1 \ra=\binom{0}{1}=|\da \ra
\end{equation}
\end{spacing}
 are in use.   The arrow notation is borrowed from physics where the two directions represents the two orientations of an electron's spin. 
Thus, all the definitions used below to denote a qubit are equivalent,
\tboxed{
\vspace{-0.1in}
\begin{equation} \label{6}
|\psi \ra=a|0 \ra+b |1 \ra=a|\ua \ra+b |\da \ra:=a\binom{1}{0}+b \binom{0}{1}:=\binom{a}{b}, \ \ |a|^2+|b|^2=1~,
\end{equation}
}
where $:=$ means, literally, {\it can also be written as}.

\begin{figure}[!h]
\centering
\includegraphics[width=6truecm, angle=0]{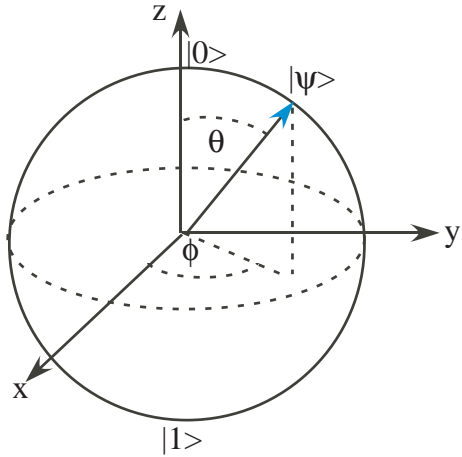}
\caption{\footnotesize{ A qubit $|\psi \ra$ is represented as a point (a tip of an arrow) on the Bloch sphere.}}
\label{Bloch}
\end{figure}

The number of degrees of freedom (parameters) 
of a qubit is 2 (two complex numbers with one constraint combined with the phase freedom). 
The phase freedom 
allows us to chose $a$ to be real and positive. An elegant way to represent a qubit is by choosing two angles $\theta$ and $\phi$ such that $a$=$\cos(\theta/2), b$=$e^{i \phi} \sin(\theta/2), 0 \le \theta \le \pi, \ \ 0 \le \phi \le 2 \pi$:
\begin{equation} \label{7}
\binom{a}{b}=    [\cos(\theta/2) | \!  \!  \uparrow \rangle + e^{i \phi}
    \sin(\theta/2) | \!  \!  \downarrow \rangle ] =\begin{pmatrix} \cos(\theta/2) \\ e^{i \phi} \sin(\theta/2)
   \end{pmatrix}
   \end{equation}
   \ \\
   The two angles $\theta$ and $\phi$ determine a point on the unit sphere (globe) with Cartesian coordinates, 
   \begin{equation} \label{8}
   x=\sin \theta \cos \phi, \ \ y=\sin \theta \sin \phi, \ \ z=\cos \theta, \ \ x^2+y^2+z^2=1~.
   \end{equation}
   Therefore, every point on the unit sphere with spherical angles $(\theta, \phi)$ uniquely define a qubit 
  $\binom {a}{b}$  according to Eq.~(\ref{7}). In physics this construction is referred to as {\em Bloch Sphere}, as displayed in Fig.~\ref{Bloch}. 
    In particular, the north pole $\theta=0$ corresponds to $|0 \ra=|\ua \ra=\binom {1}{0}$ and the south pole, $\theta=\pi$ corresponds to $|1\ra=|\da \ra=\binom {0}{1}$. 

  \subsection{Operations on a Single Qubit: Quantum Strategies}
 In Eq.~(\ref{3a})  and (\ref{3b}) we defined two classical strategies, ${\bf 1}$ and $Y$ 
 as operations on bits. According to Eq.~(\ref{3}) they are realized by $2 \times 2$ matrices ${\bf 1}=\binom {10}{01}, \ \ Y=\binom {~~0 1}{-1 0}$ and act on the bit vectors $|0\ra=\binom{1}{0}$ and 
 $|1\ra=\binom{0}{1}$.  
 In this subsection we develop the quantum analogs: We are interested in operations on qubits, (also referred to as single qubit quantum gates) that transform a qubit $|\psi \ra=a|0\ra+b|1 \ra$ into another qubit $a'|0 \ra+b'|1 \ra$. 
 
 There are some restrictions on the allowed operations on qubits. 
 First, a qubit is a vector in two dimensional Hilbert space and therefore, operations on 
 a single qubit must be realized by $2 \times 2$ complex matrices. Second, we have seen in Fig.~\ref{Bloch} that a qubit is a point on a point on the Bloch sphere and therefore, the new qubit must have 
 the same unit length (the radius of the Bloch sphere). 
 In other words the unit length of a qubit must be conserved under any operation. From what we 
 learn from Appendix \ref{Mat}, this means that
  any allowed operation on a qubit  is defined by 
  a {\it unitary $2 \times 2$ matrix}  $U$. 
  In the notation of Eq.~(\ref{5}) a unitary operation on a qubit represented as a two component vector $\binom {a}{b}$ is defined as, 
  \begin{spacing}{0.8}
  \begin{equation} \label{9}
  U \begin{pmatrix} a\\b \end{pmatrix}=\begin{pmatrix} U_{11}a+U_{12}b \\ U_{21}a+U_{22}b \end{pmatrix} \equiv \begin{pmatrix} a' \\b' \end{pmatrix}, \ \ |a|^2+|b|^2=|a'|^2+|b'|^2=1.
  \end{equation}
  \end{spacing}
  \ \\
    For reasons to become clear later on we will restrict ourselves to unitary transformations $U$ with unit determinant, Det[U]=1. The collection of all $2 \times 2$ unitary matrices with unit determinant, 
   form a group under the usual rule of matrix multiplication.   This is the $SU(2)$ group (see Appendix \ref{GT} on group theory), that plays a central role in physics as well as in abstract group theory. 
The most general 
  form of a matrix $U \in SU(2)$ is, 
  \begin{equation} \label{11}
  U(\phi,\alpha,\theta)=\begin{pmatrix} e^{i \phi}\cos \tfrac{\theta}{2} &  e^{i \alpha}\sin \tfrac{\theta}{2} \\ - e^{-i \alpha}\sin \tfrac{\theta}{2}&  e^{-i \phi}\cos \tfrac{\theta}{2} \end{pmatrix}~, \ 0 \le \phi, \alpha \le 2 \pi, \ \ 0 \le \theta \le \pi~.
  \end{equation}
 Although we have not yet defined the notion of quantum game, 
 we assert that,  in analogy with Eq.~(\ref{3a} (that defines player's classical strategies as operations on bits), the  operation on qubits 
  (such that each player acts with his $2 \times 2$ matrix on his qubit),  is an implementation of each player's quantum strategy. Thus, 
  \tboxed{
\underline{Definition} In quantum games, the (infinite number of) quantum strategies of each player $i=1,2$ is the infinite set of his $2 \times 2$ matrices $U(\phi_i, \alpha_i, \theta_i)$ as  defined in Eq.~(\ref{11}).  
The infinite collection of these matrices form the group SU(2) of unitary $2 \times 2$ matrices with unit determinat. Since the functional form of the matrix $U(\phi,\alpha,\theta)$ is given by Eq.~(\ref{11}), the strategy of player $i$ is determined by his choice of the three angles ${\bm \gamma}_i=(\phi_i,\alpha_i,\theta_i)$. Here ${\bm \gamma}_i$ is just a short notation for the three angles. The three angles $\phi, \alpha, \theta$ are referred to as the {\it Euler angles}. 
}
The quantum strategy specified by the $2 \times 2$ matrix $U(\phi, \alpha, \theta)$ as specified above has a geometrical interpretation. This is similar to the geometrical interpretation given to qubit as a point on the Bloch sphere in Fig.~\ref{Fig0}, where the two angles $(\phi,\theta)$ determine a point on the boundary of a sphere of unit radius in {\it three dimensions}. Such  a (Bloch) sphere, is 
a two dimensional surface denoted by $S^2$. On the other hand, the three angles $\phi, \alpha, \theta$ 
defining a quantum strategy determine a point on the surface of the unit sphere in {\it four dimensional} space, $\mathbb{R}^4$ (the 4 dimensional Euclidean space). The unit sphere is in this space is defined as the collection of points with Cartesian coordinates $(x,y,z,w)$ restricted by the equation $x^2+y^2+z^2+w^2=1$. This equality defines the surface of a three dimensional sphere denoted by $S^3$ (impossible to draw a figure). The equality is satisfied 
by writing the four Cartesian coordinates as, 
\begin{equation} \label{33}
 x=\sin \theta \sin \phi \cos \alpha, \ \  y=\sin \theta \sin \phi, \sin \alpha, \ \ z=\sin \theta \cos \phi, \ w=\cos \theta~. 
 \end{equation}
 An alternative definition of a player's strategy is therefore as follows:
\tboxed { \underline{Definition} A strategy of player $i$ in a quantum analog of a two-players two-strategies classical game is 
a point $ {\bm \gamma}_i=(\phi_i, \alpha_i, \theta_i) \in S^3$}
Thus, instead of a single number $0$ or $1$ as a strategy of the classical game, the set of quantum strategies has a cardinality $\aleph^3=\aleph$. 
\subsubsection{Classical Strategies as Special Cases of Quantum Strategies}
A desirable property from a quantum game is that the players can reach also their classical 
strategies. Of course, the interesting case is that reaching the classical strategies does not 
lead to Nash equilibrium, but the payoff awarded to players in a quantum game that use their classical strategies serve as a useful reference point.  Therefore, we ask the question whether, by an appropriate choice of the three angles $(\phi,\alpha,\theta)$ the quantum strategy $U(\phi,\alpha,\theta)$ is reduced to one of the two classical strategies ${\bf 1}$ or $Y$. First, it is trivially seen that 
$U(0,0,0)={\bf 1}$. 
It is now clear  why we chosen the classical  strategy that flips the state of a bit as $Y=\binom{01}{-1 0}$ and not as $\sigma_x=\binom{01}{10}$, because 
Det[$U(\phi,\alpha,\theta)$]=1$\forall \phi, \alpha,\theta$ whereas Det[$\sigma_x$]=-1. On the other hand, 
we notes that $U(0,0,\pi)=\binom{~~0~ 1}{-1~ 0}= Y$.  The quantum game procedure to be described in the next Section is such that the difference between $\sigma_x$ and $Y$ does not affects the payoff at all, and therefore, we may conclude that the classical strategies are indeed, obtained as special cases of the quantum strategies, 
\begin{spacing}{0.8}
\begin{equation} \label{quantclass}
U(0,0,0)={\bf 1}=\begin{pmatrix} 1&0\\0&1\end{pmatrix}, \ \ U(0,0,\pi)=Y=\begin{pmatrix} 0&1\\-1&0 
\end{pmatrix}.
\end{equation}
\end{spacing}
  
\subsection{Two qubit States} 
In Eqs. (\ref{2}) and (\ref{2a}) we represented two-bit states as tensor products 
of two one-bit states. Equivalently, a two-bit state is represented by a four dimensional 
vector, three of whose components are 0 and one component is 1 see Eq.~(\ref{2}). 
Since each bit can be found in one of two states $|0 \ra$ or $|1 \ra$  there are exactly four two-bit states. With two-qubit states, the situation is dramatically different in two respects. First, as noted in connection with Eq.~(\ref{6}), each qubit $a|0 \ra+b |1 \ra$ with 
$|a|^2+|b|^2=1$ can be found in an infinite number of states. 
This is easily understood by noting that, according to Eq.~(\ref{7}) 
and Fig.~\ref{Bloch}, each qubit is a point on the two-dimensional (Bloch) sphere. 
Accordingly, once we construct two-qubit states by tensor products of two one-qubit states we expect a two-qubit state to be represented by a four dimensional vector of complex numbers. Second, and much more profound, there are four dimensional vectors that are {\it not} represented as a tensor product of two 
two-dimensional vectors. Namely, in contrast with the classical two-bit states, there are two-qubit states that are not represented as a tensor product of two one-qubit states. This is referred to as {\it entanglement} and will be explained further below.   
In a two-players two-strategies classical game, 
 each player has its own bit upon which he can operate (namely, chose his strategy). Below we shall define a quantum game that is
 based on two-player two-strategies classical game. In such game, 
each player has its own qubit upon which he can operate by an $SU(2)$ matrix $U(\phi,\alpha,\theta)$ (namely, chose his quantum strategy). \\
\subsubsection{ Outer (tensor) product of two qubits} 
In analogy with Eq.~(\ref{2}) that defines the 4 two-bit states we 
define an outer (or tensor) product  of two qubits 
$|\psi_1 \ra \otimes |\psi_2 \ra \in {\cal H}_2 \otimes {\cal H}_2$ 
using the notation of Eq.~(\ref{5}) as follows: Let $|\psi_1 \ra=a_1|0  \ra+b_1 |1  \ra$ and $|\psi_2 \ra=a_2|0  \ra+b_2 |1  \ra$ be two qubits numbered 1 and 2. We define their outer (or tensor) product as, 
\begin{eqnarray}
&&|\psi_1 \ra \otimes |\psi_2 \ra=(a_1|0  \ra+b_1 |1  \ra) \otimes (a_2|0  \ra+b_2 |1  \ra)
\nonumber \\
&&=a_1a_2 |0  \ra\otimes |0  \ra+a_1b_2 |0  \ra\otimes |1  \ra+b_1a_2 |1  \ra\otimes |0  \ra+
b_1b_2 |1  \ra\otimes |1  \ra \in {\cal H}_2 \otimes {\cal H}_2~.  
 \label{12} 
\end{eqnarray}  
 In terms of 4 component vectors, the tensor products of the elements such as $|0  \ra\otimes |0  \ra$ 
are the same as the two-bit states defined in Eq.~(\ref{2}), and therefore, in this notation we have, 
 \begin{spacing}{0.8}
\begin{equation} \label{13}
|\psi_1 \ra \otimes |\psi_2 \ra=a_1a_2  \begin{pmatrix} 1\\0 \\0\\0 \end{pmatrix}+a_1b_2  \begin{pmatrix} 0\\1 \\0\\0 \end{pmatrix}+b_1a_2  \begin{pmatrix} 0\\0 \\1\\0 \end{pmatrix}+b_1b_2 \begin{pmatrix} 0\\0 \\0\\1 \end{pmatrix}=\begin{pmatrix} a_1a_2\\a_1b_2 \\b_1a_2\\b_1b_2 \end{pmatrix}~.
\end{equation} 
\end{spacing}
A tensor product of two qubits as defined above is an example of a {\em two qubit state}, briefly referred to as \underline {2qubits}. 
The coefficients of the four products in Eq.~(\ref{12}) (or, equivalently, the four vectors in Eq.~(\ref{13})), 
 are complex numbers referred to as {\em amplitudes}. Thus, we say that the amplitude of $|0  \ra\otimes |0  \ra$ in the 2qubits $|\psi_1 \ra \otimes |\psi_2 \ra$ is $a_1 a_2$ and so on. 
Using simple trigonometric identities it is easily verified that the sum of the coefficients is 1, namely,
\begin{equation} \label{14}
|a_1a_2|^2+|a_1b_2|^2+|b_1a_2|^2+|b_1b_2|^2=1~.
\end{equation}
2qubits can also be related to a Bloch sphere (but we will not do it here). 

We have seen in Eq.~(\ref{13}) that a tensor product of two qubits is a 2qubits that is written as a linear combination of the four basic 2qubit states
\begin{equation} \label{15}
|0  \ra \otimes |0  \ra \equiv |00 \ra, \ \ |0  \ra \otimes |1  \ra \equiv |01 \ra, \ \ 
|1  \ra \otimes |0  \ra \equiv |10 \ra, \ \ |1  \ra \otimes |1  \ra \equiv |11 \ra,
\end{equation}
From the theory of Hilbert spaces we know the the 2qubits defined in Eq.~(\ref{15} form a basis in ${\cal H}_4$.\\ 
This bring as to the following 
\tboxed{
\underline{Definition:} A general 2qubits $|\Psi \ra \in {\cal H}_4$ has the form, \\
$~~~~~~~~~~~~~~~~~~~~~~ |\Psi \ra = a|0  \ra \otimes |0  \ra+b|0  \ra \otimes |1  \ra+c|1  \ra \otimes |0  \ra+d|1  \ra \otimes |1  \ra$
\vspace{-0.2in}
\begin{equation} \label{2qubits}
=a|00\ra+b|01\ra+c|10 \ra+d|11 \ra, 
\ \ \mbox{with} \ \ |a|^2+|b|^2+|c|^2+|d|^2=1.
\end{equation}
}
Note the difference between this expression and the outer product of two qubits as 
defined in Eq.~(\ref{12}), in which the coefficients are certain products of the 
coefficients of the qubit factors. In the expression (\ref{2qubits}) the coefficients are 
arbitrary as long as they satisfy the normalization condition. Therefore, Eq.~(\ref{12}) is a special case of (\ref{2qubits}) but not vice-versa. This observation leads us naturally to the next topic, that is, entanglement. 
\subsection{Entanglement}
 Entanglement is one of the most fundamental concepts in quantum information and in quantum game theory. 
In order to introduce it we ask the following question: Let
\begin{equation} \label{16}
 |\Psi \ra = a|0 0\ra+b|01 \ra+c|1 0 \ra+d|11 \ra, \ \ \mbox{with} \ \ |a|^2+|b|^2+|c|^2+|d|^2=1,
 \end{equation}
 as already defined in Eq.~(\ref{2qubits}) 
denote a general 2qubits. Is it \underline {always} possible to represent it as a tensor product of two single qubit states as in Eqs.~(\ref{12}) or (\ref{13}) ?? The answer is \underline{NO}. Few counter examples with two out of the four coefficients set equal to 0 are, 
\begin{equation} \label{17}
|T \ra \equiv \frac{1}{\sqrt{2}} (|01 \ra+|10 \ra), \ \ |S \ra \equiv \frac{1}{\sqrt{2}} (|01 \ra-|10 \ra), \ \ |\psi_\pm \ra \equiv \frac{1}{\sqrt{2}} (|00 \ra\pm i |11 \ra)~. 
\end{equation}
where the notations T=triplet and S=singlet are borrowed from physics. 
These four 2qubits are referred to as {\em maximally entangles Bell states}. 
We now have,
\tboxed{
\underline{Definition} 
A 2qubits $|\Psi \ra$ as defined in Eq.~(\ref{16}) is said to be \underline {\bf entangled} iff it {\it cannot} be represented as a tensor product of two single qubit states as in Eqs.~(\ref{12}) or (\ref{13}).
}
Entanglement is a pure quantum mechanical effect that appears in manipulating 2qubits. It does not occur in manipulations of bits. There are only four 2bit states as defined in Eq.~(\ref{2}), all of them 
are obtained as tensor products of single bit states, so that by definition they are not entangled. 
The concept of entanglement is of utmost importance in many aspects of quantum mechanics. It led 
to a very long debate initiated by a paper written in 1935 by Albert Einstein, Boris Podolsky and Nathan Rosen  referred to as the EPR paradox that questioned the completeness of quantum mechanics. 
The answer to this paradox was given by John Bell in 1964. Entanglement plays a central role in 
quantum information. Here we will see that it also plays a central role in quantum game theory. 
Strictly speaking, without entanglement, quantum game theory reduces to the classical one. \\
\subsection{ Operations on 2qubits (2qubits Gates)} 
An important tool in manipulating 2qubits are operations transforming one 2qubits to another. 
Borrowing from the theory of quantum information these are called two-qubit gates. Writing a general 2qubits as defined in Eq.~(\ref{16}) in terms of its 4 vector of coefficients, 
 \begin{spacing}{0.8}
\begin{equation} \label{18}
|\Psi \ra= a|0 0\ra+b|01 \ra+c|1 0 \ra+d|11 \ra =  \begin{pmatrix} a\\b\\c\\d \end{pmatrix}~,
\end{equation}
\end{spacing}
\noindent
a 2-qubit gate is a unitary $4 \times 4$ matrix (with unit determinant) acting on the 4 vector of coefficients, in analogy with Eq.~(\ref{9}),
\begin{spacing}{0.7}
\begin{equation} \label{19}
{\cal U}  \begin{pmatrix} a\\b\\c\\d \end{pmatrix}= \begin{pmatrix} a'\\b'\\c'\\d' \end{pmatrix}~, \ \ 
{\cal U} \in SU(4), \ \ |a|^2+|b|^2+|c|^2+|d|^2=|a'|^2+|b'|^2+|c'|^2+|d'|^2=1~. 
\end{equation}
\end{spacing}
\vspace{0.2in}
In the same token as we required the matrices $U$ operating on a single qubit state to have unit determinant, that is $U \in SU(2)$, we require ${\cal U}$ also to have a unit determinant, that is, ${\cal U} \in SU(4)$, the group of $4\times 4$ unitary complex matrices with unit determinant. \\
\subsubsection{2-qubit Gates Defined as Outer Product of Two 1-qubit Gates}
 Let us recall that the two-player strategies in a {\it classical game} are defined as outer product of 
 each single player strategy (${\bf 1}$ or $Y$), defined in Eq.~(\ref{3b}) that operate on two bit states  as exemplified in Eq.~(\ref{twobit}). Let us also recall that each player in a {\it quantum game} has a strategy  $U(\phi_i,\alpha_i,\theta_i)$ that is a $2 \times 2$ matrix as defined in Eq.~(\ref{11}). 
 Therefore, we anticipate that the two-player strategies in a {\it quantum game} are defined as outer product of the two single player strategies. Thus,  a 2-qubit gate
 of special importance is the \underline {outer product operation} ${\cal U}=U_1 \otimes U_2$ where each player acts on his own qubit. Explicitly, the operation of  ${\cal U}=U_1 \otimes U_2$  on $|\Psi \ra$ 
given in (\ref{18}) is,  
\begin{equation} \label{20} 
{\cal U}|\Psi \ra=a[U_1|0  \ra] \otimes [U_2|0  \ra]+b[U_1|0  \ra] \otimes [U_2|1  \ra]+c[U_1|1  \ra] \otimes [U_2|0  \ra]+d[U_1|1  \ra] \otimes [U_2|1  \ra]~. 
\end{equation} 
Again, before defining the notion of quantum game, we assert that this operation defines the set of combined quantum strategies 
in analogy with the classical game set of combined strategies defined in Eq.~(\ref{3b}). Thus,
\tboxed{
The (infinite numbers of) elements in the set $A_1 \times A_2$  of combined (quantum) strategies are 
  $4 \times 4$ matrices,  $U(\phi_1,\alpha_1,\theta_1) \otimes U(\phi_2,\alpha_2,\theta_2)$. These $4 \times 4$ matrices act on two qubit states defined above, e.g Eq.~(\ref{18}). 
The single qubit operations are defined in Eq.~(\ref{9}). 
}
\subsubsection{Entanglement Operators (Entanglers)}
We have already underlined the crucial importance of the concept of entanglement 
in quantum games. Therefore, of crucial importance
for quantum game is an operation executed by an \underline{entanglement operator} $J$ that acts on a non-entangled 2qubits and turns it into an entangled 2qubits. 
Anticipating the importance and relevance of Bell's states introduced in Eq.~(\ref{17}) for quantum games, we search entanglement operators  $J$ that operate on the non-entangled state $|0  \ra \otimes |0  \ra=|00 \ra$ and create 
the maximally entangled Bell states such as  $|\psi_+ \ra$ or $|T \ra$ as defined in Eq.~(ref{17}). 
For reason that will become clear later we should require that $J$ is unitary, that is, 
$J^\dagger J=JJ^\dagger={\bf 1}_4$ (see Appendix \ref{Mat}). With a little effort we find, 
\begin{spacing}{0.7}
\begin{equation} \label{22} 
J_1|00 \ra=  \frac{1}{\sqrt{2}} \begin{pmatrix} 1&0&0&i\\0&1&-i&0\\ 0&-i&1&0\\i&0&0&1 \end{pmatrix} \begin{pmatrix} 1\\0\\0\\0 \end{pmatrix}=\frac{1}{\sqrt{2}} \begin{pmatrix} 1\\0\\0\\i \end{pmatrix}=\frac{1}{\sqrt{2}} (|00 \ra+i|11 \ra)=|\psi_+ \ra~.  
\end{equation}
\end{spacing}
\ \\
\begin{spacing}{0.7}
\begin{equation} \label{22a} 
J_2|00 \ra=  \frac{1}{\sqrt{2}} \begin{pmatrix} 0&1&1&0\\1&0&0&0\\ 1&0&1&0\\0&0&0&1 \end{pmatrix} \begin{pmatrix} 1\\0\\0\\0 \end{pmatrix}=\frac{1}{\sqrt{2}} \begin{pmatrix} 0\\1\\1\\0 \end{pmatrix}=\frac{1}{\sqrt{2}} (|01 \ra+|10 \ra)=|T \ra~.  
\end{equation}
\end{spacing}
\ \\
   It is straight forward to check that $J_1$ and $J_2$ as defined above are unitary and that application of $J_1^\dagger$ instead of $J_1$ on the initial state $|00 \ra$ in Eq.~(\ref{22}) 
yields the second Bell's state $|\psi_-\ra$ also defined in Eq.~(\ref{17}), while $J_2^\dagger|00\ra=|S \ra.$.  There is, however, some subtle difference between $J_1$ and $J_2$ that will surface later on. 
\subsubsection{Partial Entanglement Operators}
Intuitively, the Bell's states defined in Eq.~(\ref{17}) are {\it Maximally entangled}  because 
the two coefficients before the two bit states (say, $|00 \ra$ and $|11 \ra$) have the same 
absolute value, $1/\sqrt{2}$. We may think of an entangled state where the weights of the two 2-bit states are unequal, in that case we speak of {\it partially entangled state}. Thus, instead of the maximally entangled Bell states $|\psi_+ \ra$ and $|T \ra$ defined in Eqs.~(\ref{17}), 
(\ref{22}) and (\ref{22a}) we may consider the partially entangled state $|\psi_+(\gamma) \ra$ and $|T(\gamma) \ra$ that depend on a continuous parameter (an angle) $0 \le \gamma \le \pi$ defined as, 
\begin{spacing}{0.8}
\begin{equation} \label{22b} 
|\psi_+(\gamma) \ra=\cos \frac{\gamma}{2} |00 \ra+\sin \frac{\gamma}{2} |11\ra~,  \ \ |\psi_+(0)\ra=|00 \ra, \ |\psi_+(\pi)\ra=|11 \ra, \ |\psi_+(\frac{\pi}{2})\ra=|\psi_+ \ra~. 
\end{equation}
\begin{equation} \label{22c} 
|T(\gamma) \ra=\cos \frac{\gamma}{2} |01 \ra+\sin \frac{\gamma}{2} |10\ra~,  \ \ |T(0) \ra=|01 \ra, \ |T(\pi)\ra=|10 \ra, \ |T(\frac{\pi}{2})\ra=|T \ra~. 
\end{equation}
\end{spacing}
\ \\
The notion of partial entanglement can be put on a more rigorous basis once we have a tool to determine the degree of entanglement. Such a tool does exists, called 
{\it Entanglement Entropy} but it will not be detailed here. 
The reason for introducing partial entanglement is that it is intimately related with the 
existence (or the absence) of pure strategy  Nash equilibrium in quantum games 
as will be demonstrated below. 

In the  same way that we designed the entanglement operators $J_1$ and $J_2$ that, 
upon acting on the two-bit state $|00 \ra$ yield the maximally entangled Bell's states 
$|\psi_+\ra$ and $|T \ra$, we need to design analogous partial entanglement operators 
$J_1(\gamma)$ and $J_2(\gamma)$ that,  upon acting on the two-bit state $|00 \ra$ yield the partilly entangled  states  $|\psi_+(\gamma) \ra$ and $|T(\gamma) \ra$. With a little effort we find, 
\begin{spacing}{0.7}
\begin{equation} \label{22d}
J_1(\gamma)=\begin{pmatrix} \cos \frac{\gamma}{2}&0&0&i \sin \frac{\gamma}{2}\\
0& \cos \frac{\gamma}{2}&-i \sin \frac{\gamma}{2}&0\\
0&-i \sin \frac{\gamma}{2}&\cos \frac{\gamma}{2}&0 \\
i \sin \frac{\gamma}{2}&0&0& \cos \frac{\gamma}{2} \end{pmatrix}, \ \ J_2(\gamma)=\begin{pmatrix} 0& \cos \frac{\gamma}{2}&0&- \sin \frac{\gamma}{2}\\
 \cos \frac{\gamma}{2}&0&- \sin \frac{\gamma}{2}&0\\
 \sin \frac{\gamma}{2}&0&\cos \frac{\gamma}{2}&0 \\
0& \sin \frac{\gamma}{2}&0& \cos \frac{\gamma}{2} \end{pmatrix} ~. 
\end{equation} 
\end{spacing}
\end{spacing}
\begin{spacing}{1.5}
\begin{center}
\section{\small Quantum Games} \label{Section4}
\end{center}
We come now to the heart of our work, that is, description and search for 
pure strategy Nash equilibrium in these games. Quantum games have different structures and different rules than classical games. The skeptical reader might justly argue that introducing a quantum game with an attempt to confront it with 
its classical analogue is meaningless. It is just like inventing a 
new chess game by using a $10 \times 10$  chessboard (instead of the usual $8 \times 8$ one) and adding four more pieces to each player. 

There is, however two points that connect a classical game with its quantum analog. First, the quantum game is based on a classical game and the payoffs in the quantum game are determined by the payoff function of the classical game. 
Second, the classical strategies are obtained as a special case of the quantum strategies. Depending on the entanglement operators $J$  defined in Eq.~(\ref{22d}), the players may even reach the classical square in the game table. 
In most cases, however, this will not lead to a Nash equilibrium. 

\subsection{How to Quantize a Classical Game?} 
 With all these complex numbers running around, it must be quite hard to 
 imagine how this formalism can be connected to a 
game in which people have to take decisions and get tangible rewards that depend on 
their opponent's decisions, especially when these rewards are expressed in real numbers
(dollars or years in prison). Whatever we do, at the end of the day, a passage to real numbers must take place. To show how it works, we start with an old faithful  classical game  (e.g the prisoner dilemma) and show how to turn into into a quantum game that still ends with rewarding its 
players with tangible rewards. This procedure is referred as {\it quantization of a classical game}. 
We will carry out this task in two steps. In the first step we will consider a classical game 
and endow each player $i$ with a quantum strategy (The $2 \times 2$ matrix $U(\phi_i,\alpha_i,\theta_i)$ defined in Eq~(\ref{11}). At the same time, we will also design a new payoff system that 
translates the complex numbers appearing in the state of the system into a real reward. This first 
step leads us to a reasonable description of a game, but proves to be inadequate if we want to 
achieve a really new game, not just the classical game from which we started our journey. 
This task will be achieved in the second step. 

Suppose we start with the same classical game as described in Section \ref{Section1}, 
that is given in its normal form with specified payoff functions as,
 \begin{center}
  $~~~~~~$ Player 2   \\
  Player 1
\begin{tabular}{|l| |l| |l| |l| |l| |l|} \hline  &~ $I$ ~  & ~$Y$ \\ \hline ~$I$~ & u$_1$(I,I),u$_2$(I,I) & u$_1$(I,Y),u$_2$(I,Y,1) \\  \hline ~$Y$~ & u$_1$(Y,I),u$_2$(Y,I) & u$_1$(Y,Y),u$_2$(Y,Y) \\  \hline \end{tabular} 
\end{center}
It is assumed that  the referee already decreed that the initial state is $|00 \ra$, 
and asks the players to choose their strategies. There his, however, one difference: Instead of using the classical strategies of either leaving a bit untouched (the strategy $I$) or operating on it with the second strategy $Y$, the referee allows each player $i=1,2$ 
to use his quantum strategy $U(\phi_i, \alpha_i, \theta_i)$ defined in Eq.~(\ref{11}). Before we find out how
all this will help the players, let us find out what will happen with the state of the system 
after such an operation. For that purpose it is convenient to use the vector notations 
specified in Eq.~(\ref{1}) or (\ref{5}), (\ref{5a}), (\ref{6}) and let each player act on his own qubit 
with his own as strategy as
explained through Eq.~(\ref{20}), thereby leading the system from its initial state $|00 \ra$ 
to its final state $|\Psi \ra$ given by,  
\begin{equation} \label{classquant1} 
|\Psi \ra=U_1 \otimes U_2|00 \ra=U_1|0 \ra \otimes U_2 |0 \ra=U_1 \binom{1}{0} \otimes U_2 \binom{1}{0}
=\binom{[U_1]_{11}}{[U_1]_{21}} \otimes 
\binom{[U_2]_{11}}{[U_2]_{21}} 
=\begin{pmatrix} [U_1]_{11}[U_2]_{11} \\ [U_1]_{11}[U_2]_{21}\\ [U_1]_{21}[U_2]_{11}\\ [U_1]_{21}[U_2]_{21}
  \end{pmatrix} 
\end{equation}
With the help of Eq.~(\ref{18}) we may then write,
\begin{equation} \label{classquant2} 
|\Psi \ra= [U_1]_{11}[U_2]_{11}|00 \ra+[U_1]_{11}[U_2]_{21}|01 \ra+[U_1]_{21}[U_2]_{11}|10 \ra+[U_1]_{21}[U_2]_{21}|11 \ra \equiv a|00\ra+b|01\ra+c|10 \ra+d|11 \ra. 
\end{equation}
From Eq.~(\ref{11}) it is easy to determine the dependence of the coefficients on the angles (that is the strategies of the two players) , 
for example $a=[U_1]_{11}[U_2]_{11}=e^{i (\phi_1+\phi_2)} \cos \frac{\theta_1}{2}  \cos \frac{\theta_2}{2}$ and so on. Since $|\Psi \ra$ is a 2qubits then, as we have stressed all around, in Eqs.~\ref{2qubits} or (\ref{19}) we have $|a|^2+|b|^2+|c|^2+|d|^2=1$. This leads us naturally 
to suggest the following payoff system.
\tboxed{
The payoff $P_i$ of player $i$ is calculated similar to the calculation of payoffs in correlated equilibrium {\it classical} games, with the absolute value squared of the amplitudes $a,b,c,d$ 
(themselves are {\it complex numbers}) as the corresponding probabilities, 
\begin{spacing}{0.8}
\begin{equation} \label{25} 
P_i(\phi_1,\alpha_1,\theta_1; \phi_2,\alpha_2,\theta_2)=|a|^2u_i(0,0)+|b|^2u_i(0,1)+|c|^2 u_i(1,0)+|d|^2 u_i(1,1)~. 
\end{equation} 
\end{spacing}
}
For example, prisoner's 1 and 2 years in prison in the prisoner dilemma game table, Eq.~(\ref{PD}) are,
\begin{equation} \label{26}
P_1= -4|a|^2-6|b|^2-2|c|^2-5|d|^2,  \ \ P_2=-4|a|^2-2|b|^2-6|c|^2-5|d|^2. 
\end{equation}
The alert reader must have noticed that this procedure ends up in a classical game with mixed strategies. First, once absolute values are taken, the role of the two angles $\phi$ and $\theta$ is void because 
\begin{equation} \label{trivial}
|a|^2=\cos^2 \frac{\theta_1}{2}\cos^2 \frac{\theta_2}{2}, \ |b|^2=\cos^2 \frac{\theta_1}{2}\sin^2 \frac{\theta_2}{2}, \ |c|^2=\sin^2 \frac{\theta_1}{2}\cos^2 \frac{\theta_2}{2}, \  |d|^2=\sin^2 \frac{\theta_1}{2}\sin^2 \frac{\theta_2}{2}.
\end{equation}
What is more disturbing is that we arrive at an old format of classical games with mixed strategies. 
Since $\cos^2 \frac{\theta}{2}+\sin^2 \frac{\theta}{2}=1$, we immediately identify the payoffs  in Eq.~(\ref{25}) as those resulting from mixed strategy classical game where a prisoner $i$ chooses to confess with probability $\cos^2 \frac{\theta_i}{2}$ and to don't confess with probability $\sin^2 \frac{\theta}{2}$. In particular, the pure strategies are obtained as specified in Eq.~(\ref{quantclass}).
Thus while the analysis of the first step taught us how to use quantum strategies and how to 
design a payoff system applicable for a complex state of the system $|\Psi \ra$ as defined in 
Eq.~(\ref{classquant2}), it did not prevent us from falling into the trap of triviality in the sense that 
so far nothing is new. \\
The reason for this failure is at the heart of quantum mechanics. The initial state $|00 \ra$ upon which the players apply their strategies according to Eq.~(\ref{classquant1}) in \underline{not entangled}; 
Since it is a simple outer product of $|0\ra$ of player 1 and $|0 \ra$ of player 2, so according to the definition of entanglement given after Eq.~(\ref{17}), it is not entangled. Thus we find that,
\tboxed{ In order for a quantum game to be distinct from its classical analog, the state upon which the two players apply their quantum strategies should be entangled.
} 
That is where the entanglement operators $J$ defined in Eqs.~(\ref{22}), (\ref{22a}) and (\ref{22d}) 
come into play. Practically, we ask the referee not only to suggest a simple initial state such as $|00 \ra$ but also to choose some entanglement operator $J$ and to apply it on $|00 \ra$ as exemplified 
in Eqs.~(\ref{22}), (\ref{22a})  in order to modify it into an entangled state. Only then the players are allowed to apply their quantum strategies, after which the state of the system will be given by 
$U_1 \otimes U_2 J |00 \ra$, as compared with Eq.~(\ref{classquant1}).  
There is one more task the referee should take care of. A reasonable desired property is that 
if, for some reason the players choose to leave everything unchanged by taking  
${\bm \gamma}_i=(\phi_i,\alpha_i,\theta_i)=(0,0,0)$, namely, $U_1=U_2=I$ then the 
final state should be identical to the initial state. This is easily achieved by asking the referee to 
apply the operator $J^{-1}=J^\dagger$ on the state  $U_1 \otimes U_2 J |00 \ra$ (that was obtained 
after the players applied their strategies on the entangled state $J|00 \ra$. 
These modification change 
things entirely, and turn the quantum game into a new game with complicated strategies, 
that is, it is much richer than its classical analog. 

\begin{figure}[!h]
\centering
\includegraphics[width=6truecm, angle=0]{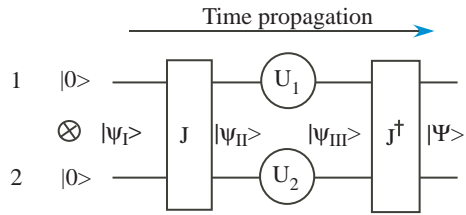}
\caption{\footnotesize{ A general protocol for a two players two strategies quantum game showing the flow of information. Besed on Eq.~(\ref{24}) to be followed on the figure from left to right. 
$U_1$ is player's 1 move, $U_2$ is player's 2 move, and $J$ is an entanglement gate.}}
\label{Fig1}
\end{figure}

Let us then organize the game protocol as explained above  by presenting a list of well defined steps. 
\begin{enumerate}
\item{} The starting point is some classical 2 players-2 strategies classical game 
given in its normal form (a table with utility functions) and a referee whose duty is 
to choose an initial two bit state and an entanglement operator $J$. 
\item{} The referee chooses a simple
non-entangled 2qubits initial state, which, for convenience,  we fix once for all to be 
 $|\psi_I \ra=|00 \ra$. As in the classical game protocol, the choice of this state does not affect 
 the game in any form, it is just a starting point. 
 \item{} The referee then chooses an entanglement operator $J$ and apply it on $|\psi_I \ra$
 to generate an entangled  state $|\psi_{II} \ra =J|\psi_I \ra $ 
 as exemplified in Eq.~(\ref{22}). This operation is part of the rules of the game, namely, it is not possible for the players to affect this choice in any way. 
\item{} At this point every player 
applies his own transformation $U_i=U(\phi_i,\alpha_i,\theta_i)$ on his own qubit. The functional dependence of $U$ on the three angles is displayed in Eq.~(\ref{11}). This is the only place where the players have to take a decision. After the players made their decisions 
the product operation is applied on $|\psi_{II} \ra$
 as in Eq.~(\ref{20}),  resulting the state $|\psi_{III} \ra=U_1 \otimes U_2|\psi_{II} \ra$.  
 \item{}  The referee then applies the inverse of $J$ (namely $J^\dagger$ since $J$ is unitary) and gets the final state 
\begin{equation} \label{24}
 |\Psi \ra=\overbrace{J^\dagger}^{\mbox {referee}} \overbrace{U_1 \otimes U_2}^{\mbox { players}} 
  \overbrace{J|00 \ra}^{\mbox {referee}}=a|00\ra+b|01\ra+c|10\ra+d|11\ra, \ \ 
\end{equation}
where the complex numbers $a,b,c,d$ with $|a|^2+|b|^2+|c|^2+|d|^2=1$  are functions of the elements of $U_1$ and $U_2$ namely, following Eq.~(\ref{11}), they are functions of the 6 angles $(\phi_1,\alpha_1,\theta_1;  \phi_2,\alpha_2,\theta_2)$. 
\item{} The players are then rewarded according to the prescription given by Eq.~(\ref{25}). 
\end{enumerate}
The set of operations leading from the initial state $|\psi_I \ra$ to the final state $|\Psi \ra$  is schematically shown in Fig.~\ref{Fig1}. 

\subsection{Formal Definition of a Two-Player Pure Strategy Quantum Game}
Based on the prescriptions given in Eq.~(\ref{24}), Fig.~(\ref{Fig1}) and Eq.~(\ref{25}) we can now give a formal definition of a  two-players two strategies quantum game that is an extension of a classical  two-players two strategies game. Necessary ingredient of a quantum game should include:
\begin{enumerate} 
\item{} A quantum system which can be analyzed using the tools of quantum mechanics, for example, a two qubits system.
\item{} Existence of two players, who are able to manipulate
the quantum system and operate on their own qubits. 
\item{} A well define strategy set for each player. More concretely, a set of 
unitary $2 \times 2$ matrices with unit determinant $U \in SU(2)$. 
\item{} A definition of the pay-off functions or utilities associated with the playersÕ strategies. More concretely, we have in mind a classical 2-player two strategies game given in its normal form ( a table of payoffs).
\end{enumerate}
\noindent
\underline{Definition} Given a classical two-players two pure strategies classical game
\begin{equation} \label{31}
 G_C=\la N=\{1,2 \}, |ij\ra, A_i=\{ I, Y \}, u_i : A_1 \otimes A_2 \to \mathbb{R} \ra.
 \end{equation} 
Its quantum (pure strategy) analog is the game,
\begin{equation} \label{32}
 G_Q=\la N=\{1,2\}, |\psi_I \ra, \{ {\cal A}_i \}, J, u_i, P_i \ra.
 \end{equation}
Here $N=\{1,2 \}$, is the set of (two) players, $|\psi_I \ra$ is the initial state suggested by the referee 
(usually a simple two-bit state such as $|00 \ra$ as in the classical game), ${\cal A}_i=U({\bm \gamma}_i) \equiv U_i$, is the infinite set quantum pure strategies of player $i$ on his qubit defined by the $2 \times 2$ matrix Eq.~(\ref{11}), 
$J$ is an entanglement operator defined along Eqs.(\ref{22}, \ref{22a}, \ref{24}) and Fig.~\ref{Fig1}, $u_i(k,\ell)$ with $k,\ell=0,1$ are the classical payoff functions of the game G and $P_i(U_1,U_2)$ are the quantum payoff functions defined in Eq.~(\ref{25}) in which the coefficients $a,b,c,d$ are complex numbers (also called amplitudes) that determine the expansion of the final state $|\Psi \ra$ as a combination of two bit states 
as in Eq.~(\ref{24}. \\
\underline{comments} \\
1)Since $U_i$ is uniquely determined by the three angles  ${\bm \gamma}_i=(\phi_i,\alpha_i, \theta_i)$ 
through Eq.~(\ref{11}) we may also regard ${\bm \gamma}_i$ as the strategy of player $i$. 
Thus, unlike the classical game where each player has but two strategies, in the quantum game the set of strategies of each player is determined by three continuous variables. As we have already mentioned, the set of strategies of a player correspond to a point on $S^3$.\\
2)  $J$ is part of the rules of the game (it is not controlled by the players). The main requirement from $J$ is that it is a unitary matrix and that after operating on the initial to bit state (taken to be $|00 \ra$ in our case) the result is an entangled 2qubits. \\
3) As we stressed in relation with Eq.~(\ref{24}), the amplitudes are functions of the two strategies 
${\bm \gamma}_i=(\phi_i,\alpha_i, \theta_i), (i=1,2)$ that are given analytically once the operations 
implied in Eq.~(\ref{24}) are properly carried out (see below).

\subsection{Nash Equilibrium in a Pure Strategy Quantum Game }
\noindent
\underline{Definition} A pure strategy Nash Equilibrium in a quantum game is a pair of strategies $({\bm \gamma}_1^*,{\bm \gamma}_2^*) \in S^3 \otimes S^3$ (each represents three angles ${\bm \gamma}_i^*=(\phi_i^*,\alpha_i^*,\theta_i^*) \in S^3$), such that 
\begin{equation} \label{34}
P_1({\bm \gamma}_1,{\bm \gamma}_2^*) \le P_1({\bm \gamma}_1^*,{\bm \gamma}_2^*) \ \forall \ \ {\bm \gamma}_1 \in S^3, \ \  P_2({\bm \gamma}_1^*,{\bm \gamma}_2) \le P_2({\bm \gamma}_1^*,{\bm \gamma}_2^*) \ \forall \ \ {\bm \gamma}_2 \in S^3.
 \end{equation}
 It is immediately realized that the concept of Nash equilibrium and its elucidation in a quantum game is far more difficult than the classical one. If each player's strategy would have been dependent on a {\it single} continuous parameter, then the use of the method of best response functions could be effective, but here each player's strategy depends on {\it three} continuous parameters, and the method of response functions might be inadequate. One of the goals of the present work is to alleviate this problem. Another important point concerns the question of cooperation.  
 In the classical prisoner dilemma game,  
a player that chooses the don't-confess strategy (Y) forces his opponent to cooperate and choose 
Y (don't confess) as well, that leads to a pure strategy Nash equilibrium (Y,Y). 
On the other hand,  in the quantum game, the situation is quite different. By looking at the payoff 
expressions in Eq.~(\ref{26}) we see that prisoner 1 wants to reach the state where $|c|^2=1$ 
and $|a|^2=|b|^2=|d|^2=0$, whereas prisoner 2 wants to reach the state where $|b|^2=1$ 
and $|a|^2=|c|^2=|d|^2=0$. Surprisingly, as we shall see below,  there are situations such that for every strategy chosen by 
prisoner 1, prisoner 2 can find a best response that makes $|b|^2=1$ 
and $|a|^2=|c|^2=|d|^2=0$ and vice versa, for every strategy chosen by 
prisoner 2, prisoner 1 can find a best response that makes $|c|^2=1$ 
and $|a|^2=|b|^2=|d|^2=0$. Since the two situations cannot occur simultaneously, 
there is no Nash equilibrium and no cooperation in this case.  \\
\subsubsection { The Role of the Entanglement Operator $J$ and Classical Commensurability}
 A desired property (although not crucial) of a quantum game is 
   that the theory as defined in Eq.~(\ref{24}) and Fig.~(\ref{Fig1}) includes the 
 classical game as a special case.  
 We already know from Eq.~(\ref{quantclass}) that the classical strategies 
 $I$ and $Y$ are obtained as special cases of the quantum ones, since $U(0,0,0)=I$ and $U(0,0, \pi)=Y$. What we require here is that by using their classical strategies, the players will be able to reach the four classical states (squares of the game table). For example, to reach the square (C,C) the coefficients 
 $a,b,c,d$ in the final state $\Psi \ra$ at the end of the game (see Eq.~(\ref{24}) should be $|a|^2=1, b=c=d=0$ and so on. For this requirement to hold, the entanglement operator $J$ 
 should satisfy a certain equality. 
 We refer to this equality to be satisfied by $J$ as {\it classical commensurability}. 
 From the discussion around Eq.~(\ref{quantclass}) we recall that 
 in a classical game, the only operations on bits are implemented either by the unit matrix $I$ (leave the bit in its initial state $|0 \ra$ or $|1 \ra$) or $Y=\binom {~0~1}{-1~0}$ (change the state of the bit from $|0 \ra$ to $|1 \ra$ or vice versa). 
 Thus, by choosing $U(0,0,0)$ or $U(0,0, \pi)$ the players virtually use classical strategies. 
 Therefore, classical commensurability implies
\begin{equation} \label{classcom}
 [Y \otimes Y,J]=0, \ \ \mbox{(Classical commensurability)}~,
\end{equation}
where we recall from Appendix \ref{Mat} that 
for two square matrices $A,B$ with equal dimensions, the commutation relation is 
defined as $[A,B]=AB-BA$. Indeed, if this condition is satisfied and both $U_1$ and $U_2$ are 
classical strategies, then $[U_1 \otimes U_2,J]=0$ because in this case $U_1 \otimes U_2=I \otimes I$ or $I \otimes Y$ or $Y \otimes I$ or $Y \otimes Y$ and as we show below, all of the four operators commute with $J$.  Consequently
\begin{equation} \label{cc}
 |\Psi \ra=J^\dagger U_1 \otimes U_2 J |00\ra=J^\dagger J U_1 \otimes U_2  |00\ra=U_1 \otimes U_2 |00 \ra,
 \end{equation}
that is what happens in a classical game as explained in connection with figure (\ref{Fig0}). 
To prove that the four two-player classical strategies listed above do commute with $J$ we note that by direct calculations 
it is easy to show that $J_1$ defined in Eq.~(\ref{22}) satisfies classical commensurability because 
an elementary manipulation of matrices shows that $J_1$
 can be written as 
 \begin{equation} \label{J1}
 J_1=e^{i\frac{\pi}{4} Y \otimes Y}=\frac{1}{\sqrt{2}}(I_4 + i Y \otimes Y),
 \end{equation}
  and this matrix naturally commutes with $Y \otimes Y$.  The first equality is derived in Appendix \ref{Mat}.
On the other hand, $J_2$ defined in Eq.~(\ref{22a}) does not satisfy classical commensurability 
as can be checked by directly inspecting the commutation relation $[Y \otimes Y,J_2] \ne 0$. 
\subsection{Absence of Nash Equilibrium for Maximally Entangled States} \label{Subsec_Absence}
After defining the notion of quantum games and their pure strategy Nash Equilibrium we approach 
the problem of finding pure strategy Nash Equilibrium. The first result in this area is negative: If the state 
$|\psi_I\ra=J|00 \ra$ is maximally entangled, (e.g, $|\psi_I \ra=|\psi^+ \ra$ (Eq.~(\ref{22})) or $|\psi_I \ra=|T \ra$  (Eq.~(\ref{22a})) the quantum game of the prisoner dilemma does not have a pure strategy Nash Equilibrium. 
Our poof of this statement will be straightforward. First we will calculate explicitly the amplitudes $a,b,c,d$ of the final wave function $|\Psi \ra$ as defined in Eq.~(\ref{24}) and in Fig.~\ref{Fig1} and then use the method of response functions and show that the two response functions $B_2({\bm \gamma}_1)$ and 
$B_1({\bm \gamma}_2)$ cannot intersect.  
\subsubsection{Calculating the Amplitudes of the Final States $|\Psi \ra$} 
In order to calculate the payoffs $P_1$ and $P_2$ according to the prescription (\ref{26}) we need 
to carry out the operations specified in Eq.~(\ref{24}) leading from the initial state $|00 \ra$ all the way to the final state $|\Psi \ra$. This is a standard manipulation in matrix multiplication that 
in the present case ends up with reasonable (not so long) expressions. 
As an example we consider the entanglement operator  
$J=J_1$ as given in Eq.~(\ref{22}) so that $J |00 \ra=|\psi^+ \ra$ 
that is a {\it Maximally Entangled State:} . Player $i$ has a {\em strategy matrix}
$U_i \equiv U({\bm \gamma}_i)=U(\phi_i,\alpha_i, \theta_i)$ as defined in Eq.~(\ref{11}). The product $U({\bm \gamma}_1) \otimes U({\bm \gamma}_2)$ acts on $|\psi_+ \ra$ according to the prescription (\ref{20}) is given explicitly as, 
\begin{equation} \label{27} 
U_1 \otimes U_2 |\psi_+ \ra=\frac{1}{\sqrt{2}}[ (U_1|0  \ra) \otimes (U_2|0  \ra)+i([U_1|1  \ra) \otimes (U_2|1  \ra)]~.
\end{equation} 
Explicitly, for a  $2 \times 2$ matrix $U=\binom{U_{11}U_{12}}{U_{21}U_{22}}$ we have, according to Eq.~(\ref{9}),  
\begin{equation} \label{28}
U|0\ra=U\binom{1}{0}=\binom{U_{11}}{U_{21}}=U_{11}|0\ra+U_{21}|1\ra, \ \ U|1\ra=U\binom{0}{1}=\binom{U_{12}}{U_{22}}=U_{12}|0\ra+U_{22}|1\ra~.
\end{equation}
Performing the outer products as in Eq.~(\ref{12}), multiplying by $J^\dagger$ we can find the corresponding amplitudes 
$a,b,c,d$ of $|\Psi \ra$ in the notation of (\ref{12}) or (\ref{18}). Straight forward but tedious calculations yield, 
\begin{eqnarray} 
&& \mbox{Coefficients of $|\Psi \ra$ for $J|00 \ra=|\psi_+\ra$} \ \mbox{(Eq.~(\ref{22}))}, \nonumber \\
&& |a|^2=\left[  \cos \tfrac{1}{2} \theta_1 \cos \tfrac{1}{2} \theta_2 \cos (\phi_1+\phi_2)-\sin \tfrac{1}{2} \theta_1 \sin \tfrac{1}{2} \theta_2 \sin (\alpha_1+\alpha_2) \right ]^2, 
\nonumber \\
&& |b|^2=\left[ \cos \tfrac{1}{2} \theta_1 \sin \tfrac{1}{2} \theta_2 \cos (\phi_1-\alpha_2)+ \sin \tfrac{1}{2} \theta_1 \cos \tfrac{1}{2} \theta_2 \sin (\alpha_1-\phi_2) \right ]^2, 
\nonumber \\
&& |c|^2=\left[\sin \tfrac{1}{2} \theta_1 \cos \tfrac{1}{2} \theta_2 \cos (\alpha_1-\phi_2)- \cos \tfrac{1}{2} \theta_1 \sin \tfrac{1}{2} \theta_2 \sin (\phi_1-\alpha_2)  \right ]^2, 
\nonumber \\
&& |d|^2=\left[  \cos \tfrac{1}{2} \theta_1 \cos \tfrac{1}{2} \theta_2 \sin (\phi_1+\phi_2)+\sin \tfrac{1}{2} \theta_1 \sin \tfrac{1}{2} \theta_2 \cos (\alpha_1+\alpha_2) \right ]^2~.
\label{29} 
\end{eqnarray} 
Compared with Eq.~(\ref{trivial}) we see that the present game 
is really novel, all the angles appear in the payoff and 
it is not reducible to any form of classical game. \\
It is instructive to check how the classical strategies are recovered as special cases of 
 the quantum ones. If both players choose $U(0,0,0)=I$ 
 then $|a|^2=1$ and  $|\Psi \ra=|00\ra$ with amplitude 1, that corresponds to the classical 
 strategy $(I,I)$ leading to the state (C,C). Similarly, if one player chooses $U(0,0,0)$ and the other chooses 
 $U(0,0, \pi)$  this leads to either $|b|^2=1$ corresponding to classical strategies $(I,Y)$ leading to the state (C,D)
 or to $|c|^2=1$ corresponding to classical strategies $(Y, I)$.  leading to the state (D,C).
Finally, if both players choose $U(0,0,\pi)$ 
 then the final state is $|\Psi \ra=|11 \ra$ with amplitude 1, that corresponds to the classical strategy (Y,Y) leading to the state (D,D). Unlike the classical game, however, this choice is, in general, not a Nash equilibrium. 
 Player 1 for example may find a strategy $U(\phi_1,\alpha_1,\theta_1)$ such that 
 $P_1[U(\phi_1,\alpha_1,\theta_1),Y] > -4$. The upshot then is that if classical commensurability is respected, then, by using classical strategies 
 the players can reach the classical positions (C,C),(C,D),(D,C) and (D,D) but the classical Nash equilibrium is not relevant for the quantum game. 
 
  \noindent
 {\bf Example 2: Triplet Bell State:} If we take $J=J_2$ as in Eq.~(\ref{22a}) we get $J |00 \ra=|T \ra$, the triplet Bell state. 
 Performing the calculations $|\Psi \ra=J^\dagger U_1 \otimes U_2 |T \ra$ we get the four probabilities, 
 \begin{eqnarray} 
 && \mbox{Coefficients of $|\Psi \ra$ for $J|00 \ra=|T\ra$=Bell's Triplet State}, \ \mbox{(Eq.~\ref{22a})} \nonumber \\
&& |a|^2=\left[ \cos \tfrac{1}{2} \theta_1 \cos \tfrac{1}{2} \theta_2 \cos (\phi_1-\phi_2)- \sin \tfrac{1}{2} \theta_1 \sin \tfrac{1}{2} \theta_2 \cos (\alpha_1-\alpha_2) \right ]^2, 
\nonumber \\
&& |b|^2=\left[ \cos \tfrac{1}{2} \theta_1 \sin \tfrac{1}{2} \theta_2 \sin (\phi_1+\alpha_2)+\sin  \tfrac{1}{2} \theta_1 \cos \tfrac{1}{2} \theta_2 \sin (\alpha_1+\phi_2) \right ]^2, 
\nonumber \\
&& |c|^2=\left[ \sin \tfrac{1}{2} \theta_1 \sin \tfrac{1}{2} \theta_2 \sin (\alpha_1-\alpha_2)- \cos \tfrac{1}{2} \theta_1 \cos \tfrac{1}{2} \theta_2 \sin (\phi_1-\phi_2) \right ]^2, 
\nonumber \\
&& |d|^2=\left[ \sin \tfrac{1}{2} \theta_1 \cos \tfrac{1}{2} \theta_2 \cos (\alpha_1+\phi_2)+ \cos \tfrac{1}{2} \theta_1 \sin \tfrac{1}{2} \theta_2 \cos (\phi_1+\alpha_2) \right ]^2~.
\label{30} 
\end{eqnarray} 
\subsubsection{Proof of Absence of Pure strategy Nash Equilibrium}
The following theorem is well known, see for example Refs.\cite{Hayden,Landsburg2}. Here we prove it directly by 
showing the the best response functions cannot intersect. \\
 \underline{Theorem} The quantum game defined as in Eq.~(\ref{32}) with $J=J_1$ as 
 given by Eq.~(\ref{22}) does not have a pure strategy Nash Equilibrium. \\
 \underline{Proof} From the expressions (\ref{29}) for the amplitudes it is evident that for any strategy $(\alpha_1, \phi_1, \theta_1)$ of player 1, player 2 can find a best response that brings him to the minimum years in prison with 
 \begin{equation} \label{B2a}
 B_2(\phi_1,\alpha_1,\theta_1)=U( \phi_2=\alpha_1-\frac{\pi}{2},  \alpha_2=\phi_1, \theta_2=\theta_1),
 \end{equation}
  because then we have 
 $|b|^2=1, |a|^2=|c|^2=|d|^2=0$.  
 Similarly, for any strategy $(\alpha_2, \phi_2, \theta_2)$ of player 2, player 1 can find a best response that brings him to the minimum years in prison with 
  \begin{equation} \label{B1a}
 B_1(\phi_2,\alpha_2,\theta_2)=U( \phi_1=\alpha_2-\frac{\pi}{2}, \alpha_1=\pi/2+\phi_2,\theta_1=\pi-\theta_2),
 \end{equation}
 because then we have, 
 $|c|^2=1, |a|^2=|b|^2=|d|^2=0$ Evidently, the two restrictions on the amplitudes cannot occur simultaneously,  and therefore, the two response functions cannot intersect. 
 Hence, there is no pure strategy Nash equilibrium $\blacksquare$

 Similarly, the quantum game defined as in Eq.~(\ref{32}) with $J=J_2$ as 
 given by Eq.~(\ref{22a}) does not have a pure strategy Nash Equilibrium. Simple manipulations based on expressions (\ref{30}) for the amplitudes lead to the following response functions, 
 \begin{equation} \label{B2b}
 B_2(\phi_1,\alpha_1,\theta_1)=U(\phi_2=\alpha_1-\frac{\pi}{2}, \alpha_2=\frac{\pi}{2}-\phi_1,\theta_2=\pi-\theta_1).
 \end{equation}  
  \begin{equation} \label{B1}
 B_1(\phi_2,\alpha_2,\theta_2)=U(\phi_1=\phi_2-\frac{\pi}{2},\alpha_1=\pi/2+\alpha_2,\theta_1=\theta_2).
 \end{equation}
It is worth emphasizing that these (negative) results are valid only if the classical game upon which the quantum game is built does not have a Pareto efficient pure strategy Nash equilibrium. If such equilibrium exists, the players will choose their quantum strategies to settle on this place. For example, if, in some special prisoner dilemma game  there is a Pareto efficient equilibrium in (C,C) then both players prefer 
$|a^2|=1, |b|^2=|c|^2=|d|^2=0$. For the first game (Eq.~(\ref{29})) they will choose $\theta_1=\theta_2, \alpha_1+\alpha_2=\pi/2, \phi_1+\phi_2=\pi$, while for the second game (Eq.~(\ref{30})) they will choose $\theta_1=\theta_2,  \phi_1=\phi_2, \alpha_1-\alpha_2=\pi$. 

{\it Starting from a non-entangled initial state (for example $|00 \ra$ and using entanglement operators $J$ as defined in Eqs.~(\ref{22}) or (\ref{22a}) leading to the maximally entangled states $|\psi_+\ra$ and $|T \ra$  respectively, the quantum game has no pure strategy NE.}\\
\ \\
The natural place to look for NE is then to consider a mixed strategy. 
Before that, however, we want to consider the concept of {\it partial entanglement}, 
since, as we shall show, it can lead to a pure strategy Nash equilibrium of the quantum game. 
\subsection{Partial Entanglement}
The states  $|\psi_+ \ra$  and $|T \ra$  defined in Eqs.~(\ref{22}) and (\ref{22a}) are ``maximally entangled" in the sense that the absolute value square of the two coefficient before the  2-bit states are equal to 1/2 so that the corresponding weights are equal. If the weights are unequal, we have partial entanglement. \\
{\bf Partial Entanglement Operator with Classical Commensurability}\\
We have already pointed out that the entanglement operator $J$ as defined in Eq.~(\ref{22}) 
satisfies classical commensurability $[Y \otimes Y,J]=0$. We now reconsider the operator $J_1(\beta)$
defined in the first equality of Eq.~(\ref{22d}). Using results from Appendix \ref{Mat}, it can be written as
\begin{equation} \label{37}
J_1(\beta)=e^{i \tfrac{\beta}{2} Y \otimes Y}=\cos \tfrac{\beta}{2} I_{4 \times 4}+ i \sin\tfrac{\beta}{2} Y \otimes Y~.
\end{equation} 
Clearly, when $\beta=0$ we have $J_1(0)=I_{4 \times 4}$ while $J_(\tfrac{\pi}{4})$ is given in Eq.~{22} that leads to the maximally entangled state $|\psi_+\ra$ on the RHS of Eq.~(\ref{22}). For $0 < \beta < \tfrac{\pi}{2}$ $J_(\beta)$ is a partial entanglement operator 
and the state $J_1(\beta)|00 \ra =|\psi_+(\beta)\ra$ defined in Eq.~(\ref{22b}) is said to be {\it partially entangled}.  When $J_1(\beta)$ is used in Eq.~(\ref{24}) it results in the final state $|\Psi \ra=a|00 \ra+b|01\ra+c|10\ra+d|11 \ra$ with complex amplitudes,
\begin{eqnarray} 
&& a=\left[ \cos \tfrac{1}{2} \theta_1 \cos \tfrac{1}{2} \theta_2 \cos (\phi_1+\phi_2)- \sin \tfrac{1}{2} \theta_1 \sin \tfrac{1}{2} \theta_2 \sin (\alpha_1+\alpha_2) \sin \beta+i \cos \tfrac{1}{2} \theta_1 \cos \tfrac{1}{2} \theta_2 \sin(\phi_1+\phi_2)\cos \beta  \right ], 
\nonumber \\
&& b=\left[ \cos \tfrac{1}{2} \theta_1 \sin \tfrac{1}{2} \theta_2 \cos (\phi_1-\alpha_2)+ \sin \tfrac{1}{2} \theta_1 \cos \tfrac{1}{2} \theta_2 \sin (\alpha_1-\phi_2)\sin \beta + i  \cos \tfrac{1}{2} \theta_1 \sin \tfrac{1}{2} \theta_2 \sin(\phi_1-\alpha_2)\cos \beta \right ], 
\nonumber \\
&& c=\left[ \sin \tfrac{1}{2} \theta_1 \cos \tfrac{1}{2} \theta_2 \cos (\alpha_1-\phi_2)- \cos \tfrac{1}{2} \theta_1 \sin \tfrac{1}{2} \theta_2 \sin (\phi_1-\alpha_2) \sin \beta-i \sin \tfrac{1}{2} \theta_1 \cos \tfrac{1}{2} \theta_2 \sin(\alpha_1-\phi_2)\cos \beta \right ], 
\nonumber \\
&& d=\left[ \sin \tfrac{1}{2} \theta_1 \sin \tfrac{1}{2} \theta_2 \cos (\alpha_1+\alpha_2)+ \cos \tfrac{1}{2} \theta_1 \cos \tfrac{1}{2} \theta_2 \sin (\phi_1+\phi_2) \sin \beta -i \sin \tfrac{1}{2} \theta_1 \sin \tfrac{1}{2} \theta_2 \sin(\alpha_1+\alpha_2) \cos \beta] \right ].
\nonumber \\
&&
\label{38} 
\end{eqnarray} 
For $\beta=\tfrac{\pi}{2}$ the squares $|a|^2,|b|^2,|c|^2,|d|^2$ are reduced to their values in Eq.~(\ref{29}). 
We will check below the existence of pure strategy Nash equilibrium for $0 < \beta < \tfrac{\pi}{2}$.  \\
\end{spacing}
\begin{spacing}{1.5}
\begin{center}
\section{\small Nash Equilibrium with Partial Entanglement} \label{Section5}
\end{center}
We have seen in subsection \ref{Subsec_Absence} that when the entanglement operator $J$ appearing in Eq.~(\ref{24})  or, alternatively, in Fig.~\ref{Fig1}, leads to a maximally entangled state $|\psi_+ \ra$ or $|T \ra$  , the quantum game  does not have pure strategy Nash equilibrium.  We also know that when $J={\bf 1}_{4 \times 4}$, 
then the classical Nash Equilibrium obtains because the state prepared by the referee for the two players to apply their strategies is just the initial state $|00 \ra$ and the players then use their classical strategies as special case of their quantum ones$|\psi_+ \ra$. This may lead to the following scenario: Suppose $J$ is classically commensurate, but displays only partial entanglement (explicitly this corresponds to $J(\beta)=J_1(\beta) $ given in Eq.~(\ref{37}) with $0<\beta<\frac{\pi}{2}$). Then there may be a threshold value $0<\beta_c<\frac{\pi}{2}$ such that 
for $0 \le \beta<\beta_c$ there is a pure strategy Nash Equilibrium  (that may coincide or may be distinct from the classical one) while for $\beta > \beta_c$ there is no pure strategy Nash Equilibrium because $J$ is close to the case of maximal entanglement. In this section we will check this hypothesis numerically using the method of response functions and show that this scenario is possible and that the quantum Nash equilibrium  might be distinct (and ameliorates) the classical one.   In the first subsection  we will explain the method of response functions, while in the second subsection the numerical algorithm will be explained. 
\subsection{Best Response Functions} \label{Sect_BRF}
The method of best response functions  is an effective method for locating Nash equilibrium in classical games with two players in which the strategy space is not complicated. Its effectiveness for the quantum game is not at all evident due to the complexity of strategy space that is a surface of the sphere $S^3$. The method that will be used below is to replace continuous variables $\phi, \alpha, \theta$ by a mesh of discrete 
points. This turns the problem to a one with finite (albeit very large) strategy space for which the method of response functions is expected to work. Therefore, we shall explain the method on the most elementary level as taught in undergraduate courses in game theory. 
\subsubsection{Finite Set of Strategies} \label{subsubsec_finite}
Let us consider a two-player classical game where each player $i$ has  K strategies, denoted as $\{ k_i \},  \ \ i=1,2, \ \ k_i=1,2,\ldots , K$. For each   strategy $k_1$ of player 1, player 2 finds a best response strategy $q_2(k_1)$ that leads him to the  highest possible payoff once $k_1$ is given (here $q_2$ is an integer between 1 and K).  
(The notation used here for the response functions is $q_i(.)$ instead of $B_i(.)$). 
Similarly, for each   strategy $k_2$ of player 2, player 1 finds a best response strategy $q_1(k_2)$ that leads him to the  highest possible payoff once $k_2$ is given. 
It should be stressed that the mapping $q_1: \{ 1,2,\ldots,K \} \to  \{ 1,2,\ldots,K \}$ 
is not necessarily one-to-one. There may be more than one response to a given strategy and there may be strategies that are not ch osen as best response. 
We can now draw two discrete "curves".  The first curve is obtained by listing $k_1$ along the $x$ axis and plotting the points $q_2(k_1)$ above the $x$ axis. The second curve  is obtained by listing $k_2$ along the $y$ axis and plotting the points $q_1(k_2)$ to the right of the $y$ axis. These discrete curves need not be monotonic, and they may not have a common point. However, if the discrete curves do have a common point 
$(q_1^*,q_2^*)$ this pair of strategies form a Nash equilibrium. The point 
$(q_1^*,q_2^*)$ can be 
found graphically or else, once the lists $q_2(k_1)$ and $q_1(k_2)$ are prepared, 
the equilibrium strategies are found by searching solution to the equation
\begin{equation} \label{k1stark2star}
|q_1^*-q_1(q_2^*)|+|q_2^*-q_2(q_1^*)|=0~.
\end{equation}
\subsubsection{Continuous Set of Strategies} \label{subsubsec_cont}
The method of best response functions is also effective when the strategy spaces are determined by {\it single} continuous parameters, $x_1 \in [a_1,b_1]$ (for player 1) and $x_2 \in [a_2,b_2]$  (for player 2). The response functions are $q_2(x_1)$ and $q_1(x_2)$ where, following the discrete case, $q_1(x_2)$ need not be one-to-one and need not be a continuous function.  Its domain is defined on $x_2 \in [a_2,b_2]$ and its target is defined in $[a_1,b_1]$.  Analogous statements hold for $q_2(x_1)$. The two functions are now plotted as explained above for the discrete case and Nash equilibrium may obtain at strategies $(q_1^*,q_2^*)\in [a_1,b_1] \times [a_2,b_2]$ such that, 
  \begin{equation} \label{q1q2star}
q_1^*=q_1(q_2^*), \ \ q_2^*=q_2(q_1^*)~.
\end{equation}

Unfortunately, this method is ineffective when each strategy space is determined by more than one continuous variable as in our quantum game where the strategy of player $i=1,2$ is determined by three angular variables, $0 \le \phi_i \le 2 \pi, \ \ 0 \le \alpha_i \le 2 \pi, \ \ 0 \le \theta_i \le \pi$ or, in short notation, ${\bm \gamma}_i=(\phi_i,\alpha_i,\theta_i)$ being a point on $S^3$. The response functions ${\bf q}_1({\bm \gamma}_2)$ and ${\bf q}_2 ({\bm \gamma_1})$ are mappings 
from $S^3$ to $S^3$. They are not necessarily one-to-one but continuous. However, any attempt to search for Nash equilibrium using the methods as described above for the simple cases is useless.  

\subsection{Quantum Game with Finite Set of Strategies} \label{Subsec_Discret}
Since it is practically useless to follow the procedure of best response functions in the 6 dimensional space of pure strategies 
${\bm \gamma}_1 \otimes {\bm \gamma_2}$ we discretize the continuous variables $\phi,\alpha,\theta$ 
in a series of steps as follows: \cite{YA}
\begin{enumerate}
\item{} The variable $0 \le \theta \le \pi$ will assume $N_\theta$ values $\theta(1)=0<\theta(2)<\theta(3) \ldots < \theta(N_\theta)=\pi$. They are assumed to be equally spaced, the spacing is then $\frac{\pi}{N_\theta-1}$. 
\item{} For every $\theta(k_\theta)$ with $1<k_\theta < N_\theta$ the variable $0 \le \phi \le 2\pi$ will assume $N_\phi$ values \\
$\phi(1)=0<\phi(2)<\theta(3) \ldots < \phi(N_\phi)=2\pi$. They are assumed to be equally spaced, the spacing is then $\frac{2 \pi}{N_\phi-1}$. For $\theta(1)=0$ and for $\theta(1)=\pi$ the variable $\phi$ assumes the single value $\phi(1)=0$. 
\item{} For every $\theta(k_\theta)$ with $1<k_\theta < N_\theta$ the variable $0 \le \alpha \le 2\pi$ will assume $N_\alpha$ values \\
$\alpha(1)=0<\alpha(2)<\theta(3) \ldots < \alpha(N_\alpha)=2\pi$. They are assumed to be equally spaced, the spacing is then $\frac{2 \pi}{N_\alpha-1}$. For $\theta(1)=0$ and for $\theta(1)=\pi$ the variable $\alpha$ assumes the single value $\alpha(1)=0$. 
\item{} The total number of strategies of each player is the $N_S=(N_\theta-2)N_\phi N_\alpha+2$. 
\item{} We can now construct a $1 \leftrightarrow 1$ lexicographic order among triples $(\phi(k_\phi),\alpha(k_\alpha),\theta(k_\theta)$ of angles, corresponds to a single integer $1 \le I(k_\phi,k_\alpha,k_\theta) \le N_S$. For example, 
\begin{equation} \label{lexico1}
 I(k_\phi,k_\alpha,k_\theta) >  I(k'_\phi,k'_\alpha,k'_\theta) \ \mbox{if} \  \begin{cases} 
 k_\theta > k'_\theta \ \mbox{or} \\ k_\theta = k'_\theta \ \mbox{but} \ k_\phi>k'_\phi \ \mbox{or} \\ k_\theta = k'_\theta \ \mbox{and} \ k_\phi=k'_\phi \ \mbox{but} \ k_\alpha > k'_\alpha~. \end{cases}
 \end{equation}
In this way a set of three continuous variables $(\phi,\alpha,\theta)$ is replaced by a single discrete variable $1 \le I \le N_S$ that uniquely determine the $N_S$ triples $[\phi(I),\alpha(I), \theta(I)]$.  
\end{enumerate} 
\subsubsection{Definition of Quantum Game with Discrete set of Strategies}
The definition (\ref{32}) of the quantum game is then modified into, 
\begin{equation} \label{44}
 G_D=\la N=\{1,2 \}, |\psi_I\ra, \{ {\cal A}_i \}=\{ 1,2,\ldots N_S \} , J, u_i, P_i \ra,
 \end{equation}
where it is understood that player $i$ choosing a strategy $I_i$ operates on his qubit 
with the matrix $U(\phi(I_i),\alpha(I_i),\theta(I_i))$ defined in Eq.~(\ref{11}). 
\subsubsection{Nash Equilibrium in Quantum Game with Discrete set of Strategies}
\label{subsubsec_QGDis_NE}
Once a mesh structure and  and lexicographic ordering procedure  are completed, 
we are in the same situation as in \ref{subsubsec_finite}. 
In this way, the problem is amenable for being treated within the best response function formalism. For each strategy $I_1$ of player 1 player 2 finds its best response $q_2(I_1)$, and vice versa,  for each strategy $I_2$ of player 2 player 1 finds its best response $q_1(I_2)$. A pure strategy Nash equilibrium occurs if there is a 
 pair of strategies $(I_1^*,I_2^*) \ni [q_2(I_1^*)=I_2^* \wedge q_1(I_2^*)=I_1^*]$.  
In analogy with the definition (\ref{34}), a pure strategy Nash equilibrium of the game (\ref{42}) is a pair of strategies 
$(I_1^*,I_2^*)$ that determines two pairs of triples
\begin{equation} \label{45}
 [\phi(I_1^*),\alpha(I_1^*),\theta(I_1^*); \phi(I_2^*),\alpha(I_2^*),\theta(I_2^*)]=[{\bm \gamma}(I_1^*),{\bm \gamma}(I_2^*)],
 \end{equation}
 such that
\begin{equation} \label{46}
 P_1[{\bm \gamma}(I_1),{\bm \gamma}(I_2^*)] \le P_1[{\bm \gamma}(I_1^*),{\bm \gamma}(I_2^*)] \ \forall I_1, \ \  P_2[{\bm \gamma}(I_1^*),{\bm \gamma}(I_2)] \le P_2[{\bm \gamma}(I_1^*),{\bm \gamma}(I_2^*)] \ \forall I_2.
 \end{equation}
 
 \subsubsection{Weak Points of the Discrete Formulation} 
Admittedly, the are at least two disadvantages with this procedure. 
First, by turning a continuous variable into a discrete and finite sequence, we throw away an infinite number of possible strategies. It might be argued that a Nash equilibrium might occur in the original game with continuous space of strategies 
and that this equilibrium is skipped in the discrete version. For that reason, we 
regard the game $G_D$ defined in (\ref{44}) as a new game, and do not claim that it is 
a bona fide representative of the original game $G_Q$ defined in (\ref{32}). However, 
since all the payoffs are continuous functions of $\phi_i, \alpha_i$ and $\theta_i$, 
it is clear that when the number $N_\phi,N_\alpha,N_\theta$ of mesh points is very large, the results pertaining to $G_D$ approach those of $G_Q$, and this include the 
existence of Nash equilibrium.  

The second disadvantage is a bit more subtle: The set of discrete strategies 
does not form a group (see Appendix \ref{GT}). We already stressed that the set of 
$2 \times 2$ unitary matrices with unit determinant form a group, called $SU(2)$. 
A product of two matrices of the form (\ref{11}) can be written as a matrix of the same form, or, explicitly,
\begin{equation} \label{nogroup1}
U(\phi, \alpha,\theta)U(\phi', \alpha',\theta')=U(\phi'', \alpha'',\theta'')~,
\end{equation}
where each angle appearing on the right and side is a function of the six angles appearing on the left hand side, (the functional form is calculable straightforwardly). 
This is not the case with discrete strategies. A strategy obtained by an application of two discrete strategies one after the other does not, in general, belong to the original set of discrete strategies. This is mathematical flaw might be relevant in 
games that require repeated applications of strategies, but in the present case of 
single and simultaneous moves, it has no effect. 
\subsection{ Concrete Examples} \label{Sec_Example}
We have already stressed that for maximally entangled states there 
is no pure strategy Nash equilibrium in the quantum game $G_Q$ if the classical game $G_C$ has a Nash equilibrium  that is not Pareto efficient. suggested at the beginning of this Section,  we would first like to check what happens for partially entangled states. This is discussed in the first example.    In the second example we consider a quantum Bayesian game (a game with incomplete information) and obtain a pure strategy NE even for a maximally entangled state under the condition that in one of the classical games there is a Pareto efficient Nash equilibrium.

 \subsection{ Nash Equilibrium in the Quantum DA Brother Game}
The classical prisoner dilemma game presented by the table (\ref{PD}) 
(the entries are years in prison) is completely symmetric. We prefer to slightly break this symmetry using 
a variant of the prisoner dilemma game, called ``The DA Brother"\cite{MWG}. In this variant, prisoner 1 is a brother of the district attorney (DA).   The DA promises his 
 felony brother that if {\it both} prisoners confess, then he (the DA) will arrange that he (his criminal  brother) will not serve in jail.  The classical game is then presented by
the following table.
  \begin{center}
$~~~~~~~~$  Prisoner 2\\
\ \\
Prisoner 1 \ \ \begin{tabular}{|l| |l| |l| |l| |l| |l|} \hline  &~ I ~(C) ~  & ~Y~(D) \\ \hline ~I~(C)~ & 0,-2 & -10,-1 \\  \hline ~Y~(D)~ & -1,-10 & -5,-5  \\  \hline \end{tabular} 
\end{center}
Recall that in the classical version, 
the initial state of the system is $|00 \ra$ or $(C,C)$, namely the referee 
(the judge in this case)  tells the prisoners that he assumes that they both confess, 
but let them decide by choosing their classical strategies ${\bf 1}$ (stay as you are) 
or $Y$ (change your decision by flipping your bit from $|0 \ra$ to $|1 \ra$. 
Unlike the familiar classical prisoner dilemma game, where both players have a dominant strategy $Y$ 
(meaning don't confess) in the DA brother game player 2 has a dominant strategy $Y$ but player 1 does not. 
However, as in the familiar game, 
there is a pure strategy Nash equilibrium $(Y,Y)$ (both players flip their bit from 
$|0 \ra$=C to $|1 \ra$=D,  with penalties 
$(P_1,P_2)=(-5,-5)$ namely, each prisoner gets 5 years in 
prison after deciding not to confess. 

Now we study the pure strategy quantum game 
where each player has finite (albeit very large) number of strategies. 
Specifically, we take $N_\theta=9, \ \ N_ \phi=N_ \alpha=17$ so, according to the 
calculation before Eq.~(\ref{lexico1}),  each player has $N_S=2025$ strategies. 
The entanglement operator, $J$ is defined in Eq. (\ref{22d}) and the amplitudes 
$a,b,c,d$ are explicitly given in Eq.~(\ref{38}), where the angles $\phi_i, \alpha_i, \theta_i, i=1,2$ covers the discrete mesh as $I_i$ runs from 1 to $N_S=2025$, and $\beta$ is the entanglement parameter as explained before Eq.~(\ref{38}). The corresponding years in prison
are specified in Eq.~(\ref{25}), and given explicitly in terms of the amplitudes 
$a,b,c,d$ and the utility functions in the table, \\
$~~~~~~~~~~~~~~~~~P_1=0 \times |a|^2-10|b|^2-1 \times |c|^2-5 \times |d|^2, \ \ P_2=-2|a|^2-1 \times |b|^2-10 |c|^2-5|d|^2.$

First we verified that in the maximally entangled case $\beta=\pi/2$ the utility functions do not coincide even at a single point. Then we decrease $\beta$ in small steps and and find that for $\gamma>1.2$  there is no pure strategy Nash equilibrium. However, for $\beta < 1.2$ we found a   pure strategy Nash equilibrium. For $\beta=1$ this is exemplified in the following three figures. \\
\begin{figure}[!h]
\centering
\includegraphics[width=6truecm, angle=90]{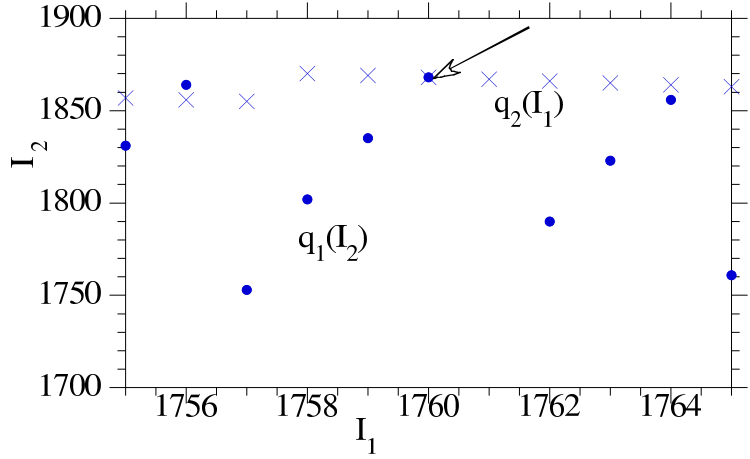}
\caption{\footnotesize{Best response functions $q_1(I_2)$ and $q_2(I_1)$ for the quantum DA brother game for entangled parameter (angle) $\beta=1$. The discretized version yields an intersection point (Nash equilibrium strategies) at
$(I_1^*=q_1(I_2^*)=1760,I_2^*=q_2(I_1^*)=1868)$. The axes domains should extend between 0 and 2025 but we focus on the region where the discrete "curves" intersect. }}
\label{f12}
\end{figure}\\
\ \\
First, in Fig.~\ref{f12},
the discrete best response functions are plotted in the small range between 1700 and 2000 in order to magnify the region where they meet at the point $I_1^*=1760, I^*_2=1868$ marked by white arrow in the figure. Due to the lexicographic ordering, the best response functions do not show any kind of regularity of course. But the coincident point is robust as is verified  in the next couple of figures, 
The Nash equilibrium for the pair of strategies $I_1^*=1760, I^*_2=1868$ is found  as an internal solution (the angles are not at the edge of their respective domains). For this value of the entanglement parameter $\beta=1$, the ``payoffs" (equql to minus number of years in jail) are\\
$~~~~~~~~~P_1=-1.45 > -5, \ \ P_2=-2.83 > -5,$\\ 
so both prisoners are much better off with the quantum version compared with the classical one. 
\begin{figure}[!h]
\centering
\includegraphics[width=6truecm, angle=90]{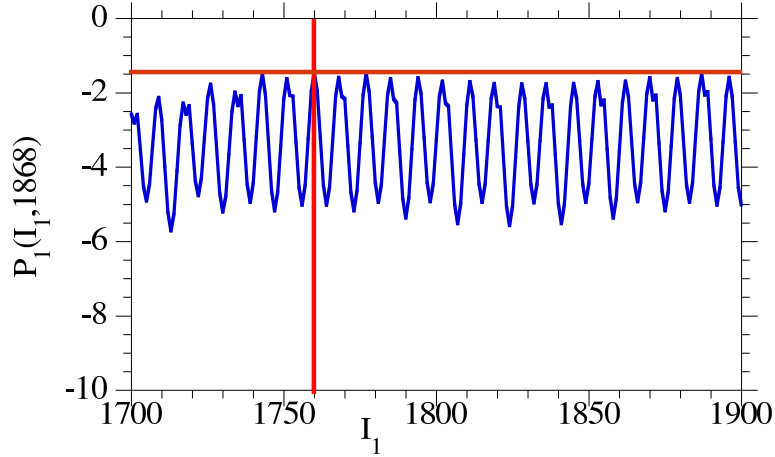}
\caption{\footnotesize{ For the same conditions of Fig.~ \ref{f12}, this figure shows the ``payoff" of the first prisoner  $P_1(I_1,I_2^*=1868)$ as function of $I_1$, showing maximum at $(I_1^*=1760)$. The payoff is negative because it is defined as
minus the number of years in jail. }}
\label{p1}
\end{figure}
\begin{figure}[!h]
\centering
\vspace{-0.1in}
\includegraphics[width=6truecm, angle=90]{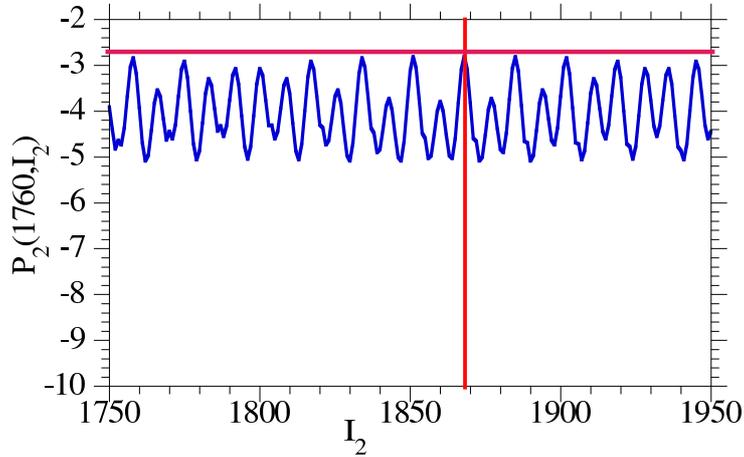}
\caption{\footnotesize{ For the same conditions of Fig.~ \ref{f12}, this figure shows the ``payoff" of the second prisoner  $P_2(I_1^*=1760,I_2)$ as function of $I_2$, showing maximum at  
$(I_2^*=1868)$. The payoff is negative because it is 
defined as minus the number of years in prison. }}
\label{p2}
\end{figure}
Let us then summarize the results as displayed in Figs.~\ref{f12}, 
\ref{p1}, \ref{p2} relevant for the quantum DA brother game at 
partial entanglement with $\beta=1$. 
\begin{enumerate}
\item{} Fig.~\ref{f12} shows that the two best response functions 
$q_1(I_2)$ and $q_2(I_1)$ intersect at $(I_1^*=1760,I_2^*=1868)$. This point defines a Nash equilibrium corresponding to pair of strategies $(I_1^*,I_2^*)$. The corresponding angles $\phi_1(I_1^*), 
\alpha_1(I_1^*), \theta_1(I_1^*)$ and $\phi_2(I_2^*), 
\alpha_2(I_2^*), \theta_2(I_2^*)$  that define the strategy matrices of players 1 and 2 according to Eq.~(\ref{11}) are not specified.  
\item{} Fig.~\ref{p1} shows that the first prisoner cannot improve his status compared with $P_1(I_1^*=1760,I_2^*=1868)$ if prisoner 2 sticks to his strategy $I_2^*=1868$, namely, $P_1(I_1,I_2^*) \le P_1(I_1^*,I_2^*), \ \forall I_1$. 
\item{} Similarly, Fig.~\ref{p2} shows that the second prisoner cannot improve his status compared with $P_2(I_1^*=1760,I_2^*=1868)$ if prisoner 1 sticks to his strategy $I_1^*=1760$, namely, $P_2(I_1^*,I_2^*) \le P_2(I_1^*,I_2), \ \forall I_2$. 
\end{enumerate}
\subsubsection{Upper Bound on the Degree of Entanglement}
\begin{figure}[!h]
\centering 
\includegraphics[width=7truecm, angle=90]{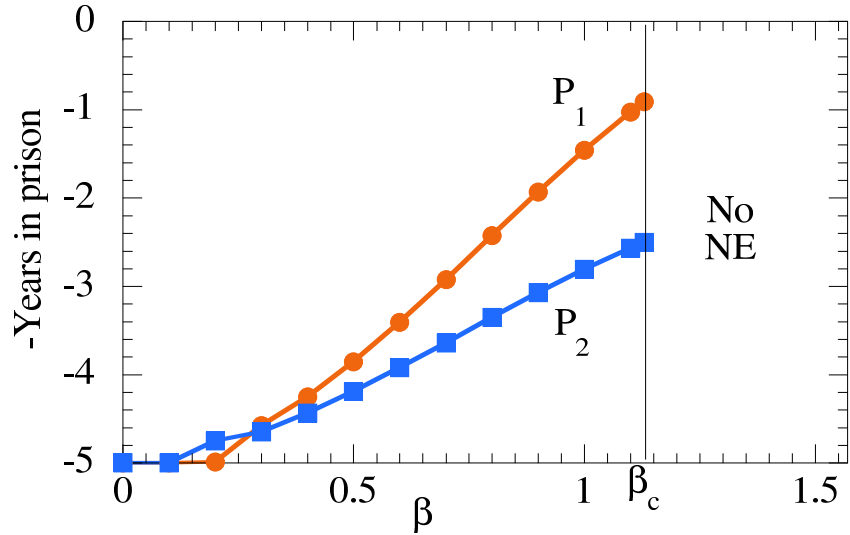}
\caption{\footnotesize{ Demonstration of threshold entanglement constant $\beta_c$ above which there is no pure strategy Nash Equilibrium in the DA brother quantum game. The figure shows the payoffs of the two prisoners (-minus number of years in prison) for each value of $\beta$ for which Nash equilibrium exists. There is no Nash equilibrium above $\beta_c=1.13$. }}
\label{Pbeta}
\end{figure}

The discussion above leads us to the following scenario: For $\beta=0$ 
there is no entanglement and the players reach the classical Nash equilibrium through the strategies $Y \otimes Y$, that entails payoffs 
(-5,-5), namely, they do not confess and get five years in jail each. On the other hand, at maximal entanglement $\beta=\pi/2$ the is no Nash equilibrium, as we have rigorously proved. We have also found Nash equilibrium in the partially entangled quantum game for $\beta=1$ with payoffs $P_1=-1.45, \ P_2=-2.83$, much better than the classical ones. 
Therefore, it is reasonable to suggest that as $\beta$ is varied continuously between 0 and $\pi/2$ the payoffs improve above the classical ones, until there is some upper bound $0<\beta_c<\pi/2$ above 
which there is no Nash equilibrium anymore.  We test this conjecture numerically by tracing the payoffs of the two prisoners as function of $\beta$. The results are displayed in Fig.~\ref{Pbeta}
The conclusions that can be drawn from figure \ref{Pbeta} are as follows:
\begin{enumerate}
\item{} There is a small region above $\beta=0$ where each player sticks to his classical strategy\cite{Flitney1}.
\item{} Pure strategy Nash equilibrium in the quantum game exists for $0 \le \beta \le \beta_c \le \pi/2$ where $\beta_c$ depends on the classical payoff functions. 
\item{} As long as pure strategy Nash equilibrium in the quantum game exists, 
(namely $\beta < \beta_c$) the payoffs are higher than the classical ones and they increase monotonically with the entanglement parameter $\beta$.     
\item{}  I speculate that the payoff curves in Fig.~\ref{Pbeta} extrapolate to 
$(P_1,P_2)=(0,2)$ which is the classical payoffs for the strategies (C,C). This means that for $\beta \le \beta_c$ higher entanglement draws people toward cooperation. 
\end{enumerate} 

\end{spacing}
\begin{spacing}{1.5}
\begin{center}
\section{\small Advanced Topics}\label{Section6}
\end{center}
In this Section we shall briefly some advanced topics. These include Mixed Strategy Quantum Games 
in section \ref{Sect_Mix}, Bayesian Quantum Games in section \ref{Sect_Bayes}, and 
quantum games based on two-player three-strategies classical games, that require the 
introductions of qutrits (an extension of the notion of qubit for the case of a three bit basis). 
 \subsection{Mixed Strategies} \label{Sect_Mix}
 In Section \ref{Section4} we used the best response functions $B_2({\bm \gamma}_1)$, Eq.~(\ref{B2a}), and 
 $B_1({\bm \gamma}_2)$, Eq.~(\ref{B1a}), and showed that Starting from a non-entangled initial state (for example $|00 \ra$ and using entanglement operators $J$ as defined in Eqs.~(\ref{22}) or (\ref{22a}) leading to the maximally entangled states $|\psi_+\ra$ and $|T \ra$  respectively, the quantum game has no pure strategy NE. This naturally motivates the quest for defining quantum games with mixed strategies that might lead to mixed strategy Nash equilibria. 
 
  In subsection \ref{Subsec_Mix_Finite} we define a  mixed strategy quantum game with finite number of pure strategies, 
  and its mixed strategy Nash equilibrium. Then, in subsection \ref{Subsec_Mix_Example}
 we give an example of the existence of a mixed strategy Nash equilibrium\cite{Chen} in a quantum game 
 with maximal entanglement, where we proved that pure strategy Nash equilibrium does not exist. 
 Finally, in subsection \ref{Subsec_Mix_Gen} we will specify the general structure of mixed strategies in quantum games 
 based on 2-players 2-strategies classical game and cite a theorem by Landsburg pertaining to their existence. 
 \subsection{Mixed Strategy Quantum Game with Finite Number of Pure Strategies} \label{Subsec_Mix_Finite}
 When the number of points in each player's strategy set is continuously infinite [such is the number of ${\bm \gamma}_i=(\phi_i,\alpha_i,\theta_i)]$ 
 the definition of mixed strategy requires the notion of distribution over a continuous space. 
 This will be briefly carried out 
 in subsection \ref{Subsec_Mix_Gen}. But it is useful to start with the simpler case where each player $i$
 has finite number $K$ of strategies, say ${\bm \gamma}_i(k), \ \ k=1,2,\ldots,K$, as 
 we discussed in our numerical approach formalism in Section \ref{Section5}.  If $K$ is very large, 
 the situation approaches the continuum limit. For each choice of strategies $({\bm \gamma}_1(k_1),{\bm \gamma}_2(k_2))$ 
 the (absolute value squared of the) amplitudes $a,b,c,d$ will depend 
 on $({\bm \gamma}_1(k_1), {\bm \gamma}_2(k_2))$ where $k_1,k_2=1,2,\ldots,K$. 
 The explicit functional relation depends on the details of the game played. For example, with maximally entangled $J$ leading to $|\psi^+\ra$ the functional form is given in Eq.~ \ref{29}, whereas for partially 
 entangled $J$ the functional form is given in Eq.~ \ref{38}. 
 For short notation we write 
 $a({\bm \gamma}_1(k_1), {\bm \gamma}_2(k_2))=a(k_1,k_2)$ and similarly for $b,c,d$. 
 
 In a mixed strategy quantum game with finite pure strategy spaces $A_i=\{ {\bm \gamma}_i(k_i), k_i=1,2,\ldots K \}$ each player chooses strategy ${\bm \gamma}_i(k_i)$ with probability $0 \le p_i(k_i) \le 1$ such that  $\sum_{k_i=1}^K p_i(k_i)=1$.  A given sequence of $K$ probabilities for player $i$ is shortly denoted as
 ${\bf p}_i= [p_i(1),p_i(2),\ldots p_i(K)] $.
 Formally, the set $\{ {\bf p}_i \}$ of all such $K$-tuples  is a set of probability distributions over the strategy set ${\cal A}_i =\{ {\bm \gamma}_i(k_i), \ k_i=1,2,\ldots K. \}$ A {\it profile} of mixed strategies ${\bf p}={\bf p}_1 \times {\bf p}_2$ 
 induces a probability distribution on  ${\cal A}={\cal A}_1 \times {\cal A}_2$. For a given strategy profile ${\bf p}={\bf p}_1 \times {\bf p}_2$, assuming independent randomization, the probability of an action profile ${\bm \gamma}_1(k_1) \times  {\bm \gamma}_2(k_2) \in {\cal A}$ is $p_1(k_1))p_2(k_2)$. 
 The payoff $P_i({\bf p}_1,{\bf p}_2)$ of player $i$ in a mixed strategy game will then be, 
 \begin{equation} \label{Paymix1}
 P_i({\bf p}_1,{\bf p}_2)=\sum_{k_1,k_2=1}^2 p_1(k_1)p_2(k_2) \left [ |a(k_1,k_2)|^2u_i(0,0)+ |b(k_1,k_2)|^2u_i(0,1)+|c(k_1,k_2)|^2u_i(1,0)+ |d(k_1,k_2)|^2u_i(1,1)) \right ]~.
 \end{equation}
We are now in a position to formulate\\
\underline{Definition:} A {\it mixed strategy quantum game $G_{\mathrm{Q,mixed}}$} based on two-player 2-strategy classical game $G_C$ is the collection (in all places $i \in N$), 
\begin{equation} \label{GQmixed}
G_{\mathrm{Q,mixed}}=\left \la N=\{1,2 \}, |\psi_I \ra, \{ {\cal A}_i \}, \{ {\bf p}_i \}, J, P_i \right \ra,
\end{equation}
 where $\{ {\bf p}_i \}$  is the set of probability distributions over the strategy set ${\cal A}_i =\{ {\bm \gamma}_i(k_i), \ k_i=1,2,\ldots K. \}$, and $P_i: {\bf p}_1 \times {\bf p}_2 \to \mathbb{R}$ assign to each player the payoff according to the prescription (\ref{Paymix1}). The other entries are as defined in the pure strategy game back in Eq.~(\ref{32}). \\
 \underline{Definition:} A mixed strategy Nash Equilibrium of the quantum game $G_{\mathrm{Q,mixed}}$ is a 
 pair of strategies $({\bf p}_1^*,{\bf p}_2^*)$ such that, 
 \begin{equation} \label{MSNE}
 P_1({\bf p}_1,{\bf p}_2^*) \le P_1({\bf p}_1^*,{\bf p}_2^*) \ \forall {\bf p}_1, \ \ P_2({\bf p}_1^*,{\bf p}_2) \le P_2({\bf p}_1^*,{\bf p}_2^*) \ \forall {\bf p}_2~.
 \end{equation}
 
 \subsection{Simple Example of Mixed Strategy Nash Equilibrium in Quantum Games} \label{Subsec_Mix_Example}
\begin{figure}[!h]
\centering
\includegraphics[width=6truecm, angle=0]{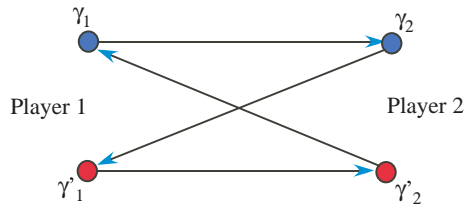}
\caption{\footnotesize{Mixed strategy Nash equilibrium where the ``ping-pong" exchange of best response functions is closed (see text for details).  }}
\label{Fig_6.1}
\end{figure}

 The fact that in the pure strategy game with maximal entanglement each player 
 has a best response that forces his opponent to cooperate while he does not 
 prevents the occurrence of pure strategy Nash equilibrium but seems to be useful in 
 searching an example for mixed strategy Nash equilibrium. 
 The analysis below can be followed by looking at Fig.~\ref{Fig_6.1}. 
Suppose player 1 choses his strategy randomly as ${\bm \gamma}_1=(\phi_1,\alpha_1, \theta_1)$.
 If  player 2 knows that, he (player 2) choses his best response 
 ${\bm \gamma}_2({\bm \gamma}_1)=(\phi_2,\alpha_2,\theta_2)$ according to the prescription 
 specified in Eq.~(\ref{B2a}). This will lead to the case $|b|^2=1$ and $|a|^2=|c|^2=|d|^2=0$ in which case 
 prisoner 1 will spend 6 years in prison and prisoner 2 will spend only two years in prison. 
If player 1 knows that, he will chose the corresponding best response to ${\bm \gamma}_2$ as
${\bm \gamma}_1' [{\bm \gamma}_2({\bm \gamma}_1)]$ according to the prescription of 
Eq.~(\ref{B1a}). This will lead to the case $|c|^2=1$ and $|a|^2=|b|^2=|d|^2=0$ in which case 
 prisoner 1 will spend only 2 years in prison and prisoner 2 will spend 6 years in prison. 
 As a response, player 2 choses his best response 
${\bm \gamma}_2'\{ {\bm \gamma}_1' [{\bm \gamma}_2({\bm \gamma}_1)] \}$ again according to the prescription 
 specified in Eq.~(\ref{B2a}). 
By inspecting the best response functions in Eqs.~(\ref{B2a}) and (\ref{B1a}), however, it is not difficult to show that the best response of player 1 to the final move ${\bm \gamma}_2'\{ {\bm \gamma}_1' [{\bm \gamma}_2({\bm \gamma}_1)] \}$ of player 2 is, according to  the prescription 
 specified in Eq.~(\ref{B1a}), simply ${\gamma}_1$, and the chain is hence closed. 

Once the strategy ${\bm \gamma}_1$ is chosen by player 1, all the other three strategies 
${\bm \gamma}_2, {\bm \gamma}_1'$ and ${\bm \gamma}_2'$ are 
uniquely determined. Let us consider the quantum prisoner dilemma based on the classical 
game presented by table (\ref{PD}). 
Suppose now that player $i=1,2$ chooses the strategy ${\bm \gamma}_i$ with probability 1/2 and the strategy ${\bm \gamma}'_i$ with probability 1/2. Then, prisoner 1 has a 50\% chance that the final state 
will be $|\Psi \ra=|10 \ra$ and thereby get a penalty of two years in prison and 50\% chance that the final state  will be $|\Psi \ra=|01 \ra$ and thereby get a penalty of six years in prison. The converse is with prisoner 2. Thus, on the average, each one gets four years in prison, better than the classical 
result of five years in prison. The fact that the strategies are determined as best responses and that the game is symmetric
guarantee that this is indeed a mixed strategy Nash equilibrium. 

It is useful to stress that although each players chooses to bet on two strategies, the game as described above is 
not a quantum game with finite number of strategies in the sense defined in Eq.~(\ref{GQmixed})  because because 
in $G_{\mathrm{Q,mixed}}$ the strategies $\{ {\bm \gamma}_i(k_i) \}$ are fixed a-priori, and cannot be adjusted. 
Thus, every player must have the capability of choosing whatever strategy point he wishes. However, based on our results  with the numerical algorithm with finite but large number of strategies, this difficulty can be alleviated. 
\subsection{General Form of a Mixed Strategy Quantum Game} \label{Subsec_Mix_Gen}
In subsection \ref{Subsec_Mix_Finite} we discussed mixed strategy quantum game with finite number of 
quantum strategy. 
In the previous subsection \ref{Subsec_Mix_Example} we gave a particular example of mixed strategy 
that also proved to lead to a mixed strategy Nash equilibrium for a quantum game where each player 
has all the allowed quantum strategies $\{ {\bm \gamma}_i \}$, but he chooses but two strategies 
with probabilities with probabilities $p_i$ and $1-p_i$. We need to formulate a possible mixed strategy 
where each player can choose every subset out of all possible strategies with whatever probability 
he likes. In that case,    
the most general form of mixed strategy for a player  is determined by a distribution function 
$\rho(\phi, \alpha, \theta)$ such that the strategy is given by 
\begin{equation} \label{42} 
\mbox{Mixed Strategy}=M=\int \rho (\phi,\alpha, \theta)U(\phi,\alpha, \theta) d\phi d \alpha \sin 2 \theta d \theta
\end{equation}
The product $d \phi d \alpha \sin 2 \theta d \theta$
 is the surface element on the sphere $S^3$ (remember that 
a given strategy ${\bm \gamma}=(\phi, \alpha, \theta)$ is a point on the sphere $S^3$). 
Its integral gives the surface of $S^3$, which, for radius $R=1$ gives $S_3=2 \pi^2$ (recall that 
the surface of $S^2$ (our usual sphere, the globe) is $4 \pi$. Thus, if a player prefers a 
uniform distribution, he chooses $\rho (\phi,\alpha, \theta)U(\phi,\alpha, \theta)=\frac{1}{2 \pi^2}$ 
(but it is easy to show that it does not lead to mixed strategy Nash equilibrium). 
This formalism includes the strategies used
in the game discussed in subsection \ref{Subsec_Mix_Example} as a special case.
If a player wants to chose a strategy $[\phi(1),\alpha(1), \theta(1)]$ with probability $p$ and 
another strategy $[\phi(2),\alpha(2), \theta(2)]$ with probability $1-p$ he takes
\begin{equation} \label{rhospecial}
\rho (\phi,\alpha,\theta)=\frac{1}{\sin 2 \theta}[p \delta(\phi-\phi(1))\delta(\alpha-\alpha(1))\delta(\theta-\theta(1))+(1-p) \delta(\phi-\phi(2))\delta(\alpha-\alpha(2))\delta(\theta-\theta(2))],
\end{equation} 
where $\delta(.)$ is the Dirac delta function. 

To compute the payoffs in a mixed strategy game with mixed strategy profile $\rho_1(.) \times \rho_2(.)$ 
we assume that   
Player $i$ chooses the strategy  $(\phi_i,\alpha_i, \theta_i)$ 
and find the final states $|\Psi \ra$ as in Eq.~(\ref{24}), 
where each complex amplitude $a,b,c,d$ depends on $(\phi_1,\alpha_1, \theta_1;\phi_2,\alpha_2, \theta_2)$
Then, instead of Eq.~(\ref{25}), 
the expected payoff of player $i$ is then,
\begin{eqnarray}
&& P_i(\rho_1,\rho_2)=\int \rho_1 (\phi_1,\alpha_1, \theta_1)\rho_2 (\phi_2,\alpha_2, \theta_2)
\nonumber \\
&& = \left [|a|^2u_i(0,0)+|b|^2u_i(0,1)+|c|^2 u_i(1,0)+|d|^2 u_i(1,1) \right ]
[d\phi_1 d \alpha_1 \sin 2 \theta_1 d \theta_1]
[d\phi_2 d \alpha_2 \sin 2 \theta_2 d \theta_2]~.
 \label{43} 
\end{eqnarray}
The formal definition of a mixed strategy quantum game with infinitely continuous strategy sets 
is a direct extension of the definition  (\ref{GQmixed} ) with $\rho_i$ instead of ${\bf p}_i$, and similarly, 
the definition of a mixed strategy Nash equilibrium follows from that of Eq.~(\ref{MSNE}). 

At first sight, the quest for finding mixed strategy Nash equilibria for this general case is virtually hopeless, due to the complexity of the strategy spaces. However, in a recent paper \cite{Landsburg2}, Landsburg proved that the set of possible mixed strategy Nash equilibria is remarkably simple. The conditions for the 
theorem and the detailed results will not be specified here, but the main result is that the corresponding strategies (distributions) $\rho_1^*$ and $\rho_2^*$  are supported at a small number (3 or 4) of isolated points 
on $S^3$. Namely, $\rho_1^*$ and $\rho_2^*$ have the structure displayed in Eq.~(\ref{rhospecial}) 
except that the number of terms might be 3 or 4 instead of 2 in Eq.~(\ref{rhospecial}). 

%

\subsection{Bayesian Quantum Games} \label{Sect_Bayes}
While the topic of
 quantum games with  full information received considerable attention, the topic of  quantum games with incomplete information is less studied\cite{IqbalBG}. In this Section we extend the game procedure 
 developed in Section \ref{Section4}\cite{EWL} to include also quantum Bayesian games. 
 We shall carry it out by following a simple example derived from the full information DA brother game. 
 Following the protocol suggested by Harsanyi\cite{Harsanyi} for classical games with incomplete information, we will analyze a quantum Bayesian game with two types of prisoners denoted as 2I and 2II facing the DA brother prisoner.  It will be shown that when the game played between the DA brother and prisoner type 2II has a Pareto efficient Nash equilibrium, the quantum Bayesian game in which both types 
 face the DA brother have a pure strategy Nash equilibrium even with maximal entanglement. 

 \subsection{Example: Two Types of Prisoners Facing the DA Brother}   
 In order to introduce quantum games with imperfect information we will 
start with a simple classical game and quantize it. In the DA brother game discussed above, 
prisoner 1 (the DA brother) might now face two types of prisoner 2: Type 2I (probability $\mu$) is the same prisoner 2 from the previous game. He is sure that if he does not confess he will get either one year or five  years in prison 
depending on whether prisoner 1 confesses or not. But type  type 2II (probability $1-\mu$) is afraid that by not confessing he will get six more  years in prison. The game table then looks as follows.  
\begin{center}
$~~~~~~~~$  Prisoner 2I $~~~~~~~~~~~~~~$  Prisoner 2II  \\
Prisoner 1 \ \ \begin{tabular}{|c| c| c| c| c| c|} \hline  &~ I(C) ~  & ~Y(D) \\ \hline ~I(C)~ & 0,-2 & -10,-1 \\  \hline ~Y(D)~ & -1,-10 & -5,-5  \\  \hline \end{tabular} 
\ \ \  \ \ \begin{tabular}{|c| c| c| c| c| c|} \hline  &~ I(C) ~  & ~Y(D) \\ \hline ~I(C)~ & 0,-2 & -10,-7 \\  \hline ~Y(D)~ & -1,-10 & -5,-11  \\  \hline \end{tabular} 
\end{center}
\subsubsection{The Classical Version \cite{MWG}}
The classical Nash equilibrium  is simple to find. Player type 2I has a dominant strategy of not confessing while type 2II has a dominant strategy to confess. Note also that in the game played between prisoner 1 and type 2II of prisoner 2, the strategy 
(I,I) $\to $ (C,C) is Pareto efficient. Assuming the two types of player 2 stick to their dominant strategies then if player 1 confesses he gets $-10 \mu+0(1-\mu)$ while if he does not he gets $-5 \mu-1(1-\mu)$. Therefore, player 1 strategy is 
\begin{equation} \label{47}
\mbox{player 1 strategy}=\begin{cases}  I \ \ \mbox{(confess) if} \ \ -10 \mu+0(1-\mu)> -5 \mu-1(1-\mu) \ \ \Rightarrow \mu < \frac{1}{6}, \\
 Y \ \  \mbox{(don't confess) if} \ \ -10 \mu+0(1-\mu)< -5 \mu-1(1-\mu) \ \ \Rightarrow \mu >\frac{1}{6}, \\ 
 \mbox{indifferent if} \ \ \mu=\frac{1}{6}~. \end{cases}
 \end{equation}
 The Nash equilibrium and the corresponding ``payoffs" are, 
 \begin{equation} \label{48} 
 \mbox{Nash equilibrium}=\begin{cases} (IYI) \to (CDC) \ \ (-10 \mu, -1,-2) \ \ \mu<\frac{1}{6}, \\
(YYI) \to  (DDC) \ \ (-1-4 \mu, -5,-10)  \ \ \mu> \frac{1}{6}. \end{cases}
 \end{equation}
 Henceforth, the game as defined above is referred to as 
 {\it The classical DA brother Bayesian game}. 
 \subsubsection{Definition of a Pure Strategy Quantum Bayesian Game}
 \vspace{-0.0in}
 A formal definition of a quantum Bayesian game is now in order. Since we limit our formulation of 
 classical games in terms of bits (and remembering that each bit can get two values $0$ or $1$), we will 
 limit our discussion to quantum Bayesian games in which analogous classical games there are two possible decisions.   Let 
 \begin{equation} \label{CB}
  G_{CB}=\la N,\{S_i \}, \{ u_i(.) \}, {\bm M},F(.)], \ \ N=\{1,2,\ldots n \}, i=1,2,\ldots,n, F(.) \ra~,
  \end{equation}
 denote a classical Bayesian game.  Here $S_i=(I,Y)$ is the classical strategy set for player $i$, $u_i(s_i,s_{-i}, \mu_i)$ is a payoff function of player $i$ where $\mu_i \in M_i$ is a random variable generated by nature that is observed only by player $i$. The joint probability distribution of $\mu_i$, $F(\mu_1,\mu_2,\ldots \mu_n)$ is a common knowledge and ${\bf M}=\times_{i=1}^n M_i$. Then a pure strategy quantum Bayesian game is defined as, 
  \begin{equation} \label{QB}
  G_{QB}=\la N,\{ {\bm \gamma}_i \}, \{ u_i(.) \}, {\bm M},F(.), J, P_i, \ \ N=\{1,2,\ldots n \}, i=1,2,\ldots,n, F(.) \ra~,
  \end{equation}
 where the definitions to be modified compared with $G_{CB}$ are as follows: 1) ${\bm \gamma}_i=(\phi_i,\alpha_i, \theta_i)$ is the set of angles that determine the the quantum strategy $U(\phi_,\alpha_i, \theta_i)$ 
 according to Eq.~(\ref{11}). 2) $J$ is the entanglement operator fixed by the referee.
  3) $P_i({\bm \gamma}_i, {\bm \gamma}_{-i}, \mu_i)$ are the payoff of player $i$ determined by the 
  quantum rules, see Eq.~(\ref{25}) for $P_i$ defined for the full information game.  A modification 
  required for the Bayesian game is explicitly given below in Eq.~(\ref{49}). 
 \subsubsection{The DA brother Quantum Bayesian Game}
 Now let us concentrate on the quantum version of the 
 DA brother Bayesian game. Our discussion here will focus on the general formulation and will not enter the discretization and numerical 
  formalism. The strategies are determined 
  by the three angles chosen by each player 1,2I and 2II 
  \begin{equation} \label{gammaBayes} 
  {\bm \gamma}_1=(\phi_1, \alpha_1, \theta_1),  \ {\bm \gamma}_{2 I}=(\phi_{2 I}, \alpha_{2 I}, \theta_{2 I}),  \ {\bm \gamma}_{2 II}=(\phi_{2 II}, \alpha_{2 II}, \theta_{2 II}).
  \end{equation} 
  That leads according to Eq.~(\ref{11}) to the three matrices $U({\bm \gamma}_1), U({\bm \gamma}_{2I}), U({\bm \gamma}_{2II}) $. 
Each type of player 2,  namely, 2I and 2II faces player 1 and the quantum game between them is conducted according to the rules specified in Section \ref{Section3} especially Fig.~\ref{Fig1}. Each game results in the corresponding  final state (the subscripts should include also payer 1 but it is omitted for convenience)
  \begin{equation} \label{Eq_Psi12I2II}
  |\Psi_{ 2I} \ra= a_{ 2I}|00 \ra+b_{ 2I}|01 \ra+c_{ 2I}|10 \ra+d_{ 2I}|11 \ra, \  
   |\Psi_{ 2II} \ra= a_{ 2II}|00 \ra+b_{ 2II}|01 \ra+c_{ 2II}|10 \ra+d_{ 2II}|11 \ra.
   \end{equation}
   As we shall see below, these two final states determine the payoff of all three players, including player 1.  
   The coefficients in the expression for  $ |\Psi_{ 2I} \ra$ depend on ${\bm \gamma}_1, {\bm \gamma}_{2I}$  
   and the coefficients in the expression for  $ |\Psi_{ 2II} \ra$ depend on ${\bm \gamma}_1, {\bm \gamma}_{2II}$. Explicit expressions for the coefficients depend on the entanglement operator $J$ that is used by the referee. Below we will concentrate on the case of maximal entanglement matrix $J=J_1$, with $J|00 \ra=|\psi^+\ra$ as defined in Eqs.~(\ref{22} and (\ref{J1}). The coefficients are given in Eq.~(\ref{29}) 
   wherein for player of type 2I the angles  $\phi_2,\alpha_2,\theta_2$ 
   in Eq.(\ref{29}) are to be replaced by $\phi_{2I},\alpha_{2I},\theta_{2I}$  and 
   for player of type 2II the angles  $\phi_2,\alpha_2,\theta_2$ 
   in Eq.(\ref{29}) are to be replaced by $\phi_{2II},\alpha_{2II},\theta_{2II}$. 
    Following the expressions for the 
 payoff function as in Eq.~(\ref{25}) and the present game 
 tables, the corresponding payoffs are, 
  \begin{eqnarray} \label{49} 
 && P_{2I}({\bm \gamma}_1; {\bm \gamma}_{2I})=|a_{2I}|^2\times(-2)+|b_{2I}|^2\times(-1)+|c_{2I}|^2 \times(-10)+|d_{2I}|^2 \times(-5)~\nonumber \\
  && P_{2II }({\bm \gamma}_1; {\bm \gamma}_{2II })=|a_{2II }|^2\times(-2)+|b_{2II }|^2\times(-7)+|c_{2II }|^2 \times(-10)+|d_{2II }|^2 \times(-11)~\nonumber \\
   && P_1({\bm \gamma}_1; {\bm \gamma}_{2I},{\bm \gamma}_{2II })=\overline {|a|^2} \times 0+\overline{|b|^2}\times(-10)
   +\overline{|c|^2} \times(-1)+\overline{|d|^2} \times(-5)~\nonumber \\
   &&\overline {|a|^2}= \mu |a_{2I}|^2+(1-\mu) |a_{2II}|^2, \ \overline{|b|^2}= \mu |b_{2I}|^2+(1-\mu) |b_{2II}|^2, 
   \nonumber \\ 
 && \overline{|c|^2}= \mu |c_{2I}|^2+(1-\mu) |c_{2II}|^2, \ \overline{|d|^2}= \mu |d_{2I}|^2+(1-\mu) |d_{2II}|^2. \
\end{eqnarray} 
\subsubsection{ Definition of a Pure Strategy Nash Equilibrium in  Quantum Bayesian Game}
We will define the pure strategy Nash equilibrium for the specific game under study, but a generalization to an arbitrary game as defined in Eq.~(\ref{QB}) is straightforward. 
A pure strategy Nash equilibrium for the quantum Bayesian game 
derived from the classical DA brother Bayesian game 
is  the triple of strategies  $({\bm \gamma}_1^*, {\bm \gamma}_{2I}^*,{\bm \gamma}_{2II}^*)$ (if it exists, recall that each ${\bm \gamma}$ stands for three angles $\phi, \alpha, \theta$ 
as in Eq.~(\ref{gammaBayes})) that satisfies, 
 \vspace{-0.1in}
 \begin{eqnarray} \label{52}
&& P_1({\bm \gamma}_1,{\bm \gamma}_{2I}^*,{\bm \gamma}_{2II}^*) \le P_1({\bm \gamma}_1^*,{\bm \gamma}_{2I}^*,{\bm \gamma}_{2II}^*) \ \forall {\bm \gamma}_1, \ \  P_{2I}({\bm \gamma}_1^*,{\bm \gamma}_{2I}) \le P_{2I}({\bm \gamma}_1^*,{\bm \gamma}_{2I}^*) \ \forall {\bm \gamma}_{2I},\nonumber \\
&& P_{2II}({\bm \gamma}_1^*,{\bm \gamma}_{2II}) \le P_{2I}({\bm \gamma}_1^*,{\bm \gamma}_{2II}^*) \ \forall {\bm \gamma}_{2II}.
 \end{eqnarray}
 \vspace{0.in}
 \subsubsection{Nash Equilibrium Despite Maximal Entanglement}
 We have seen in subsection \ref{Subsec_Absence} that when $J$ is leads to maximally  entangled state $|\psi^+\ra$ there is no pure strategy Nash equilibrium in 
 the (full information) game played between player 1 and player 2I. One of the conditions for the proof of this negative result is that there is no Pareto efficient 
 pure strategy Nash equilibrium in the classical game, which is indeed the case 
 as far as the game between player 1 and player 2I is concerned. On the other hand, in the classical game played between players 1 and 2II the profile of strategies ${\bf 1} \otimes {\bf 1}$ (both confess) is a Pareto efficient pure strategy Nash equilibrium. What can be said about the Quantum version? Intuitively, we expect that for small $\mu$, player 2II will dominate and the game will have a pure strategy Nash equilibrium, but at some value of $\mu$ player 2I will dominate and 
 there will be no equilibrium. We show below that this is indeed what happens, and that the critical value of $\mu$ is 1/6, that is exactly the value where, in the classical game, player 1 changes his strategy from ${\bf 1}$ (confess) to $Y$ (don't confess). 
      
From Eqs.~(\ref{49}) it is clear that for every strategy ${\bm \gamma}_1$ player type 2I will seek his classical strategy 
and try to arrive at the situation where $|b_{2I}|^2=1, \ |a_{2I}|^2=|c_{2I}|^2=d_{2I}|^2=0,$ while player type 2II will seek his classical strategy 
and try to arrive at the situation where $|a_{2II}|^2=1, \ |b_{2I}|^2=|c_{2I}|^2=d_{2I}|^2=0.$ First let us check with the help of Eqs.~(\ref{29}) if they can indeed achieve it 
and than check the response of player 1. \\
{\bf Player 2I best response (he wants $|b_{2I}|^2=1$):} According to Eq.~(\ref{29}) we have, \\
 $|b_{2I}|^2=\left[ \cos \tfrac{1}{2} \theta_1 \sin \tfrac{1}{2} \theta_{2I} \cos (\phi_1-\alpha_{2I})+ \sin \tfrac{1}{2} \theta_1 \cos \tfrac{1}{2} \theta_{2I} \sin (\alpha_1-\phi_{2I}) \right ]^2.$ Recall that for $\theta=0$ $\alpha$ is not defined and 
 conventionally assumes the value 0. Similarly, for $\theta=\pi$ $\phi$ is not defined and 
 conventionally assumes the value 0. Therefore, by choosing \\
 $~~~~~~~~~~~~~~~~~~~~~~~~~~~~~~~\theta_{2I}=\pi-\theta_1, \ \alpha_{2I}=\phi_1, \ \ \phi_{2I}=\alpha_1-\frac{\pi}{2} \ \mbox{modulo} \ 2 \pi$\\
 player 2I gets, $|b_{2I}|^2=|\sin \frac{1}{2}(\theta_1+\theta_{2I})|^2=\sin^2 \frac{\pi}{2}=1$. 
 The modulo $2 \pi$ is optional in order to keep $0 \le \phi_{2I}< 2 \pi$.
 Therefore the best response function of player 2I (that is a triple functions) is,
 \begin{equation} \label{BR2I}
 q_{2I}({\bm \gamma}_1)=q_{2I}(\phi_1,\alpha_1, \theta_1)=(\alpha_1-\frac{\pi}{2}, \phi_1,\pi-\theta_1)~.
 \end{equation}
 {\bf Player 2II best response (he wants $|a_{2II}|^2=1$):} According to Eq.~(\ref{29}) we have, \\
 $|a_{2II}|^2=\left[  \cos \tfrac{1}{2} \theta_1 \cos \tfrac{1}{2} \theta_{2II} \cos (\phi_1+\phi_{2II})-\sin \tfrac{1}{2} \theta_1 \sin \tfrac{1}{2} \theta_{2II} \sin (\alpha_1+\alpha_{2II}) \right ]^2, $ Therefore, by choosing \\
 $~~~~~~~~~~~~~~~~~~~~~~~~~~~~~~~\theta_{2II}=\theta_1, \ \phi_{2II}=-\phi_1 \ \mbox{modulo} \ 2 \pi, \ \ \alpha_{2II}=-(\alpha_1+\frac{\pi}{2}) \ \mbox{modulo} \ 2 \pi$\\
 player 2II gets, $|a_{2II}|^2=|\cos \frac{1}{2}(\theta_1-\theta_{2II})|^2=1$. 
 Therefore the best response function of player 2II (that is a triple functions) is,
 \begin{equation} \label{BR2II}
q_{2II}({\bm \gamma}_1) =q_{2II}(\phi_1,\alpha_1, \theta_1)=(- \phi_1,-(\alpha_1+\frac{\pi}{2}),\theta_1)~.
 \end{equation}
 
 Finding the best response function of player 1, $q_1({\bm \gamma}_{2I}, {\bm \gamma}_{2II})$ is virtually hopeless. However, guided by the classical game 
 results, we are tempted to test whether, for small $\mu$, the first player will 
 choose his classical strategy ${\bm \gamma}_1=(0,0,0)$ which means 
 that his $2 \times 2$ strategy matrix is ${\bf 1}$. In that case, the best response 
 functions of players 2I and 2II are $q_{2I}(0,0,0)$ and $q_{2II}(0,0,0)$ where the 
 functions are defined in Eqs.~(\ref{BR2I}) and (\ref{BR2II}). In other words, we have\\
 \underline{Proposition:} For $\mu \le 1/6$, a pure strategy Nash equilibrium of the Bayesian quantum game is the "triple of triples" 
 \begin{equation} \label{NEQBayesian}
( {\bm \gamma}_1^*,{\bm \gamma}_{2I}^*,  {\bm \gamma}_{2II}^*)=[(0,0,0),q_{2I}(0,0,0),q_{2II}(0,0,0)]~.
 \end{equation}
 \underline{Proof} By construction, ${\bm \gamma}_{2I}^*$ and ${\bm \gamma}_{2II}^*$ are best responses to ${\bm \gamma}_1^*$ and therefore,\\
 $P_{2I}({\bm \gamma}_1^*,{\bm \gamma}_{2I}) \le P_{2I}({\bm \gamma}_1^*,{\bm \gamma}_{2I}^*) \ \forall \ {\bm \gamma}_{2I}$ \ \
 $P_{2II}({\bm \gamma}_1^*,{\bm \gamma}_{2II}) \le P_{2II}({\bm \gamma}_1^*,{\bm \gamma}_{2II}^*) \ \forall \ {\bm \gamma}_{2II}$\\
 To check for $P_1$ we use Eq.~(\ref{49}) and recall the expression for the coefficients from Eq.~(\ref{29}). For any ${\bm \gamma_1}=(\phi_1, \alpha_1, \theta_1)$ we find, after some calculations, 
 \begin{eqnarray} 
 && P_1(\phi_1,\alpha_1,\theta_1, {\bm \gamma}_{2I}^*,{\bm \gamma}_{2II}^*)=
 -10 \left [ \mu (\cos \frac{1}{2}\theta_1 \cos \phi_1)^2 +(1-\mu) (\sin \frac{1}{2} \theta_1 \sin \alpha_1)^2 \right ] \nonumber \\
 && -\left[\mu (\cos \frac{1}{2}\theta_1 \sin \phi_1)^2+(1-\mu) (\sin \frac{1}{2} \theta_1 \cos \alpha_1)^2 \right ] \nonumber \\
 && -5\left[\mu (\sin \frac{1}{2}\theta_1 \cos \alpha_1)^2+(1-\mu) (\cos \frac{1}{2} \theta_1 \sin \phi_1)^2 \right ]~.
 \label{P1QBayesian}
 \end{eqnarray}
 Although finding a global maximum of a function of three continuous variables is 
 not an easy task, all my numerical test indicates that 
 for $\mu \le \frac{1}{6}$, the payoff $P_1(\phi_1,\alpha_1,\theta_1, {\bm \gamma}_{2I}^*,{\bm \gamma}_{2II}^*)$ of player 1 has a global maximum at $(\phi_1,\alpha_1,\theta_1)=(0,0,0)$ with payoff value $-10 \mu$. Thus, for $\mu \le 1/6$ the classical and quantum payoffs are identical, $(P_1,P_2,I,P_{2II})=(-10 \mu,-1,-2)$. On the other hand, for $\mu>1/6$ it is easy to show that $P_1$ as defined in Eq.~(\ref{P1QBayesian}) does not have a global maximum at $(0,0,0)$ and as my numerical algorithm indicates, the quantum version of the DA brother Bayesian game with maximal entanglement does not have a pure strategy Nash equilibrium. 
 
\subsection{Two-Players Three Strategies Games} \label{Section2P3S}
So far, all our analysis was constructed upon classical games with two strategies per each player. These two strategies are represented by $2 \times 2$ matrices ${\bf 1}=\binom {1 0}{0 1}$  and $Y=\binom{~0~1}{-1~0}$ that operate on the two bit states represented as vectors $|0\ra=\binom{1}{0}$ and $|1 \ra=\binom{0}{1}$. 
In this section we will briefly touch upon sthe topic of quantum games based on 2-players 3-strategies classical game. The reason for carrying out this analysis is to check whether, in these structures there are special interesting features whose elucidation makes it  worth to study despite the augmented complication. 
First, in subsection \ref{Subsec_Trits} we will describe how to cast 
the classical game in quantum information format, and define the notion of trits and classically non-commuting strategies. Then, in subsection \ref{Subsec_Qtrits} we will analyze the construction of the quantum game and define the notion of qutrits and quantum strategies as $3 \times 3$ matrices forming the group SU(3). 
\subsection{Two Players Three Strategies Classical Games: Trits} \label{Subsec_Trits}
Now consider a two-players classical game with three strategies for each player. 
For example, prisoners may have three options, C,S and D for confess, Stay quiet (or Shut up) or Don't confess. In analogy with the bit notation 0,1, these three options are marked by 1,2,3 respectively, 
or in our ket notation $|1 \ra, |2 \ra, |3 \ra$. \\
\underline{Definition:} On object or a state that can  assume three values is 
referred to as {\em trit}. \\
In analogy with the two-component vector notation for bits, we have,
\begin{spacing}{0.8}
\begin{equation} \label{4a}
\mbox{trit state 1} = \begin{pmatrix} 1\\0 \\0 \end{pmatrix}=|1\ra, \ \  \mbox{ trit state 2} = \begin{pmatrix} 0 \\ 1 \\0 \end{pmatrix}=|2 \ra, \ \  \mbox{ trit state 3} = \begin{pmatrix} 0 \\ 0 \\1 \end{pmatrix}=|3 \ra. 
\end{equation}
\end{spacing}
\ \\
Similarly to Eqs.~(\ref{2}) and (\ref{2a}) we define two trit states as nine components vectors. 
In the the nine two-trit states are denoted as, $|i j \ra, \ \ i=1,2,3 \ \ j=1,2,3$. 
These nine two trit states correspond to the nine squares of the game table. 

The protocol of the classical game with 2-player and 3-strategies is similar to that of 
the 2-players 2-decision game. The judge calls the prisoners and tells them he assume that they are in a 
two-trit state $|1 1 \ra$ meaning (C,C) namely both confess. He then asks them to 
decide whether to leave their trit state as it is on $|1 \ra$ or to change it either to $|2 \ra$ (meaning S) or to $|3 \ra$ (meaning D). These replacement operations are the players strategies. 
Since trits are represented by  three component vectors, operation on trits 
are represented by $3 \times 3$ matrices. Unlike the case for bits, where the operations are 
exhausted by ${\bf 1}_2$ (the unit $2 \times 2$ matrix) and $Y$, 
the strategies of the two players in the present game include ${\bf 1}_3$ (the $3 \times 3$ unit matrix 
leaving the trit as it is, 
$S_{12}$ (swapping of $|1 \ra $ and $|2 \ra$ namely, replacing C by S) and $S_{13}$ (swapping $|1 \ra$ and $|3 \ra$ namely replacing $C$ by $D$. Note that altogether we have four swapping operations 
that leave at least one component of the trit untouched, the fourth one is $S_{23}$ that swaps 
$|2 \ra$ and $|3 \ra$ but is not used in a simultaneous game with a single move if the initial state suggested by the judge is $|11\ra=(CC)$. The four operations ${\bf 1}_3, S_{12}, S_{13}, S_{23}$ form a subset of $S_3$, the set of permutations on three objects that has six operations. The other two operations 
are such that all three elements change their place but they are not used in a simultaneous game with a single move. The six elements of $S_3$ form a group, but the four operations ${\bf 1}_3, S_{12}, S_{13}, S_{23}$ do not. 
 
In matrix notations the four operations are, 
\begin{spacing}{1.1}
\begin{equation} \label{4b} 
{\bf 1}_3=\begin{pmatrix} 1&0&0\\ 0&1&0\\0&0&1 \end{pmatrix}, \ \ S_{12}=\begin{pmatrix} 0&1&0\\1&0&0\\0&0&1 \end{pmatrix} , \ \ S_{13}=\begin{pmatrix} 0&0&1\\0&1&0\\1&0&0 \end{pmatrix} , \ \ S_{23}=\begin{pmatrix} 1&0&0\\0&0&1\\0&1&0 \end{pmatrix}
\end{equation}
\end{spacing}
\vspace{0.2in}
\noindent
Unlike the case of two-players two-strategies game where the two operations ${\bf 1}_2$ and $Y$  commute with each other, $[{\bf 1}_3,Y]=0$, here the strategies $S_{ij}$ do not commute with each other, for example, 
\begin{equation} \label{4c}
[S_{12},S_{13}]= S_{12}S_{13}-S_{13}S_{12} \ne 0, \ \ \mbox{etc}.
\end{equation}

 \subsection{Qutrits} \label{Subsec_Qtrits}
 
 In this section we will briefly introduce the notion of quantum trits (qutrits) and quantum strategies in the quantum version of a 
 two-players three-decisions game. This will mainly include a few definitions and 
 some basic properties, since the analysis of such quantum game is too complicated and naturally falls beyond the scope of this thesis. 
 
 In order to define a qutrit we consider a three dimensional Hilbert space ${\cal H}_3$ in which the three states of a trit, $|0 \ra, |1 \ra$ and $|2 \ra$ form an ortho-normal basis. This means that 
 1) Every element (vector, or ket ) $|\psi \ra \in {\cal H}_3$ is expressible as a linear combination 
 \begin{equation} \label{qutrits1}
 |\psi \ra=v_1|1\ra+v_2|2 \ra+v_3|3 \ra, \ \mbox{where} \ v_1,v_2,v_3 \in \mathbb{C}~. 
 \end{equation}
 2) Orthogonality relations $\la i|j \ra=\delta_{ij}, \ \ i,j=1,2,3$. From our knowledge of the properties of a Hilbert space, we can define the norm (squared) , or the length squared of $|v \ra$ by taking the inner product of $|v \ra$ with itself (this is exactly analogous to taking a scalar product of a vector with itself (for real vectors, economists use the notation 
 $v'v$). In Hilbert space we use the Dirac notation and write it as $\la v|v \ra$. 
 The result is of course $\la \psi | \psi \ra=|v_1|^2+|v_2|^2+|v_3|^2$ that is a real non-negative number. As in the case of qubits, we are interested in vectors of unit length. Whilst a qubit is a vector $a|0\ra+b|1 \ra \in {\cal H}_2$ of unit norm, $|a|^2+|b|^2=1$ we have,\\
 \underline{Definition:}  A qutrit is a vector $|\psi \ra=v_1|1\ra+v_2|2 \ra + v_3|3 \ra \in {\cal H}_3$ of unit norm, $|v_1|^2+|v_2|^2+|v_3|^2=1$. In analogy with Eq.~(\ref{7}), the representation of qutrits in terms of angular variables reads,
 \begin{equation} \label{qutritsangles}
 |\psi \ra=e^{i \alpha} \sin \theta \cos \phi |1\ra+e^{i \beta} \sin \theta \sin \phi |2\ra+\cos \theta |3 \ra~.
 \end{equation} 
 
 \subsubsection{Operations on Qutrits: Strategies}
 Instead of the $2 \times 2$ unitary matrix $U$ defined in Eq.~(\ref{11}) as a players strategy in the two-decision game, a strategy of a player in a three decision game is a unitary $3 \times 3$ complex matrix $U$ with unit determinant det[$U$]=1. The (infinite) set of all these matrices form a group, referred as the $SU(3)$ group. It plays a central role in physics especially in the classification of elementary particles. Unlike the two-strategies game where the 
 $2 \times 2$ strategy matrices $U(\phi, \alpha, \theta) \in SU(2)$ defined in Eq.~(\ref{11}) depend on three Euler angles, the quantum strategies $ U \in SU(3)$ in the three strategies game depend on eight Euler angles,
  $U(\alpha_1,\alpha_2, \ldots, \alpha_8) \in SU(3)$. That turns any attempt to use numerical approach virtually useless.  Instead, we will list a few properties of the pertinent quantum game that indicates that it is principally different from the quantum game based on 2-players 2-strategies classical games. 
 
 \subsubsection{Entanglment of Two Qutrit States}
 Like in the simpler quantum games based on 2-players 2-strategies, entanglement plays a 
 crucial role also in 2-players 3-strategies games. 
 First, let us define a two qutrit state and then define entanglement. 
 A general two qutrit state (an element in ${\cal H}_9$ can be written as, 
 \begin{equation} \label{2qutritgen}
 |\Gamma \ra=\sum_{i,j=1}^3 v_{ij}|i j \ra , \ \ \sum_{i,j=1}^3 |v_{ij}|^2=1~.
 \end{equation}
Consider now two qutrits, 
 \begin{equation} \label{2qutrits} 
 |\psi_1\ra=\sum_{i=1}^3 a_i |i \ra, \ \ a_i \in \mathbb{C}, \ \sum_{i=1}^3|a_i|^2=1, \ 
 |\psi_2\ra=\sum_{j=1}^3 b_j |j \ra, \ \ b_j \in \mathbb{C}, \ \sum_{j=1}^3|b_j|^2=1~.
 \end{equation} 
 Their outer (or tensor) product is defined as, 
 \begin{equation} \label{2qutrittensor}
 |\psi_1\ra \otimes |\psi_2 \ra=\sum_{i,j=1}^3 a_ib_j |ij\ra~.
 \end{equation} 
 Then we have:\\
 \underline{Definition:} A general 2 qutrits state $| \Gamma \ra$ as defined in Eq.~(\ref{2qutritgen}) is said to be entangled if it {\it cannot } be written as an outer product of two qutrits as in Eq.~(\ref{2qutrittensor}). 
 We give (without proof) an example of maximally entangled
two-qutrit states, 
\begin{equation} \label{Max_2qutrit}
|\Psi \ra_{\mathrm {ME}}=\frac{1}{\sqrt{3}} (u_1|1 1 \ra+u_2|22 \ra+u_3|3 3 \ra), \ \ u_i \in \mathbb{C}, \ \ |u_i|=1~.
\end{equation} 
\subsubsection{New Elements in Quantum Games based on 2-Players 3-Strategies Classical Games}
Suppose we try to organize the conduction of a quantum game based on a classical two-players three strategies game as a straightforward extension of the procedure used to quantize a   two-players two strategies quantum game as displayed in Fig.~(\ref{Fig1}).  
 The first move by the referee, that is, fixing an initial 
 two-qutrit state (usually the classical two trit states $|11 \ra=(C,C)$) is indeed identical.   But the second 
 operation, namely, operating by the entanglement operator is less straightforward because we first 
 have to identify the maximally and partially entangled states in ${\cal H}_3 \otimes {\cal H}_3$ and then to design  the $9 \times 9$ matrix $J$ that turns the non-entangled two qutrit initial state $|11 \ra$ into a maximally entangled state in analogy with Eq.~(\ref{22}) or Eq.~(\ref{22a}). The maximally entangled state 
$|\Psi \ra_{\mathrm {ME}}$ has already been identified in Eq.~(\ref{Max_2qutrit}). In searching for an entanglement operator $J$ such that $J|11 \ra=|\Psi \ra_{\mathrm {ME}}$ we recall an important 
and desirable property that we want to be satisfied by $J$, namely, classical commensurability. Mathematically it means that $J$ should commute with all outer products of the classical strategies 
(see Eqs.~(\ref{classcom}) and  (\ref{J1}) for the 2-players 2-strategies case). The reason for demanding 
classical commensurability is to assure that classical strategies are a special case of the quantum strategies as explained in connection with Eq.~(\ref{cc}). 

Concentrating on the quantum game with initial  state, $|11 \ra$ we then require 
\begin{equation} \label{ccqutrits}
 J|11 \ra=|\Psi \ra_{\mathrm {ME}}, \ \ [J, S_{12} \otimes S_{13}]=0,
\end{equation}
where the classical strategies are defined in Eq.~(\ref{4b}). 
A necessary and sufficient condition for satisfying  Eq.~(\ref{ccqutrits}) will be composed of outer products of non-trivial $3 \times 3$ matrices that commute with {\it both} $S_{12}$ and $S_{13}$. But that is impossible as we now prove,\\
\underline{Theorem:} In a  quantum game based on 2-players 3-strategies classical game 
conducted as in Fig.~\ref{Fig1} with a non-trivial entanglement operator $J$ (Namely, the state 
$J|11 \ra$ is entangled)   
there is no classical commensurability, namely, the classical strategies are not achieved as a special case of the quantum strategies. \\
\underline{Proof} A necessary and sufficient condition for classical commensurability is a relaxed version of Eq.~\ref{ccqutrits}, namely,
\begin{equation} \label{ccqutrits1}
 J|11 \ra=|\mbox{Entangled State}\ra, \ \ [J, S_{12} \otimes S_{13}]=0.
\end{equation}
The second equality is possible only if $J$ is a function of $A \otimes A$ where $A$ is a $3 \times 3$ matrix
 satisfying $[A,S_{12}]=[A,S_{13}]=0$, and the first equality requires that $A$ is not simply a multiple of the unit matrix ${\bf 1}_3$. Therefore we need to prove the following \\
 \underline{lemma:} If a $3 \times 3$ matrix $A$ satisfies $[A,S_{12}]=[A,S_{13}]=0$ the $A=C {\bf 1}_3$ 
 where $C \ne 0$ is a number. \\
 \underline{Proof of the lemma:} Although the lemma can be proved by brute force writing down the 
 equation implied by the commutation relations, we choose a more elegant way mainly because it can be easily generalized to games with any finite number of strategies. The lemma will be proved in steps.
 \begin{enumerate}
 \item{} If $[A,S_{12}]=[A,S_{13}]=0$ then $A$ commutes with any monomial of $S_{12}$ and $S_{13}$. For example, $[A,S_{12}^2]=AS_{12}^2-S_{12}AS_{12}=AS_{12}^2-AS_{12}^2=0$  and so on. 
 \item{} We have already stated that $S_{12}$ and $S_{13}$ are the matrices representing permutations on three elements, specifically, 
 \begin{spacing}{0.9}
 \begin{equation} \label{Sperm}
 S_{12}\begin{pmatrix}1\\2\\3\\ \end{pmatrix}=\begin{pmatrix} 0&1&0\\1&0&0\\0&0&1 \end{pmatrix} \begin{pmatrix}1\\2\\3\\ \end{pmatrix}=\begin{pmatrix}2\\1\\3\\ \end{pmatrix}, \ \ \ S_{13}\begin{pmatrix}1\\2\\3\\ \end{pmatrix}=\begin{pmatrix} 0&0&1\\0&1&0\\1&0&0 \end{pmatrix} \begin{pmatrix}1\\2\\3\\ \end{pmatrix}=\begin{pmatrix}3\\2\\1\\ \end{pmatrix}.
 \end{equation}
 \end{spacing}
 \vspace{0.2in}
 \noindent
 It is easily verified that simple monomials of $S_{12}$ and $S_{13}$ generate {\it all the other permutations}  of three objects,  altogether 6 elements (including the permutation ${\bf 1}_3: (123) \to (123)$. Explicitly, 
  \begin{spacing}{0.9}
 \begin{equation} \label{Sperm1}
 S_{12}S_{13}=\begin{pmatrix} 0&1&0\\0&0&1\\1&0&0 \end{pmatrix}: (123) \to (231); \ \ S_{13}S_{12}=\begin{pmatrix} 0&0&1\\1&0&0\\0&1&0 \end{pmatrix}: (123) \to (312).
 \end{equation}
 \end{spacing}
  \begin{spacing}{0.9}
 \begin{equation} \label{Sperm2}
 S_{12}S_{13}S_{12}=\begin{pmatrix} 1&0&0\\0&0&1\\0&1&0 \end{pmatrix}=S_{23}: (123) \to (132).
 \end{equation}
 \end{spacing}
 Therefore, according to 1), {\it the matrix $A$ commutes with  all the six matrices representing the set  $S_3$ of all permutations of three objects. }
 \item{} From Appendix \ref{GT} we know that $S_3$ is a (non-commutative) group containing 6 elements, 
and that the five matrices $S_{12},S_{13},S_{23}, S_{12}S_{13}, S_{13}S_{12}$ listed above together with the $3 \times 3$ unit matrix ${\bf 1}_3$ form an {\it irreducible representation of $S_3$}.  
\item{} The Schure lemma in group theory states that if a matrix commutes with all the matrices that form an irreducible representation of a group, then this matrix is a multiple of the unit matrix. Hence $A=C {\bf 1}_3$ 
and the lemma is proved $\blacksquare$
\end{enumerate}
\subsubsection{Designing $J_1$}
A natural question is whether we can find an entanglement operator $J$ such that when it acts on the two qutrit state 
$|00 \ra$ it yields $|\Psi \ra_{\mathrm {ME}}$ of Eq.~(\ref{Max_2qutrit}) as specified in Eq.~(\ref{ccqutrits})
For two qubit states we defined the corresponding 
entanglement operator $J(\beta)$ in Eq.~(\ref{37}). When it acts on two qubit state $|00 \ra$ it gives the  state $|\psi_+(\beta) \ra$ which, for $\beta=\pi/4$ gives the maximally entangled Bell state $\psi_+(\frac{\pi}{4})$ of Eq.~(\ref{22b}). But with 
qutrits the design of $J$ (a $9 \times 9$ matrix) is more complicated. To find it we note that in order to get the 
two qutrit state $|11 \ra$ from the qutrit state $|00 \ra$ we have to operate on $|00 \ra$ with $X \equiv [S_{12}S_{13}] \otimes [S_{13}S_{12}]$, see definition in Eq.~(\ref{Sperm1}), whereas in order to get the 
two qutrit state $|22 \ra$ from the qutrit state $|00 \ra$ we have to operate on $|00 \ra$ with $X^T \equiv [S_{13}S_{12}] \otimes [S_{12}S_{13}]$, see definition in Eq.~(\ref{Sperm1}). 
 Consider the $9 \times 9$ matrix 
\begin{equation} \label{qutrits_Z}
Z \equiv X+X^T, \ \ \Rightarrow \ \ Z|00 \ra=|11 \ra+|22 \ra, \ \ Z^2=Z+2 \times {\bf 1}_{9 \times 9}~,
\end{equation} 
where the last equality holds because $[X,X^T]=0$ and $XX^T= {\bf 1}_{9 \times 9}$. Now let us define, 
\begin{equation} \label{J_qutrits}
J(\beta)=e^{i \beta Z}=a(\beta)+b (\beta)Z,
\end{equation}
where the equality holds because $Z^2=Z+2 \times {\bf 1}_{9 \times 9}$ and the expansion of the exponent as 
we learn from Eq.~(\ref{expA}) in section \ref{Mat} yields only linear expression with $Z$. To get the coefficients 
$a(\beta)$ and $b(\beta)$ we perform derivative of both sides of Eq.~(\ref{J_qutrits}) and obtain, 
\begin{equation} \label{qutrit_derJ}
J'=a'+b'Z=i Z e^{i \beta Z}=i Z(a+bZ)=iaZ+ib Z^2=iaZ+ib (Z+2) =i(a+b)Z+2 i b.
\end{equation} 
Equating powers of $Z$ we get a set of differential equations, 
\begin{spacing}{0.8}
\begin{equation} \label{diffeq_J}
a'=2 i b, \ \ b'=i(a+b), \ \ a(0)=1, \ \ b(0)=0, \ \ \Rightarrow \ \ \begin{pmatrix} a(\beta) \\ b(\beta) \end{pmatrix}=e^{i \beta \binom{0 2}{1 1}}\begin{pmatrix} 1\\0 \end{pmatrix}~.
\end{equation}
\end{spacing}
The calculations of the exponent are easily done with the results 
\begin{equation} \label{qutrit_ab}
a=\frac{1}{3}e^{-i \beta}(e^{3 i \beta}+2), \ \ b=\frac{1}{3}e^{-i \beta}(e^{3 i \beta}-1),
\end{equation} 
Inserting these results in Eq.~(\ref{J_qutrits}) and using Eq.~(\ref{qutrits_Z}) we get, 
\begin{equation} \label{qutrits_J1}
J(\beta)|00 \ra=\frac{1}{3} e^{-i \beta} \left [ (e^{3 i \beta}+2)|00 \ra+(e^{3 i \beta}-1)(|11 \ra+|22 \ra) \right ]~. 
\end{equation} 
Maximal entanglements obtains when the absolute values of all three coefficients are equal, namely,
\begin{equation} \label{JME}
|e^{3 i \beta}+2|=|e^{3 i \beta}-1| \ \ \Rightarrow \ \ \beta=\frac{2 \pi}{9}=40^0~.
\end{equation}
Thus we have found the desired entanglemnt operator. 

\subsubsection{Qutrits Summary}
Introducing the quantum game based on 2-player 3-classical strategies brings a few new elements, 
\begin{enumerate}
\item{} Richer quantum information content through the interaction of qutrits. 
\item{} Richer strategy content encoded by SU(3) matrices that depend on eight Euler angles. 
\item{} Intricate entanglement pattern in 2-qutrit states. 
\item{} Absence of classical commensurability. The classical strategies are not achieved as a 
special case of the quantum strategies because it is impossible to design an entanglement operator $J$ that commutes with all the classical strategies. 
\item{} A non-trivial entanglement operator $J(\beta)$ acting on two qutrits states such that for $\beta=\frac{2 \pi}{9}$ the two qutrit state $J(2\pi/9)|00 \ra$ is maximally entangled. 
\end{enumerate} 
 \end{spacing}
\begin{spacing}{1.5}
\begin{center}
\section{\small Appendices}\label{Section7}
\end{center}
In this section we will present some of the mathematical tools required for 
handling the material exposed in this thesis. First, in Section \ref{CN} we 
define complex numbers, since they are the ground for 
defining Hilbert space upon which 
quantum mechanics is based. Then in Section \ref{Fields} we will 
introduce the notion of Fields and Linear Vector Spaces above the field of complex numbers. 
As we have seen, in our formulations of quantum games in terms of qubits and qutrits, 
complex matrices are all around us, so the introduction to matrices and their algebra 
is exposed in section \ref{Mat}.  We have also employed the notion of group theory along the 
thesis and in Subsection \ref{Subsec_Qtrits} we have based our proof of the theorem on absence of classical commensurability on group theory. Therefore, some basic aspects of group theory are 
listed in Section \ref{GT}. As a last preparation for discussing quantum mechanics we 
explain the concept of Hilbert space in Section \ref{HS}. With all these tools at hand, we are ready to 
introduce the basic principles of quantum mechanics in the last Section \ref{HS}.   

The exposition below is relatively short compared to the 
huge amount of the pertinent material. It will include basic definitions and a few important properties relevant for the material presented in the preceding Sections. No proofs will be given as they 
can be found elsewhere.  Yet these appendices are not copy-past production. 
I could also direct the reader to internet sites where this material is presented but chosen to 
avoid it and present it in a more concerted manner commensurate with the material presented in this thesis. 

\noindent
{\bf The level of presentation is rather elementary, because the target audience may be   wide (as I hope). A great deal of effort is therefore made to 
make the material understandable to non-experts.}

\subsection{Complex Numbers} \label{CN}
Complex numbers are all around us, although we are not always aware of that. Electrical engineering is 
heavily based on the notion of complex numbers. 
One of the basic entities  of quantum mechanics is called wave function, and it is described by a complex number.  It should be stressed right at the onset that measurable quantities are expressed in terms of real numbers. The relevance of quantum mechanics to real life and outcome of laboratory experiment 
is guaranteed by the fact that after the wave function is manipulated and interfere with another wave functions, a real number is generated by taking an absolute value of the wave function results, that can be directly measured. This is why we are obliged to study the algebra of complex numbers, a task that we take up now without further apology.  
\subsubsection{The Field of Complex Numbers} 
The natural need to extend the  notion of real numbers as we use in everyday life is our frustration to realize that there are simple equations like $x^2+1=0$ that have no real solutions. So why not include in what we regards as "numbers" also the solutions of such equations? 
If we do it we should take care that all the 
beautiful structure (axioms) that governs the realm of real numbers should be valid also for the wider class of numbers referred to as {\em complex numbers}. Thus, if we add the solution of the above equation and call it $i$ such that $i^2=-1$ we should be able to perform the operation of addition, multiplication, of number that include $i$ as well. This brings us to the notion of Field. \\
\ \\
\underline{Definition} A collection $\la S + \bullet \ra$ where $S$ is a set and $+$ and $\bullet$ are two mappings from $S \times S \to S$ is called a field if the following axioms are satisfied: 
\begin{enumerate} 
\item{} S is non-empty, and contains at least two elements denoted as $0$ and $1$.
\item{} $s_1, s_2 \in S \ \Rightarrow s_1+s_2 \in S, \ \ s_1+s_2=s_2+s_1, \ \ s_1+(s_2+s_3)=(s_1+s_2)+s_3, \ \ s+0=s$.
\item{} For every $s \in S$ there is an element $\bar {s} \in S, \ \ni s+\bar{s}=0$. For obvious reasons 
$\bar{s}$ is also denoted as $-s$. 
\item{} $s_1, s_2 \in S \ \Rightarrow s_1 \bullet s_2 \in S, \ \ s_1 \bullet s_2=s_2 \bullet s_1, \ \ s_1\bullet(s_2\bullet s_3)=(s_1\bullet s_2)\bullet s_3, \ \ s \bullet 0=0, \ s \bullet 1= s$.
\item{} For every $s \in S, \ s \ne 0$ there is an element $ s^{-1} \in S, \ \ni s  s^{-1}=1$. 
\end{enumerate} 
The first example that comes to our mind is the set of real numbers with the two operations of addition and multiplication. The corresponding field is denoted as $\mathbb{R}$. Then we may think of the field of rational numbers, these are all numbers that can be 
written as a quotient $r=p/q$ where $p$ and $q$ are integers, and $q \ne 0$. But we may think of other examples. The case $S=\{0,1 \}$ with addition modulo 2 (namely $1+1=0$) and the usual multiplication is a field containing two elements. 

When we extend the set of real numbers such that it will include also complex numbers we 
should assure that the axioms formulated above are satisfied. In that case we speak about the field $\mathbb{C}$ of complex numbers. 
\subsubsection{Complex Numbers: Definition  and Algebraic Properties} 
A complex number $z$ and its complex conjugate $z^*$ (that is also a complex number) are defined as 
 \vspace{-0.2in}
 \begin{equation} \label{53}
 z=x+iy, \ \ z^*=x-iy,  \ \ x,y \in \mathbb{R}, \ \ i=\sqrt{-1}~.
 \end{equation}
$x$ is called the {\it real part of $z$}, denoted as $x$=Re$(z)$ and $y$ is called the {\it imaginary part of $z$}, denoted as $y$=Im$(z)$. If $y=0$,  $z$ is said to be real and if $x=0$, $z$ is said to be purely imaginary. 
If both $x=y=0$ we write simply $z=0$.  Note that by definition, the powers of $i$ are such that
\begin{equation} \label{ipowers}
i^{4n}=1, \ \ i^{4n+1}=i, \ \ i^{4n+2}=-1, \ \ i^{4n+3}=-i, \ \ n=0,1,2,\ldots.
\end{equation}

Complex numbers can be added, subtracted multiplied and divided. At the end of each such operation, we would like to write the resulting complex numbers in the standard form of Eq.~(\ref{53}). Thus, if $z_1=x_1+i y_1$ and $z_2=x_2+i y_2$ we have 
\vspace{-0.2in}
 \begin{equation} \label{54}
z_1 \pm z_2=(x_1\pm x_2) +i(y_1 \pm y_2)~, z_1z_2=(x_1+i y_1)(x_2+i y_2)=(x_1 x_2-y_1 y_2)+i(x_1y_2+x_2 y_1)~,
 \end{equation}
where the distributive law has been used and the product of the two pure imaginary numbers $(iy_1)(i y_2)=-y_1y_2$ yields a real number since $i^2=-1$. After the addition and subtraction of complex numbers have been defined, we may write
 \begin{equation} \label{54a}
\mbox{Re}(z)=x=\frac{1}{2}(z+z^*), \ \ \mbox{Im}(z)=y=\frac{1}{2 i}(z-z^*).
\end{equation} 
Before working out the quotient of two complex numbers, we note that the product of a complex number $z$ with its complex conjugate number $z^*$ is real and positive, (unless $z=0$),  
\begin{eqnarray}
&& zz^*=(x+iy)(z-iy)=x^2+y^2 \equiv |z|^2 \ge 0, \ \ (\mbox{equality only if \ \ } z=0). \nonumber \\
&& |z|=\sqrt{zz^*}=\sqrt{[\mathrm{Re}(z)]^2+[\mathrm{Im}(z)]^2}= \sqrt{x^2+y^2} = \mbox{absolute value of z} \ge 0~.
 \label{55}
\end{eqnarray}
The absolute value of a complex number is then real and positive (except when $z=0$ in which case 
$|0|=0$). It is the absolute value of wave functions (that appears in quantum mechanics as complex numbers) that makes the connection between the abstract theoretical quantities (wave functions or amplitudes) and measurable quantities (such as electrical currents, light intensity and information stored in quantum computers). While complex numbers cannot be compared (we cannot say that $z_1 > z_2$), their absolute values are comparable as real numbers, and we can say that $|z_1| > |z_2|$ (or vice versa).  

With the multiplication rules (\ref{54}) and (\ref{55}) we are now in a position to 
derive an expression for the quotient of two complex numbers,
\begin{equation} \label{56}
\frac{z_1}{z_2}=\frac{x_1+iy_1}{x_2+iy_2}=\frac{(x_1+iy_1)(x_2-iy_2)}{z_2z_2^*}=\frac{x_1x_2+y_1y_2}{x_2^2+y_2^2}+i\frac{y_1x_2-x_1y_2}{x_2^2+y_2^2}~,
\end{equation}
which is casted in the form (\ref{53}) of [Real part] + $i$[Imaginary part]. A particular case is 
\begin{equation} \label{57}
\frac{1}{z}=\frac{z^*}{zz^*}=\frac{x-iy}{x^2+y^2}=\frac{x}{x^2+y^2}-i \frac{y}{x^2+y^2}~.
\end{equation} 
We list below a few properties of complex conjugation and absolute values 
that are easily proved from the definitions and the addition, multiplications and quotient rules:
\begin{equation} \label{58}
(z_1+z_2)^*=z_1^*+z_2^*, \ \ (z_1z_2)^*=z_1^*z_2^*, \ \ \left ( \frac{z_1}{z_2} \right )^*= \frac{z_1^*}{z_2^*}, \ \ (z^n)^*=(z^*)^n, 
\end{equation}
\begin{equation} \label{59}
|z_1z_2|=|z_1||z_2|, \ \ \left | \frac{z_1}{z_2} \right |=\frac{|z_1|}{|z_2|}, \ \ (z_2 \ne 0), \ \ |z_1 \pm z_2| \le |z_1|+|z_2|~.
\end{equation} 
\subsubsection{Geometrical  Representation of Complex Numbers: The Complex Plane} 
The fact that a complex number is represented by two components (its real and imaginary parts) reminds us of a point in the plane where each point is determined by its two coordinates once a Cartesian frame is 
given. This leads to a geometric representation of complex numbers that is very elegant and convenient for illustration of manipulation of complex numbers. The pertinent plane is referred to as the {\it Complex Plane}. Like in ordinary plane geometry, there are two basic representations, the Cartesian and the Polar. In both of them the starting point is to draw two perpendicular lines on the plane: The horizontal line called the real axis (denoted as $x$) and the vertical line called the imaginary axis (denoted as $iy$ or sometimes just $y$). As in the ordinary plane geometry we assign a real number on any axis such that the intersection point corresponds to the real number 0 on both axes and is referred to as the origin.
\subsubsection{Cartesian Representation}
\begin{figure}[!h]
\centering
\includegraphics[width=9truecm, angle=0]{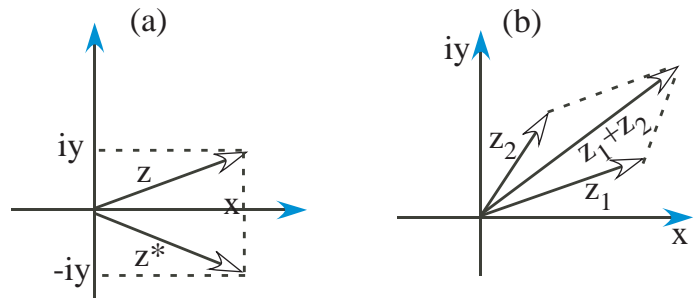}
\caption{\footnotesize{ Cartesian representation of complex numbers in the complex plane. (a) 
the complex number $z$ and its complex conjugate $z^*$. (b) Addition of two complex numbers is similar to addition of physical vectors (e.g forces) in mechanics. }}
\label{Fig7_1}
\end{figure}

In the Cartesian representation a point in the plane has two coordinates obtained by projecting it on the real and imaginary axes. In Fig.~\ref{Fig7_1}(a) we draw the complex number $z$ as an arrow from the origin to a point with coordinates $(x,iy)$ and 
its complex conjugate as an arrow from the origin to a point with coordinates  $(x,-iy)$.  By writing $z=x+iy$ we perceive each projection as a vector along the corresponding axis and then perform vector addition 
as in classical mechanics (the parallelogram of forces, but here it is a rectangle). 
This construction is very convenient for illustrated the addition of complex numbers written algebraically in Eq.~(\ref{54}). As we see in Fig.~\ref{Fig7_1}(b) the two complex numbers $z_1$ and $z_2$ are represented by two vectors, and their sum, $z_1+z_2$ is obtained by adding the two vectors, (this time it is a generic parallelogram).  
\subsubsection{Polar Representation: The unit Circle}
\begin{figure}[!h]
\centering
\includegraphics[width=9truecm, angle=0]{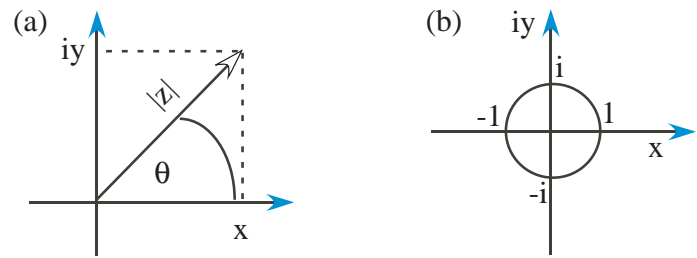}
\caption{\footnotesize{ Polar representation of complex numbers in the complex plane. (a) 
the complex number $z$ is represented by a vector of length $|z|$ and an angle $\theta$ that 
the vector is rotated counterclockwise with respect to the real axis $x$. (b) The unit circle in the complex plane contains the tips of all complex numbers of length 1. Special points on the unit circle and the corresponding angle theta are $(1,0), (i, \frac{\pi}{2}),(-1,\pi), (-i, -\frac{\pi}{2})$.}}
\label{Fig7_2}
\end{figure}

While the Cartesian representation is convenient for the description of adding complex numbers, it is not so practical in describing algebraic operations like $z^2$ or $\sqrt{z}$. For these cases we use the {\it polar representation of complex numbers}. Instead of representing the complex number by its projections $x$ and $iy$ as in the Cartesian representation, it is represented by its length $|z|=\sqrt{x^2+y^2}=\sqrt{zz^*}$ 
and the angle $\theta$ it generates with the real axis ($x$) as is shown in Fig.~\ref{Fig7_2}a.   
The angle $\theta$ should cover an interval of $2 \pi$ and for convenience we take it to vary between 
$- \pi$ and $\pi $. 
Simple trigonometric considerations lead to the relation between between the two representations 
\begin{eqnarray} 
&& x=|z| \cos \theta, \ \ y=|z| \sin \theta, \ \ z=x+iy=|z|(\sin \theta + i \sin \theta)~,\ \ \mbox{(from polar to Cartesian)}, \nonumber \\
&& |z|=\sqrt{x^2+y^2}, \ \ \tan \theta=\frac{y}{x}~ \ \ \mbox{(from Cartesian to polar)}~. 
\label{cartpolar1}
\end{eqnarray} 
of special importance is the {\it unit circle in the complex plane} shown in Fig.~\ref{Fig7_2}b. It contains all the points for which $|z|=1$ while $\theta$ varies continuously between $-\pi$ and $\pi$. The advantage of 
using the polar representation will become clear after we express the the sum $(\sin \theta + i \sin \theta)$ 
appearing in Eq.~(\ref{cartpolar1}) in term of exponential. 
\subsubsection{Complex Functions of Real Variable}
We will often encounter the need to evaluate complex function of a real variable, 
$f: \mathbb{R} \to \mathbb{C}$. Namely for real number $x$, $f(x)$=Re$[f(x]$+$i$Im$[f(x)]$ is a complex number and we are required to determine the the two {\it real functions}  Re$[f(x]$ and Im$[f(x)]$. In simple cases this task is trivial, for example, for $f(x)=(1+ix)^2=1+2ix-x^2$ we immediately find Re$[f(x]$ =$1-x^2$ and Im$[f(x)]$=$2x$. But as $f(x)$ is more complicated, the evaluation may be much more difficult. 
For example, how can we calculate $f(x)=\sqrt{1+ix}$?, or $f(x)=e^{ix}$? or $f(x)=\tan(ix)$? 
Fortunately, there is a central equation that tells us how to calculate 
$e^{ix}$ and that turns out to simplify such calculations tremendously.  

Consider the Taylor (power) expansion of the three {\it real} functions,  $e^x$, $\sin x$ and $\cos x$. 
\begin{equation} \label{expx}
e^x=\sum_{n=0}^\infty \frac{x^n}{n!}, \ \ \sin x=\sum_{n=0}^\infty (-1)^n\frac{x^{2n+1}}{(2n+1)!}, \ \ 
\cos x=\sum_{n=0}^\infty (-1)^n\frac{x^{2n}}{(2n)!}. 
\end{equation}
Now let us inspect the expansion of $e^x$ and replace $x$ by $ix$. Let us inspect the cases $n$ even and $n$ odd separately. The even powers with $n$ odd, namely $2n=2,6,10,\ldots$ will change sign because 
$i^{2 n}=-1$ for odd $n$. The even powers with $n$ even, namely $2n=0,4,8,\ldots$ will stay intact because $i^{2 n}=1$ for even $n$. Compared with the expansion of $\cos x$ the sum of all the even powers gives us $\cos x$. Now let us inspect the odd powers $n=1,3,5,\ldots$ in the expansion of 
$e^x$ (after replacing $x$ by $ix$). Here our work is simple because $i^n=i i^{n-1}$ so that if $n$ is 
odd then $n-1$ is even and we can use our earlier result. A little inspection shows that the sum of all odd powers is pure imaginary (namely it is an $i$ times a real number) and that the real number multiplying $i$ is equal to the expansion of $\sin x$. Hence we arrive at the fundamental relations, 
\tbox{
\begin{equation} \label{expix}
e^{ix}=\cos x+i \sin x~, \ \ e^{-ix}=(e^{ix})^*=\cos x - i \sin x, \ \ |e^{ix}|=\sqrt{e^{ix}e^{-ix}}=1 \ \ \forall \ x \in \mathbb{R}~. 
\end{equation} 
}
The function $e^{ix}$  is of extreme importance in science and engineering.  
In electrical engineering, the the time dependence of currents and voltages is often containing a factor
 $e^{i \omega t}$ where $t$ is the time and $\omega$ is the frequency. At the end of calculations, the 
passage to trigonometric functions is carried out according to Eqs.~(\ref{expix}). The function $e^{i x}$ is sometimes referred to as "phase" or "phase factor".  Since the trigonometric functions are {\it periodic}, 
namely $\sin x=\sin (x + 2 m \pi), \ m=0, \pm 1, \pm 2, \ldots$ so is the function $e^{ix}$. In particular, $e^{i 2 m \pi}=1$.

Equations (\ref{expix}) can be inverted to express the trigonometric functions of a real variable $x$ in terms of the two exponentials, 
\begin{equation} \label{sincosexp}
\cos x=\frac{1}{2}(e^{ix}+e^{-ix}), \ \  \sin x=\frac{1}{2i}(e^{ix}-e^{-ix}).
\end{equation} 
Applications of the above formulae go much beyond the algebra of complex numbers. Suppose we are asked to express $\cos^n x$ as a combination of functions of the form $\cos k x$, for example, 
$\cos^2 x=\frac{1}{2}(1+\cos 2x)$. Then, \\
$\cos^n x$=$\frac{1}{2^n}(e^{ix}$+$e^{-ix})^n$=$\frac{1}{2^n} \sum_{k=0}^n \binom {n}{k} e^{i (n-k)x}e^{-i k x} 
$=$\frac{1}{2^n} \sum_{k=0}^n \binom {n}{k} e^{i (n-2 k)x}$=$\frac{1}{2^n} \sum_{k=0}^n \binom {n}{k} [ \cos (n$-$2 k)x$+$i  \sin (n$-$2 k)x]. $\\
Now in any equation involving complex numbers, the real and imaginary parts on both sides must be equal. In the above equation, the LHS is real, and therefore, the imaginary part on the RHS must vanish. 
This leads us to the identity $\cos^nx$=$\frac{1}{2^n} \sum_{k=0}^n \binom {n}{k}  \cos (n$-$2 k)x$. 
\subsubsection{Back to the Polar Representation: Algebraic Manipulations on Complex Numbers} 
Now we go back to the first equation in (\ref{cartpolar1}) and use Eq.~(\ref{expix}) to get the polar representation of $z$ and its natural power as,
\begin{equation} \label{polarexp}
z=|z|(\cos \theta+i \sin \theta)=|z|e^{i \theta} \ \ \Rightarrow \ \ z^n=|z|^n e^{i n \theta}~.
\end{equation}
Thus, the polar representation of $z^n$ is obtained by drawing an arrow of length $|z|^n$ and stretch it along an angle $n \theta$. Similarly, we can now compute 
\begin{equation} \label{sqrtz}
\sqrt{z}=\pm \sqrt{|z|}e^{i \theta/2}. 
\end{equation}
For example, we notice from Fig.~\ref{Fig7_2}b that $i=e^{i \pi/2}$, and therefore, $\sqrt{i}=\pm e^{i \pi/4}=\pm(\cos \frac{\pi}{4}+i \sin \frac{\pi}{4})=\pm \frac{1+i}{\sqrt{2}}$. Indeed, if we square this expression and remember that $i^2=-1$ we get $i$.

\subsubsection{Roots of Unity} 
Suppose we want to find all the solutions of the equation $z^n=1, \ \ n \in \mathbb{Z}_+$ (a positive integer). 
For $n$ even we know two real solutions $z=\pm 1$
 and for $n$ odd we know only one real solution, $z=1$. 
All the others are complex. The collection of all solutions (real and complex) are referred to as {\it roots of unity}. From our discussion of the polar representation we infer that $|z|=1$ so that all the roots of unity lie on the unit circle in the complex plane. In order to find them we write $z=e^{i \theta}, \ \Rightarrow \ z^n=e^{i n \theta}$.  We also found after Eq.~(\ref{expix}) the $e^{2 i m \pi}=1$. Therefore we have,
\begin{equation} \label{Root1}
z^n=1 \ \Rightarrow \ e^{i n \theta} =e^{2 i m \pi}, \ \Rightarrow \ \theta=\frac{2 m \pi}{n} , m=0,1,2,\ldots, n-1~.
\end{equation} 
This gives us the $n$ roots of unity $\{ e^{2 \pi i m/n} \} $ all of them lie on the unit circle. Thus, by extending our field of numbers from real to complex we found all the $n$ solution of the equation $z^n-1=0$. The fact that there are $n$ solutions to and equation involving a highest power $z^n$ is not accidental. In fact we have,\\
\underline{The Fundamental Theorem of Algebra:} Let $P_n(z)=\sum_{k=0}^n a_k z^k, \ \ a_k \in \mathbb{C}, \ a_n \ne 0$ be a polynom of degree $n$.  Then the equation $P_n(z)=0$ has $n$ solutions 
$z_1,z_2,\ldots,z_n$ (some of them might be equal). If all the coefficients are real , $a_k \in \mathbb{R}$ then the complex solutions comes in pairs of conjugate numbers, $z_i=x_i+i y_i, \ \ z^*_i=x_i-i y_i$. 

By this we conclude our survey of complex number theory. Unfortunately, the most beautiful and exciting 
part of it all, complex functions of {\it complex variable} falls out of our scope. 

\subsection{ Linear Vector Spaces} \label{Fields}
The basic mathematical structure of quantum mechanics involves state
vectors and linear operators that transform one vector into another.
The vectors, which represent quantum states, are members of a complex
vector space, endowed with an additional property 
(an inner product) that will be introduced later on. 
 In this appendix we review the key elements of 
 linear vector spaces. More advanced topics such as operators and matrices 
 are introduced in subsequent sections. 
\subsubsection{Definitions and Basic Properties}
We begin by defining a {\em vector space} over the field $\mathbb{C}$
of complex numbers as a set
\marginpar{linear \\ vector \\ space}%
$V$ of elements, called {\em vectors}, along with two operations `$+$'
and `$\cdot$' called {\it vector addition} and {\it scalar
multiplication} satisfying the following properties:
\begin{enumerate}
    \item If ${\bf u}$, ${\bf v} \in V$, their {\em vector sum} ${\bf
    u} + {\bf v}$ is an element of $V$.
    \item
    If ${\bf u}$, ${\bf v} \in V$ then ${\bf u} + {\bf v} =
    {\bf v} + {\bf u}$.
    \item
    If ${\bf u}$, ${\bf v}$, ${\bf w} \in V$ then $({\bf u} + {\bf v})
    + {\bf w} = {\bf u} + ({\bf v} + {\bf w})$.
    \item
    There is a {\em zero vector} ${\bf 0} \in V$ such that ${\bf u} +
    {\bf 0} = {\bf u}$ for all ${\bf u} \in V$.
    \item
    Each vector ${\bf v} \in V$ has an {\em additive inverse} ${\bf w}
    \in V$ such that ${\bf u} + {\bf w} = {\bf 0}$.  The inverse of a
    vector ${\bf v}$ is often denoted by $-{\bf v}$.
    \item 
    If $r \in \mathbb{C}$ and ${\bf u} \in V$, then $r \cdot {\bf u}
    \in V$.  Henceforth this operation of scalar multiplication will
    be written simply as $r {\bf u}$.
    \item
    If $r, s \in \mathbb{C}$ and ${\bf u} \in V$ then $(r + s){\bf u}
    = r {\bf u} + s {\bf u} \in V$.  Here the $+$ on the left hand
    side (LHS) of the equation is addition in $ \mathbb{C}$ and the
    $+$ on the RHS is vector addition in the vector space $V$.
    \item
    If $r \in \mathbb{C}$ and ${\bf u}, {\bf v} \in V$, then $r 
    ({\bf u} + {\bf v}) = r {\bf u} +  r  {\bf v}$.
    \item
    If $r$ and $s$ are any scalars, and ${\bf u} \in V$, $(rs)
    {\bf u} = r (s  {\bf u})$.
    \item
    $1 {\bf u} = {\bf u}$, and $0 {\bf u} = {\bf 0}$.
\end{enumerate}

The simplest examples of vector spaces come from plane and 3D Euclidean
geometry.  All vectors lying in the plane that originate from the same
point (say, the origin) form a two-dimensional Euclidean vector space
over the field of real numbers.  A similar construction holds in three
(and higher) dimensions.

It is obvious that if ${\bf v}_1, {\bf v}_2 \in V$ and $c_1,c_2 \in
\mathbb{C}$ then ${\bf w} \equiv c_1{\bf v}_1+c_2{\bf v}_2 \in V$;
${\bf w}$ is then said to be a {\em linear combination} of ${\bf v}_1$
and ${\bf v}_2$ with coefficients $c_1$ and $c_2$.  This construction
is easily extended to form linear combinations of $n$ vectors where
$n$ is a positive integer.  The $N$ vectors ${\bf v}_1,{\bf
v}_2,\ldots , {\bf v}_N \in V$ with ${\bf v}_i \ne {\bf 0}$ for all
($\forall$) $i$ are said to be {\em linearly independent} if and only
if (iff)
\begin{equation} \label{Eq:1.LA_a}
    \sum_{i=1}^N c_i {\bf v}_i={\bf 0} ~, \ \ \Leftrightarrow \ \ \
    c_i=0 \ \forall \ i~.
\end{equation}
I.e., it is not possible to express ${\bf 0}$ as a linear combination
of linearly independent $\{ {\bf v}_i \}$ except when all complex
coefficients $c_i \in \mathbb{C}$ are $0$.  If $N$ is maximal in the
sense that there is at least one set of $N$ linearly independent
vectors ${\bf v}_1,{\bf v}_2,\ldots , {\bf v}_N \in V$ with ${\bf v}_i
\ne {\bf 0} \ \ \forall \, i$, but there is no set of $N+1$ linearly
independent vectors, then $V$ is said to be a vector space of
dimension $N$, or equivalently, $N$ is the dimension of $V$; this is
denoted by $V^{(N)}$.  A set of linearly independent vectors can be
used to represent every vector in $V^{(N)}$.  Any set of linearly
independent vectors ${\bf v}_1,{\bf v}_2,\ldots , {\bf v}_N \in
V^{(N)}$ can be considered a ``coordinate system'' and is called a
{\em basis}, i.e., a {\em complete set of basis vectors}.  In this
basis, we can write an arbitrary vector ${\bf u}$ as ${\bf u} =
\sum\limits_{j=1}^N c_j {\bf v}_j$ and it can be represented by
\begin{equation} \label{Eq:1.LA_basis}
    {\bf u} = \sum\limits_{j=1}^N c_j  {\bf v}_j \, := \left( \! \!
    \begin{array}{c}
        c_1  \\
        \vdots  \\
	c_N
    \end{array}
   \! \! \right) ~,
\end{equation}
where $:=$ means ``can be represented as''.  A basis is not unique;
generally, there are an infinite number of bases that can be used.  It
should be stressed that vectors can be defined without specifying a
particular basis.  
%
%
\subsubsection{Dirac Notation}   \label{SubSec:App_LA_Dirac}

In quantum mechanics, {\em Dirac notation} is often used as a
powerful tool to treat vector spaces.  The use of Dirac notation is
not limited to quantum mechanics.  Among its other virtues, it
significantly simplifies manipulations in vector spaces with inner
products, such as Hilbert spaces (see below), which form the
mathematical basis for quantum mechanics.  In this sense, Dirac
notation is much more than just notation, and serves as a conceptual
framework for dealing with state space in quantum mechanics.

Vectors in Dirac notation are written as $|u \ra$, $|v \ra$ (instead
of ${\bf u}$, ${\bf v}$ or $\vec{u}$, $\vec{v}$), $|\psi \rangle$, $|
\phi \rangle$, etc., and are called {\em kets vectors}.  Note that
Greek letters are often used in representing state-vectors of a
quantum system.  When Dirac notation is used, the vector space $V$ is
also called a {\em ket space} (the dimension $N$ is often not
explicitly specified).  Any vector $|\psi \rangle \in V$ is
expressible as a linear superposition of basis vectors \{$|\phi_j
\rangle\} \subset V$.  Thus, in Dirac notation,
Eq.~(\ref{Eq:1.LA_basis}) is written:
\marginpar{linear \\ superposition}%
\begin{equation} \label{Eq:1.LA_1}
  |\psi \rangle  = \sum\limits_{j=1}^N {c_j |\phi_j \rangle } 
  \, := \left( \! \! \begin{array}{c}
        c_1  \\
        \vdots  \\
	c_N
    \end{array}
   \! \! \right) ~,
\end{equation}
where, as in Eq.~(\ref{Eq:1.LA_basis}), the symbol $:=$ means that
the ket $| \psi \rangle$ is represented as a column vector of the
coefficients $\{ c_j \}$ (in quantum mechanics, these are sometimes
called amplitudes) that multiply the basis kets $\{ |\phi_j \ra \}$.

To properly understand Dirac notation for quantum mechanics we need to
define another vector space, the {\em dual space}, or the {\em bra
space}, $V^\dagger$, that is directly related to $V$.\marginpar{dual
space} Vectors in $V^\dagger$ are called {\em bras} and are written as
$\la u |$, $\la v |, \la \chi |, \la \psi | \ldots$.  The two vector
spaces $V$ and $V^\dagger$ have the same dimension $N$ and basically
have identical structure.  Vectors and operations in $V$ are in
one-to-one correspondence with those in $V^\dagger$ (in mathematical
parlance, one says that the two spaces are isomorphic, meaning that
there is a mapping between elements of the two spaces that preserves
the structure of the vector space).  At this stage, the Hermitian
adjoint superscript $\dagger$ is used just to distinguish between the
ket space $V$ and the dual or bra space $V^\dagger$, and to map kets
onto bras and vice versa.  Thus, $|\psi \ra \in V \Leftrightarrow \la
\psi| \in V^\dagger$, is compactly written, $|\psi \ra ^\dagger =\la
\psi|$.  Note that $(c|\psi \ra )^\dagger =c^* \la \psi|$, and the
correspondence between linear combinations in $V$ and $V^\dagger$ is,
\begin{equation} \label{Eq:1.LA_6'}
    (c_1 |\phi_1\rangle + c_2 |\phi_2\rangle)^\dag = c_1^* \langle
    \phi_1 | + c_2^* \langle \phi_2 | ~.
\end{equation}
The expansion of a bra vector $\la \psi| \in V^\dagger$ in a given
basis $\{ \la \phi_j| \}$ and its representation in terms of the
coefficients (the image of Eq.~(\ref{Eq:1.LA_1}) in $V^\dagger$) is
given by,
\begin{equation} \label{Eq:1.LA_1'}
    \langle \psi | = \sum\limits_{j=1}^N {\langle \phi_j | c_j^* } \,
    := \left( c_1^* \, \ldots \, c_N^* \right) ~.
\end{equation}
Hence, bra vectors are represented as row vectors whose coefficients
(amplitudes) are complex-conjugated.
\subsubsection{Inner Product Spaces}  \label{Inner_Prod}

\begin{figure}
\centering
\includegraphics[scale=0.6]{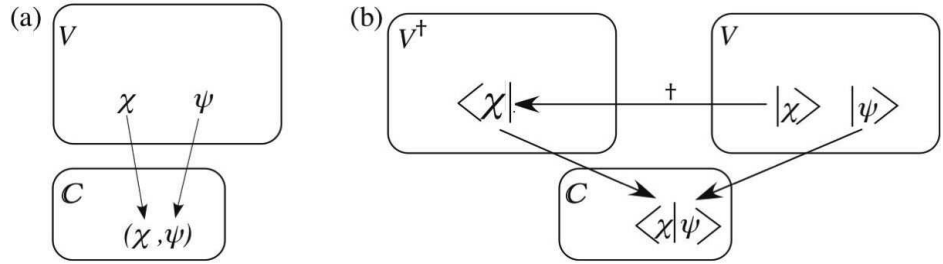}
\caption{An inner product associates a complex number $\in \mathbb{C}$
to any ordered pair of vectors in a vector space $V$.  (a) Inner
product without using Dirac notation.  (b) Inner product using Dirac
notation.}
\label{Fig_LA_Inner}
\end{figure}

Just as in Euclidean geometry, in quantum mechanics we need to specify
the notion of length of a vector (also referred to as {\em norm}), and
the notion of projection of a vector onto another vector.  This
requires the introduction of a new binary operation on the vector
space $V$, called an {\em inner product} (or sometimes a {\em scalar
product}, not to be confused with multiplication by a scalar), which
associates a complex number to any ordered pair of vectors in $V$.  If
Dirac notation is not used, the inner product of the vectors $\chi,
\psi \in V$ is written, $(\chi,\psi) \in \mathbb{C}$ [see
Fig.~\ref{Fig_LA_Inner}(a)].  When the Dirac notation is used, the
inner product of the kets $|\chi \ra, |\psi \ra \in V$ is written,
$\langle \chi | \psi \rangle \in \mathbb{C}$.\marginpar{inner product}
Note that $\la \chi| = |\chi \ra^\dagger \in V^\dagger$.  In Dirac
notation, this binary operation can be viewed as follows: first map
the ket $|\chi \ra \in V$ onto its bra image $\la \chi| \in V^\dagger$
using $|\chi \ra^\dagger = \la \chi|$.  Then associate a complex
number $\la \chi|\psi \ra \in \mathbb{C}$ with $\la \chi|$ and $|\psi
\ra$.  The inner product $\la \chi|\psi \ra$ is called a {\em
bracket}, a composition of bra and ket.  This view of inner product is
pictorially illustrated in Fig.~\ref{Fig_LA_Inner}(b).

The inner product is required to satisfy the following properties:
\begin{enumerate}
\item
$\langle \chi | \psi \rangle$ is a complex number, independent of the
basis in which the vectors are expanded.
\item
$\langle \chi | \psi \rangle = \langle \psi | \chi \rangle^*$.
\item
For any complex numbers $c_1$ and $c_2$,
$$\langle \chi | (c_1 |\psi_1
\rangle + c_2 |\psi_2 \rangle) = c_1 \langle \chi | \psi_1 \rangle +
c_2 \langle \chi |\psi_2 \rangle ~,$$
$$(c_1^* \la  \psi_1| + c_2^* \la \psi_2 |)
|\chi \rangle = c_1^* \langle \psi_1 | \chi \rangle + c_2^* \langle
\psi_2 | \chi \rangle ~.$$
\item
$\langle \psi | \psi \rangle \ge 0$, with equality if and only if $|
\psi \rangle$ is the zero vector, $| \psi \rangle = |0 \rangle$.
\end{enumerate}

\subsubsection{Properties of Inner Products}

Two kets $| \psi \rangle$ and $| \chi \rangle$ are said to be {\em
orthogonal} if\marginpar{orthogonality} 
\begin{equation} \label{Eq:1.LA_orthogonal}
   \langle \chi |\psi \rangle = 0 ~.
\end{equation}
The square of the length of the vector $| \psi \rangle$ is defined as
the inner product of the vector with itself, $\| \psi \|^2 = \langle
\psi | \psi \rangle$, i.e., the length (norm) of $| \psi \rangle$ is
\marginpar{norm}%
\begin{equation} \label{Eq:1.LA_norm1}
    \| \psi \| \equiv \sqrt{\langle \psi | \psi \rangle} \ge 0 ~.
\end{equation}
The length of any vector is real and non-negative, by virtue of the
property, $\langle \chi | \psi \rangle = \langle \psi | \chi
\rangle^*$.  Only the null vector has zero length.

Just like for ordinary vectors in 3D Euclidean space, the {\em
triangle inequality} holds for vectors in an inner product space,
\begin{equation} \label{Eq:1.LA_triangle}
    \| \psi + \phi \| \le \| \psi \| + \| \phi \| ~, \ \ \mbox{triangle inequality}
\end{equation}
with equality if and only if one of the vectors is a non-negative
scalar multiple of the other one.  The {\em Cauchy-Schwarz
inequality} holds,
\begin{equation} \label{Eq:1.LA_Schwartz}
    | \langle \phi | \psi \rangle| \le \| \psi \| \, \| \phi \| ~, \ \ \mbox{Cauchy-Schwarz  inequality}
\end{equation}
with equality only for vectors that are scalar multiples of one
another.

Other useful identities include the {\em parallelogram} and {\em
polarization identities},
\begin{equation} \label{Eq:1.LA_parallelogram}
    ||\phi+\psi||^2 +||\phi-\psi||^2 = 2(||\phi||^2+||\psi||^2) ~,
\end{equation}
\begin{equation}
    \langle \phi|\psi \rangle = \frac{1}{4}(||\phi+\psi||^2 - ||\phi -
    \psi||^2)-\frac{i}{4}(||\phi+i\psi||^2 -||\phi-i \psi||^2) ~.
\end{equation}
Given two ket vectors, $| \psi \rangle$ and $| \chi \rangle$, written
in terms of the same basis \{$|\phi_j \rangle$\}, and represented as
[see (\ref{Eq:1.LA_1})]
\begin{equation} \label{Eq:1.LA_7}
    | \psi \rangle = \left( \! \!
    \begin{array}{c}
	a_1  \\
	\vdots  \\
	a_N
    \end{array} \! \! \right) \quad \quad 
    | \chi \rangle = \left( \! \!
    \begin{array}{c}
	b_1  \\
	\vdots  \\
	b_N
    \end{array}
     \! \! \right) ~,
\end{equation}
their inner product is given by the complex number,
\begin{equation} \label{Eq:1.LA_8_gen}
    \langle \chi | \psi \rangle = \sum_{i,j=1}^N b_i^* a_j \langle
    \phi_i | \phi_j \rangle ~.
\end{equation}
Basis vectors \{$|\phi_j \rangle$\} are {\em orthonormal} if they
satisfy the conditions
\marginpar{orthonormality}%
\begin{equation} \label{Eq:1.LA_orthonormal}
  \langle \phi_i |\phi_j \rangle = \delta_{ij} = \begin{cases} 1 \, \,
  {\mathrm{ for }} \, \, i = j \\ 0 \, \, {\mathrm{ for }} \, \, i \ne
  j \end{cases}~,
\end{equation}
where $\delta_{ij}$ is called the {\em Kronecker delta
function}.\marginpar{Kronecker \\ delta function} For an orthonormal
basis
\begin{equation} \label{Eq:1.LA_8}
  \langle \chi | \psi \rangle = \left(  b_1^* \, \ldots \, b_N^* 
  \right) \left( \! \! \begin{array}{c}
  a_1  \\
  \vdots  \\
  a_N \end{array} \! \! \right) = \sum_{j=1}^N b_j^* a_j ~.
\end{equation}
If two kets $| \psi \rangle$ and $| \chi \rangle$ are orthogonal, and
the basis functions are orthonormal, Eq.~(\ref{Eq:1.LA_orthogonal})
reads, $\langle \chi |\psi \rangle = \sum_{j=1}^N b_j^* a_j = 0$.
Moreover, using (\ref{Eq:1.LA_8}), $\| \psi \| = \sqrt{\sum_{j=1}^N
|a_j|^2}$. Orthonormal bases are almost always used
in quantum mechanics.

The expansion coefficients $c_j$ in Eq.~(\ref{Eq:1.LA_1}) can be
computed by taking the inner product of (\ref{Eq:1.LA_1}) with a
complete set of basis vectors $|\phi_i \rangle$:
\begin{equation} \label{Eq:1.LA_2}
  \langle \phi_i |\psi \rangle = \sum\limits_j {c_j \langle \phi_i
  |\phi_j \rangle }  \ \forall \, i~.
\end{equation}
Equation (\ref{Eq:1.LA_2}) can be inverted to find the coefficients
$c_j$.  If the basis vectors $|\phi_j \rangle$ are {\em othonormal},
as is the case if these vectors are eigenvectors of a Hermitian (or
self-adjoint) operator [see below and
Secs.~(\ref{Subsec:Intro_Observable}) and
(\ref{Sec:FormalQM_Hermitian})], we obtain
\begin{equation} \label{Eq:1.LA_3}
  c_i  =  \langle \phi_i |\psi  \rangle ~,
\end{equation}
and thus we can write Eq.~(\ref{Eq:1.LA_1}) as:
\begin{equation} \label{Eq:1.LA_4}
  |\psi \rangle = \sum\limits_j { \langle \phi_j |\psi \rangle \, |
  \phi_j \rangle } ~.
\end{equation}
Sometimes it is convenient to write (\ref{Eq:1.LA_4}) as $|\psi
\rangle = \sum_j {| \phi_j \rangle \, \langle \phi_j |\psi \rangle}$ 
because this makes clear the insertion of the unit operator in the 
basis set expansion.
\subsection{Matrices} \label{Mat}
The use of matrices in Economics is ubiquitous. Therefore, all the details of matrix algebra are assumed to be known. Moreover, the basics of linear algebra 
taught in all Economic undergraduate programs use the notion of real matrices. 
Here we will briefly review the topic of matrices in general, including mainly complex matrices. All matrices discussed here are square matrices. 
The following section cannot be read before the reader is familiar with the algebra of complex number discussed in section \ref{CN}. 

A complex square matrix of order $N$ is an array of $N \times N$ complex numbers.  
All the manipulations known from linear algebra pertaining to real matrices are naturally extended 
to complex matrices, such as addition, multiplication, inverse, transposition, degree, determinant, trace, etc. 
Some properties peculiar to complex matrices will be discussed in this section, but we are mainly interested in unitary, special unitary and hermitian matrices to be defined below. These are the main 
type of matrices that are utilized in the theory of quantum games discussed in this thesis.  

\subsubsection{Some Operation On Matrices} 
Let us first define some operations on matrices that will be useful later on, commencing by 
operations performed on a single matrix.
The first operation that should be defined on complex matrix is complex conjugation. \\
\underline{Definition} Let $A$ be an $N \times N$ complex matrix with elements $A_{ij}$. 
We define its complex conjugate matrix $A^*$ as an $N \times N$ complex matrix with elements $A_{ij}^*$ (see Eq.~(\ref{53})). For example, $A=\binom{1+2i  ~~~ 2-i}{e^{4i}~~~  1}, \ \ A^*=\binom{1-2i ~~~ 2+i}{e^{-4i}~~~  1}$.\\
Another useful operation combines transposition and complex conjugation, \\
\underline{Definition} Let $A$ be an $N \times N$ complex matrix with elements $A_{ij}$. 
We define its hermitian adjoint matrix $A^\dagger$ as an $N \times N$ complex matrix with elements $A^\dagger_{ij}=A_{ji}^*$ or simply $A^\dagger=[A^T]^*$. For example,\\ 
$A=\binom{1+2i  ~~~ 2-i}{e^{4i}~~~  1}, \ \ A^T=\binom{1+2i  ~~~ e^{4i}}{2-i~~~  1}, \ \ A^\dagger=[A^T]^*=\binom{1-2i  ~~~ e^{-4i}}{2+i~~~  1} $. \\
Another important quantity defined for every $N \times N$ matrix is its trace, \\
\underline{Definition} Let $A$ be an $N \times N$ complex matrix with elements $A_{ij}$. 
We define its trace as Tr$[A]=\sum_{i=1}^N A_{ii}$, namely, the trace of a square matrix is the sum of its diagonal elements. For example, $A=\binom{1+2i  ~~~ 2-i}{e^{4i}~~~  1},$ Tr$[A]=2+2i$. 
If $A,B,C$ are $N \times N$ matrices then Tr$[AB]$=Tr$[BA]$ and 
Tr$[ABC]$=Tr$[BCA]$=Tr$[CAB]$. \\
To appreciate the significance of the trace, we define the notion of similarity. \\
\underline{Definition} Let $A$ be an $N \times N$ complex matrix and let $D$ be an arbitrary 
non-singular $N \times N$ matrix. We say that the matrix $A$ and the matrices $B \equiv DAD^{-1}$ are similar and denote it by $A \sim B$ (this is of corse a reflexive relation). A central problem in many branches of science (in particular in quantum mechanics) is, given a matrix $A$, to find a matrix $D$ such that $B = DAD^{-1}$ is diagonal, namely $B_{ij}=b_i \delta_{ij}$.\\
\underline{Theorem} If $A \sim B$ then det$[A]$=det$[B]$ and Tr$[A]$=Tr$[B]$. \\
Now we define an important operation on two matrices that yield a single matrix. \\
\underline{Definition} Let $A$ $B$ be two $N \times N$ complex matrices. Their {\it commutation relation} is defined as the matrix, $[A,B]=AB-BA=-[B,A]$.   If $[A,B]=0$ we say that $A$ and $B$ commute. As an example, consider the three $2 \times 2$ Pauli matrices that play a central role in quantum mechanics but also in quantum game theory, 
\begin{spacing}{0.9}
\begin{equation} \label{Paulimat1}
\sigma_1=\begin{pmatrix} 0&1\\1&0 \end{pmatrix}, 
\ \ \sigma_2=\begin{pmatrix} 0&-i \\ i &0 \end{pmatrix}, 
\ \ \sigma_3=\begin{pmatrix} 1&0 \\ 0&-1 \end{pmatrix}.
\end{equation}
\end{spacing}
\vspace{0.2in}
Then, with a little effort we can establish the following identities, 
\begin{equation} \label{Paulimat2}
[\sigma_1,\sigma_2]=i \sigma_3, \ \ [\sigma_2,\sigma_3]=i \sigma_1, \ \ [\sigma_3,\sigma_1]=i \sigma_2 ~.
\end{equation} 
\subsubsection{Cross Product} 
The operation of {\it Cross Product} (equivalently {\it Outer Product}) is extensively used in quantum games, this thesis included. We exemplify it here for the case of two $2 \times 2$ matrices but this operation is defined also for non-square matrices and there should be no restriction on the dimensions of the two matrices in the product. \\
\underline{Definition} Let $A$ $B$ be two $2 \times 2$ (complex) matrices. Their {\it cross product} is defined as the $4 \times 4$ matrix,
\begin{equation} \label{cross1}
A \otimes B=\begin{pmatrix} A_{11}B & A_{12}B\\ A_{21}B & A_{22} B \end{pmatrix}, \ \ A_{ij}B=A_{ij} \begin{pmatrix} B_{11}& B_{12} \\ B_{21} & B_{22} \end{pmatrix}. 
\end{equation} 
Note that $A \otimes B \ne B \otimes A$. In general, a cross product $A_{N_1N_2} \otimes B_{N_3N_4}$ yields a matrix $C_{N_1 \times N_3,N_2 \times N_4}$. 
\subsubsection{Definition of Eigenvalues and Eigenvectors}
Let us return to the analysis of vector spaces with inner product discussed previously. 
We learn from Linear Algebra that when a matrix $A_{NN}$ operates on a matrix $u_{N1}$ (a vector $|u \ra$) it yields a matrix $v_{N1}$ (a vector $v \ra$), namely $A|u\ra=|v \ra$. An important question is whether we can find a vector $|u \ra$ and a complex number $\lambda$ such that $|v\ra=\lambda |u \ra$.
If we find them we refer to $\lambda$ as an {\it eigenvalue of $A$} and to the vector $|u_\lambda \ra$ (that clearly depends on $\lambda$) as an 
{\it eigenvector of $A$ that belongs to the eigenvalue $\lambda$}. 
The algorithm for finding eigenvalues and eigenvectors is straightforward,   
\begin{equation} \label{Eigen1}
A|u \ra=\lambda |u \ra \ \Rightarrow (A-\lambda {\bf 1})|u \ra=0. \ \mbox{(The eigenvalue equation.)}
\end{equation}
The matrix $H(\lambda) \equiv (A-\lambda {\bf 1})$ is an $N \times N$ matrix, and the eigenvalue equayion is a set of $N$ linear homogeneous equations for the unknown components of the vector $|u_\lambda \ra$. 
We know from Linear Algebra that in order to get a solution for this set of equation the determinant of $H(\lambda)$ should vanish. From its definition, det$H(\lambda) \equiv P(\lambda)$ is a polynomial  of degree 
$N$ in the variable $\lambda$, referred to as the {\it characteristic plynomial}.  From the fundamental theorem of algebra we know that the equation 
$P(\lambda)=0$ has $N$ solutions (not all of them distinct). Solving the eigenalue equation then proceeds in two steps: 1) Find the $N$ roots $\lambda_1, \lambda_2, \ldots, \lambda_N$ of the equation $P(\lambda)=0$, and 2) Put $\lambda_i$ in the eignevalue equation and find the eigenvector $|u_{\lambda_i} \ra$ by solving the set of linear equations whose determinant is set to vanish. If we arrange the $N$ eigenvectors one near the other to form a matrix $D$ and arrange the eigenvalues as entries of a diagonal matrix $\Lambda$ we get, 
\begin{equation}  \label{Eigen2}
AD=D \Lambda, \ \ D=\left [|u_{\lambda_1} \ra,|u_{\lambda_2} \ra, \ldots, |u_{\lambda_N} \ra \right ], \ \ 
 \Lambda=\begin{pmatrix} \lambda_1&0&0&0&0\\0 & \lambda_2 &0&0&0\\.&.&.&.&. \\ 0&0&0&0&\lambda_N \end{pmatrix}
\end{equation} 
If the matrix $D$ is not singular, namely it has an inverse $D^{-1}$ we can multiply both sides by $D^{-1}$ and obtain 
\begin{equation} \label{Eigen3}
D^{-1}AD=\Lambda \sim A, \ \ \ \Rightarrow \ \mbox{Tr}A= \mbox{Tr}\Lambda=\sum_{i=1}^N \lambda_i, \ \  \mbox{det}A= \mbox{det}\Lambda=\prod_{i=1}^N \lambda_i~. 
\end{equation} 
This procedure is referred to as {\it diagonalization of the matrix $A$}. Matrix diagonalization is a central problem in quantum mechanics.  For large matrices It requires high computation resources. 
\subsubsection{Functions of Matrices} Matrices can be raised to a non-negative integer powers 
(with the definition $A^0={\bf 1}_{NN}, \ A^1=A$ and so on. Therefore, any expression 
involving a polynomial in the matrix $A$ can principally be evaluated, for example, 
$P(A)=A^3+2 A^2-A+{\bf 1}$ is a polynomial  of degree  3 in the matrix $A$. However, for more complicated 
functions, the precise definition is not always possible. In some cases the definition is straightforward, although the evaluation is not always easy. This is the case, for example, for the exponential function 
$e^{xA}$ where $x$ is a real or complex number. Using the Taylor expansion we may formally write, 
\begin{equation} \label{expA}
e^{xA}=\sum_{n=0}^\infty \frac{(xA)^n}{n!}={\bf 1}+xA+\frac{(xA)^2}{2!}+\frac{(xA)^3}{3!}+\ldots
\end{equation}
This series converges quickly and if only an approximate estimate is required it can be cutoff at some power and we are back at a polynomial function. O the other hand, there are cases where non-trivial matrices square to ${\bf 1}$, namely $A^2={\bf 1}_N$. For example, the Pauli matrices defined in Eq.~(\ref{Paulimat1}) satisfy $\sigma_i^2={\bf 1}_2$. In that case we have, 
\begin{equation} \label{expxA}
e^{xA}=\cosh x {\bf 1}_N+\sinh x A, \ \ \ (A^2={\bf 1}_N)~. 
\end{equation}
So far we discussed functions of a single matrix $A$. We can also use the above definitions for 
investigating function of two (and more) matrices, $A,B$ (both of dimensions $N \times N$). Here, however, extra care is required because $A$ and $B$ need not commute, that is, generically, $[A,B] \ne 0$. Thus, 
or example, if $[A,B] \ne 0$ we have,
\begin{equation} \label{fAB}
(A+B)^2=A^2+AB+BA+B^2 \ne A^2+2 AB+B^2, \ \  e^{A+B} \ne e^A e^B, \ \ \log AB \ne \log A+\log B~. 
\end{equation} 
\subsubsection{Transformation of Bases: Unitary Matrices} 
It is convenient to start the discussion on matrices from their role in the study of finite dimensional linear inner product vector spaces above the field of complex numbers. 
Consider a finite dimensional inner product space ${\cal V}$ of dimension $N$.  
We can express  an arbitrary vector (or equivalently state)  $|\psi \ra \in {\cal V}$ 
as a linear combination of 
$N$ linearly independent orthonormal vectors $\{ |\phi_i \ra \}$ that form a {\it basis}. 
By this we mean that every vector in the space can be expanded similarly and that 
$\la \phi_i|\phi_j \ra=\delta_{ij}$. The coefficients $\{ c_i \}$ in this expansion are complex numbers. But there are many bases in this vector space. The choice of $\{ |\phi_i \ra \}$ is somewhat arbitrary, and the expansion can be carried out using any other base, say  $\{ |\phi_i \ra \}$. The {\it same} vector $|\psi \ra$ can be expanded as
\begin{equation} \label{Mat1}
 |\psi \ra= \sum_{i=1}^N b_i |\alpha_i \ra = \sum_{j=1}^N c_j |\phi_j \ra. 
 \end{equation}
 The natural question is how the $N$ complex numbers (coefficients) $\{ b_i \}$ are related to the 
 coefficients $\{ c_j \}$. To answer this question we apply the bra $\la \alpha_k|$ on both sides 
 of Eq.~(\ref{Mat1}) and perform the corresponding inner product using the fact that $\la \alpha_k| \alpha_i \ra=\delta_{ik}$. This gives  (note the power of the Dirac notation), 
 \begin{equation} \label{Mat2}
 \la \alpha_k |\psi \ra=  b_k  = \sum_{j=1}^N \la \alpha_k|\phi_j \ra c_j \equiv \sum_{j=1}^N U_{kj} c_j \ \ 
 \Rightarrow  \bfb=U \bfc ~,
 \end{equation}
where $U$ is an $N \times N$ complex matrix with entries $U_{ij}=\la \alpha_i|\phi_j \ra$. It is easily verified that if we want to express the coefficients $\{ c_j \}$ in terms of $\{ b_i \}$ we get, after applying $\la \phi_j |$ on both sides of Eq.~(\ref{Mat1}) 
\begin{equation} \label{Mat3}
 \la \alpha_j |\psi \ra=  c_j  = \sum_{i=1}^N \la \phi_j|\alpha_i \ra b_j \equiv \sum_{i=1}^N V_{ji} b_i \ \ 
 \Rightarrow  \bfc=V \bfb ~,
 \end{equation}
 where $V$ is another $N \times N$ complex matrix with entries $V_{ji}=\la \phi_j|\alpha_i \ra$. From property 2 in the list of property of inner product we immediately see that $U_{ij}^*=\la \phi_j|\alpha_i \ra=V_{ji}$. In words: To get the matrix $V$, take the matrix $U$, transpose it to get $U^T$ (or $U'$ in the nomenclature of Economists) and then complex conjugate all its elements (see Eq.~(\ref{53})) to obtain $V=[U^T]^*$. \\
 \underline{Definition}  Let $U$ be a complex $N \times N$ matrix with elements $\{ U_{ij} \}$. We define its {\it Hermitian Adjoint Matrix} $U^\dagger$ as a complex $N \times N$ matrix 
 with elements $U^\dagger_{ij}=U_{ji}^*$ or compactly $U^\dagger=[U^T]^*$. 
 
Returning to the transformations (\ref{Mat2}) and (\ref{Mat3}) above we see that $V=U^\dagger$. 
Moreover, the relations $ \bfb=U \bfc $ and $\bfc=V \bfb$ imply that $V=U^\dagger=U^{-1}$.  This immediately implies that $U$ is not singular because it has an inverse.\\
\underline{Definition} Let $U$ be a complex $N \times N$ matrix and let $U^\dagger$ be its Hermitian adjoint. Then if $U^\dagger=U^{-1}$ We say that $U$ is a {\it Unitary matrix}. Of course, in that case, $U^\dagger$ is also unitary. We write these two statements compactly as, 
\begin{equation} \label{Mat4} 
UU^\dagger=U^\dagger U={\bf 1}_N~.
\end{equation} 
The determinant of a unitary matrix is a complex number $z$ with $|z|=1$, namely, $z$ is on the unit circle.  
Note that in Eqs.~ (\ref{Mat2}) and (\ref{Mat3}) we related the {\it expansion coefficients} of $| \psi \ra$ in the two bases. The same matrices $U$ and $U^\dagger$ are also used in the {\it transformation of bases}, expressing the vectors in one basis in terms of those from the other basis. Using the same techniques as  that used above we easily find, 
\begin{equation} \label{Mat5} 
 |\alpha_i \ra=\sum_{j=1}^N U^\dagger_{ij} |\phi_j \ra, \ \ |\phi_i \ra=\sum_{j=1}^N U_{ij} |\alpha_j \ra~. 
 \end{equation}
 Summarizing all the results in a compact form we may write,
 \begin{equation} \label{Mat6} 
 \bfb=U\bfc, \ \ |{\bm \alpha} \ra=U^\dagger |{\bm \phi} \ra, \ \ \bfc=U^\dagger \bfb, \ |{\bm \phi} \ra= U|{\bm \alpha} \ra~. 
 \end{equation}
 \subsubsection{Hermitian Matrices}
 Hermitian matrices play a cenral role in quantum mechanics. They establish the connection between the 
 abstract notion of operators in Hilbert space and measurable quantities that we encounter in everyday life such as energy, current, light etc. Here we define the notion of Hermitian matrix and list the main properties of these matrices. \\
 \underline{Definition:} A  complex $N \times N$ matrix $H$ is said to be Hermitian if 
 \begin{equation} \label{Herm1}
 H=H^\dagger=[H^T]^*. \ \ \mbox{for example} \ \ H=\begin{pmatrix} 1&1+i\\1-i&2 \end{pmatrix}~.
 \end{equation} 
 The main properties of a Hermitian matrix $H$ are listed below. 
 1) The diagonal element of$H$ are real, $H_{ii}=H^\dagger_{ii} \ \Rightarrow H_{ii}=H_{ii}^*, \ \ \Rightarrow \ H_{ii}$ is real. \\
 2) The eigenvalues $\{ \lambda_i \}$ are real. \\
 3) The eigenvectors of $H$ are orthogonal and by proper normalization can be made orthonormal: 
 \begin{equation} \label{Herm2} 
 H|u_{\lambda_i} \ra=\lambda_i |u_{\lambda_i} \ra, \ \ \Rightarrow \ \ \la u_{\lambda_i}|u_{\lambda_j} \ra=\delta_{ij}~.
 \end{equation}
 4) The eigenvectors of $H$ form a complete basis in the corresponding linear vector space. An arbitrary vector $|\psi \ra$ in this space can be expanded as,
 \begin{equation} \label{Herm3}
 |\psi \ra=\sum_{i=1}^N c_i |u_{\lambda_i} \ra~. 
 \end{equation} 
 5) For a Hermitian matrix, the matrix $U$ whose columns are composed of the eigenvectors $\{ u_{\lambda_i} \}$ of $H$ 
 (that is the matrix $D$ in Eq.~(\ref{Eigen2})) is unitary. Therefore, following Eq.~(\ref{Eigen3}), a Hermitian matrix $H$ is diagonalized by a unitary matrix $U$ composed of its eigenvectors, namely, $U^\dagger H U=\Lambda$. \\
6)  If $\alpha$ is a real number and $H$ is a Hermitian matrix then an exponent of $i \alpha H$ yields a unitary matrix, namely,
 \begin{equation} \label{Herm4} 
 U=e^{i \alpha H}, \ \ U^\dagger=[U^T]^*=e^{-i \alpha H}, \ \ UU^\dagger=U^\dagger U={\bf 1}_{N \times N}~.
 \end{equation}
 For example, the Pauli matrices defined in Eq.~(\ref{Paulimat1}) are Hermitian. Following Eq.~(\ref{expxA}) we have, 
 \begin{equation} \label{Herm5} 
U= e^{i \alpha \sigma_k}=\cos \alpha {\bf 1}_{2 \times 2}+i \sin \alpha \sigma_k, \ \ U^\dagger= e^{-i \alpha \sigma_k}=\cos \alpha {\bf 1}_{2 \times 2}- i \sin \alpha \sigma_k, \ \ (k=1,2,3)~.
\end{equation} 
7) A function $f(H)$ of a Hermitian matrix $H$ (that is a matrix in itself) can be evaluated once $H$ is diagonalized, 
\begin{equation} \label{Herm6}
f(H)=\sum_{i=1}^N |u_{\lambda_i} \ra f(\lambda_i) \la u_{\lambda_i}|,
\end{equation} 
where $\la u_{\lambda_i}|=|u_{\lambda_i} \ra^\dagger$ is a row vector ($u'^*$ in the language of mathematical economy). 
\subsection{Group Theory} \label{GT}
In our analysis of classical and quantum games we have already encountered the notion of group 
without actually realizing it. The reason  that this concept is not extremely vital in simultaneous 
one-move games.  But it is important to be familiar with this mathematical discipline since, as we perceive, it will play an important role in the analysis of more complicated games. As we shall see, strategies in classical games sometimes form a {\it discrete group} (mostly with finite number of elements) while strategies in quantum games 
form a {\it continuous group}. The theory of groups is intimately related with the notion of symmetry 
and symmetries are all around us. In this section we explain what is a group, discuss some of the basic properties and give some examples. 

\subsubsection{Discrete Groups of Finite Order}
\begin{figure}
\centering
\includegraphics[scale=0.6]{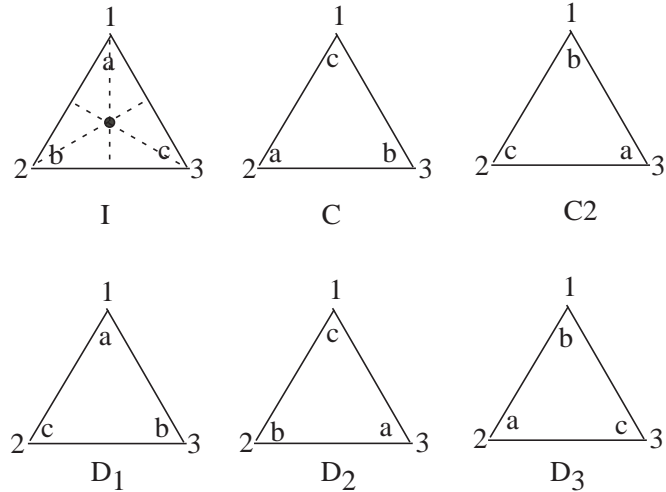}
\caption{Symmetry operations on a triangle. $I$ Unit operation leaves it as it is. $C$=120$^0$ rotation counterclockwise. $C2$=240$^0$ rotation counterclockwise. $D_{1,2,3}$=180$^0$ rotation around axes 1,2,3 (dashed lines). }
\label{Fig7_3}
\end{figure}

Instead of jumping into formal definitions and axioms, it is more useful (and more amusing) 
to become acquainted with the concept of group through a simple example. Assume there are three points fixed in the plane denoted as 1,2,3 that form an equilateral triangle. Then take such a perfect and clean triangle and put it such that each apex coincides with one point, as in Fig.~\ref{Fig7_3}.  Now close your eyes and ask your friend to perform some operation on the triangle that still leaves each apex near some fixed point. When you open your eyes you will not notice any difference. 
In fact, you may justly think that nobody touched the triangle at all. 
This means that your friend performed a {\it symmetry operation} on the triangle. 
This simple example contains all the ingredient needed to understand the concept of (discrete) groups. Now we would like to establish a minimal list of different operations your friend could have applied. For this purpose, we assign a letter to each apex of the triangle (we could as well 
paint every apex by different color). 
Simple inspection shows that your friend could have used one of the following six operations.\\
1) The unit operation $I$ which means, leave the triangle as it is, don't touch it.\\
2) The operation $C$ which means, rotate 120$^0=2 \pi/3$ counterclockwise around a perpendicular axis passing through the center of the triangle. \\
3) The operation $C2$ which means, rotate 240$^0=4 \pi/3$ counterclockwise around a perpendicular axis passing through the center of the triangle. We write it as $C2$ because it is obtained by application of the operation $C$ twice, one after the other. It is evident that performing two symmetry operations one after the other is, by itself, a symmetry operation. Some it is called 
multiplication, although we {\it do not multiply numbers}, we just {\it combine two symmetry operations}, where the order which operation is performed first is important.  \\
4-6) The three operations $D_{1,2,3}$ which mean 180$^0$ rotation around axes 1,2,3 (dashed lines in Fig.~\ref{Fig7_3}). 

Thus, we get a list of six different operations. Since the number of different positions of the triangle is exactly six, we conclude that the list of symmetry operations appearing Fig.~\ref{Fig7_3} is complete. Any composition of symmetry operation from this list will again give a symmetry operation from this list. Thus, performing $C$ twice yields $C2$, and performing $C$ followed by $D_1$ 
yields $D_2$. However performing $D_1$ followed by $C$ yields $D_3 \ne D_2$. 
It is also evident, by construction, the performing three operations one after the other is 
associative. 
Note also that after performing a symmetry operation, your friend might want to have the triangle back at its initial position. For achieving that, he does not have to turn it backward, he can always chose some operation from the list and perform it accordingly. For example, applying $C$ after applying $C2$ brings the triangle back to its initial position, which means, doing nothing (denoted as $I$). 

It is now possible to slightly formalize the structure of symmetry operations on an equilateral triangle. We denote by $G=\{I,C,C2,D_1,D_2,D_3 \}$,  the set of six symmetry operations 
 and by $g_2 \bullet g_1$ the 
symmetry operation obtained by applying $g_2 \in G$ {\it after} $g_1 \in G$. 
 We have established the following properties of the set $G$ endowed with the operation 
 $\bullet$, that can be regarded as group axioms. 
 \tboxed{
 \begin{center}
 {\Large Group Axioms}
 \end{center}
 \begin{enumerate} 
 \item{} \underline{Closeness:} $G$ is closed under $\bullet$, namely $g_1, g_2 \in G \ \Rightarrow \ g_1 \bullet g_2 \in G$. 
 Thus $\bullet:  G \times G \to G$ is a mapping. 
 \item{} \underline {Associativity:} The operation $\bullet$ is  associative, $g_3 \bullet (g_2 \bullet g_1)=(g_3 \bullet g_2) \bullet g_1, \ \ \forall \ g_1,g_2,g_3 \in G$.
 \item{} \underline{Unit Element:} There is an element $I \in G. \ni \ g \bullet I=I \bullet g=g, \ \forall \ g \in G$. $I$ is called the unit element. 
 \item{} \underline{Inverse Element: } For every $g \in G$ there exists an element $g^{-1} \in G$ such that $g\bullet g^{-1}=g^{-1} \bullet g=I.$  
 \end{enumerate} 
 }
 \subsubsection{ \underline{More Definitions and Basic Properties} }
 Now forget for a moment on our triangle and think of a set $G$ endowed with an operation $\bullet$ in general. We then formulate the following
 \begin{enumerate}
\item{}  {\bf Group:} A structure $\la G, \bullet \ra$ satisfying the four axioms listed above is 
 called a group. 
 \item{} {\bf Commutative and Non-Commutative Groups:} If $g_1 \bullet g_2=g_2 \bullet g_1$ the group is said to be {\it commutative or Abelian}. 
 Otherwise the group is said to be {\it non-commutative or non-Abelian}. 
 (As we have seen, the triangle group $G(\triangle)$ studied above is non Abelian). 
\item{} {\bf Order of A Group:} The number of elements in the group is called {\it the group's order} (the order of $G(\triangle)$ is 6). If the order is finite, the group is said to be finite. Finite order may be very small 
(for example the {\it trivial group} $G= \la \{ I \}, \bullet \ra$ is a finite group containing one element). The order may also be be very very large. The number of symmetry operations on Rubik's $3 \times 3 \times 3$ cube is 43252003274489856000. Every group of order $n<6$ is commutative (Abelian). 

\item{} {\bf Multiplication Table:} The information on any group of finite order $n$ can be encoded 
within a {\it group's multiplication table}, that is an $n \times n$ list with the element $g_i \bullet g_j \in G$  in entry $(i,j)$ (the order is important if the group in not commutative). An example of a multiplication table for the (Abelian) group $\la G=\{ 1,i,-1,-i \}, \bullet=\times \ra$ of order 4 is,
\begin{center}
\begin{tabular}{|l||l||l||l||l||l||l||l||l||l||l||l||l||l||l||l||l||l|} 
\hline  ~~ ~~ & ~~1~~& ~~i~~&~~-1~~&~~-i~~ \\ 
\hline  ~~1~~ &~~1~~ &~~i~~&~~-1~~&~~-i~~   \\ 
\hline  ~~i~~ &~~i~~ &~~-1~~&~~-i~~&~~1~~   \\ 
\hline  ~~-1~~ &~~-1~~ &~~-i~~&~~1~~&~~i~~   \\ 
\hline  ~~-i~~ &~~-i~~ &~~1~~&~~i~~&~~-1~~   \\ 
\hline
\end{tabular} 
\end{center}
We see that every row and every column contain all the group elements each element just once, and that for an Abelian group the table is symmetric around the main diagonal. 
\item{} {\bf Subgroups:} Let $\la G, \bullet \ra$ be a group and let $H \subseteq G, \ \ H \ne \emptyset$. 
Then if $\la H, \bullet \ra$ is a group, we say that {\it $H$ is a subgroup of $G$}. For example, 
$H=\{ I, C, C2 \}$ (of order 3) is an (Abelian) subgroup of $G(\triangle)$ because it satisfies the four group axioms.  We are mainly interested in the cases that $H$ is not trivial and $H \subset G$. 
Other subgroups of $G(\triangle)$ are $\{ I, D_i \}$ each of order 2. The order of a subgroup is a divisor of the order of the group, for example, Order [$G(\triangle)$/Order[$H$]=6/3=2. 
\item{} {\bf Normal Subgroup:} A subgroup $H$ of a group $G$ is said to be a {\it normal subgroup in $G$} iff $ghg^{-1} \in H, \ \ \forall g \in G$ and $\ \forall  \ h \in H$. For example $H=\{ I, C, C2 \}$ is normal subgroup of $G(\triangle)$ because $D_i h D_i^{-1}= D_i h D_i \in H$. For example, 
$D_1CD_2=C2$.  The element $ghg^{-1}$ is said to be {\it conjugate to h}. 
\item{} {\bf A simple group:} A group $G$ is said to be {\it simple} if it DOES NOT have a proper normal subgroup (here proper means neither $H=\la \{ I \}, \bullet \ra$ nor  $H=G$). The simple non-abelian group with smallest order is of order 60. 
\item{} {\bf Group Generators and Cyclic Groups:} A minimal set of elements $g_1, g_2, \ldots g_K \in G$ whose monomials $g_{i_1}^{k_1}g_{i_2}^{k_2}\ldots g_{i_K}^{k_K}$ exhausts all the group elements is called the set of {\it group generators}. For example, the generators of $G(\triangle)$ are $(C,D_1$. The other elements are obtained as $C2=C^2, \ I=C^3, \ D_2=C D_1, \ D_3=D_1 C$.  A group that has a single generator is called a {\it cyclic group}. For example, the subgroup\\
 $H \subset G(\triangle)=\la \{ I, C, C2 \}, \bullet \ra$ is a cyclic group generated by $C$. A cyclic group is obviously Abelian. 
\end{enumerate} 
\subsubsection{The Permutation Group $S_n$}
Let us now have a somewhat different look at the triangular setup. Instead of a triangle we just have 
three white balls located inside the space fixed three baskets 1,2,3 and your friend is asked to juggle their places while you close your eyes. Again if the balls are not marked, you will not notice any change after opening your eyes. Your friend executed a {\it permutation operation}
 summarized as 
follows: Take the ball from basket 1 and put it in basket $i_1$, the ball from basket 2 to 
basket $i_2$ and the ball from basket 3 to basket $i_3$. Here $i_1,i_2, i_3 \in \{1,2,3 \}$ and are distinct. Since you could not judge whether he touched the balls or not it is a symmetry operation, now refers to as 
{\it permutation symmetry}.  We have used the concept of permutation symmetry in Subsection \ref{Subsec_Trits}. In classical games a player's strategy consists of replacing the state of a bit 
(or trit) by another state, an operation that is a special case of permutation. Permutations also play a central role in social choice and in the proof of Rubinstein theorem.  Here we  will analyze this structure and expose the main definitions and the important properties. We will also map the elements of the permutation symmetry group on a set of matrices. The 
set of permutation of three balls (objects) as described above is denoted as $S_3$ and 
by simple counting we find that it contains $3!=6$ 
elements $\{s_1,s_2,\ldots, s_6\}$. Once it is established that the set $S_3$ endowed with the operation $\bullet$ of performing two successive permutations $s_2 \bullet s_1$ satisfies the 
group axioms we will call it {\it the permutation group $S_3$} whose order is 6. 
The alert reader might already guess that $S_3$ and $G(\triangle)$ are, in some sense, identical. That is, in fact, true and we will come to it soon. 
To keep the mathematics simple, we will use permutation groups with 
 small number of elements 
in our illustrative examples below, (e.g $S_3$ or $S_5$), but 
every statement  can easily be examined for $n>3$, in 
which case the permutation group $S_n$ ($n$ balls and $n$ baskets) is of order $n!$. 

A permutation (i.e., rearrangement) of $3$ balls between 3 baskets
 as described above  is represented as
 \begin{spacing}{0.8}
\begin{equation} \label{Eq:9.exch.20}
    P = \left(\! \! \begin{array}{*{20}c} 1 & 2 &3 \\
  i_1 & i_2& i_3  \\
   \end{array}  \! \! \right) ~,
\end{equation} 
\end{spacing}
\ \\
\noindent
where $i_1,i_2, i_3 \in \{1,2,3 \}$ and are distinct.  The symbol on the RHS of
Eq.~(\ref{Eq:9.exch.20}) should be read as follows: Take the ball from basket $k$ and put 
it in basket $i_k$ for $k=1,2,3$. 
Since combining two such operations one after the other 
leads us to some third operation of $3$ objects 
we conclude that $\la S_3, \bullet \ra$ is a group.
The operation $\bullet$  will be defined more precisely below together with all 
the group axioms. 
\vspace{0.2in}
\noindent 
{\bf Two cycle permutations} It is possible (and useful) to decompose a complicated permutation 
requiring to move many balls into many new baskets into a combination 
(product) of simple permutations called two-cycles $P_{ij} \equiv \binom{i \; j}{j \; i}$ or 
even simply $(i,j)$. 
This means a swap: "put ball $i$ in basket $j$ and ball $j$ in basket $i$ with all other balls untouched". Examples are 
\begin{spacing}{0.9}
\begin{equation} \label{Eq:9.exch.21}
   P = \left(\! \! {\begin{array}{*{20}c}
   1 & 2 & 3 & 4  \\
   2 & 1 & 4 & 3  \\
  \end{array} } \! \! \right) = (12)(34) ~, \ \     P = \left(\! \!  {\begin{array}{*{20}c}
    1 & 2 & 3 & 4 & 5 \\
    2 & 3 & 4 & 1 & 5 \\
   \end{array} } \! \! \right) = (14)(13)(12)(5) ~.
\end{equation}
\end{spacing} 
\noindent
Note that
the two-cycle $(34)$ on the right is applied first and $(12)$
afterward. The one-cycle $(5)$ (meaning do not move the ball in basket 5) 
is often left out for concise notation.  
The first permutation is {\em even} since can be written as a product of an
even number of two-cycles, while the second permutation is {\em odd}. 
 The structure \\
 $\la \{$ Even Permutations $\} \ra$ is a proper subgroup of S$_n$ 
 referred to as {\it the
alternating group}, denoted as $A_n$ that has an order $n!/2$. In fact, $A_5$ is a 
simple group with the smallest possible order (equal to 60). 
 A {\em cyclic permutation} $P = (1 2 \ldots N)$ takes the form
\begin{equation} \label{Eq:9.exch.23}
    P = (12\ldots N) \equiv \left(\! \! {\begin{array}{*{20}c} 
    1 & 2 & \ldots & N-1 & N \\
    2 & 3 & \ldots & N & 1  \\
    \end{array} } \! \! \right) ~.
\end{equation}
A cyclic permutation can be written as a product of two-cycles in the
form, $(a_1 a_2 \ldots a_N) = (a_1 a_N)(a_1 a_{N-1})\ldots(a_1 a_2)$.
A three-cycle is a cyclic permutation of three numbers, e.g., $(123)$,
or $(6,9,13) = \left(\!  \!  {\begin{array}{*{20}c} 6 & 9 & 13 \\ 9 &
13 & 6 \\ \end{array} } \!  \!  \right)$.  A permutation is said to be
in disjoint cycle form if it is written so that the various pairs of
cycles which define it have no number in common.

\subsubsection{Checking the Group Axioms} 
\begin{figure}[!ht]
\centering
\includegraphics[scale=0.4]{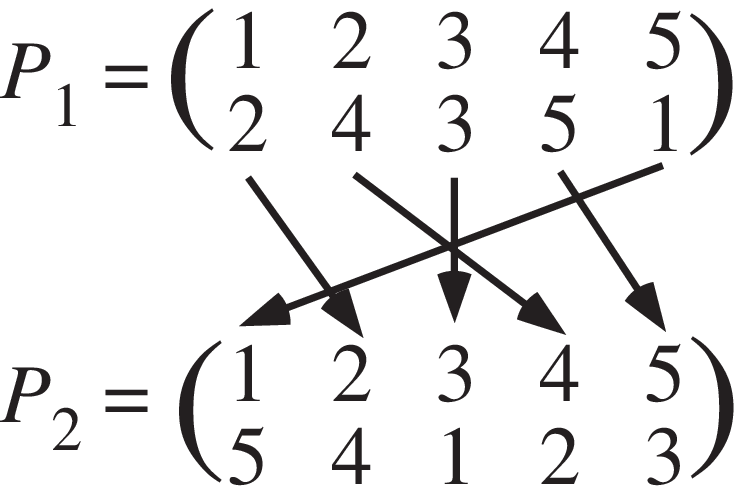}
\caption{The product, $P_2 P_1$, of the permutations in
(\ref{Eq:9.exch.24}) is obtained by following the arrows, e.g., $1 \to
2 \to 4$, $2 \to 4 \to 2$, etc, to obtain the last equation in (\ref{Eq:9.exch.24}).}
\label{Fig_9.Permutation_product}
\end{figure}

{\bf Closeness} Let us use a specific example to 
explain how the multiplication  operation$\bullet$  between two permutations is carried out. 
\begin{equation}  \label{Eq:9.exch.24}
    P_1 = \left(\! \! {\begin{array}{*{20}c} 1 & 2 & 3 & 4 & 5 \\
    2 & 4 & 3 & 5 & 1 \\
    \end{array} } \! \! \right) ~, \quad
    P_2 = \left(\! \! {\begin{array}{*{20}c}
    1 & 2 & 3 & 4 & 5 \\
    5 & 4 & 1 & 2 & 3 \\
  \end{array} } \! \! \right)  ~, \ \ \     P_2 P_1 = \left(\! \! {\begin{array}{*{20}c} 
    1 & 2 & 3 & 4 & 5 \\
    4 & 2 & 1 & 3 & 5 \\
    \end{array} } \! \! \right) ~.
\end{equation}
Note that first $P_1$ is applied, then $P_2$.  For example $P_1$ says "take the ball that is now in basket 1 and put it in basket 2,  while $P_2$ says "take the ball in basket 2 and put it in basket 4. The combination $P_1P_2$ then says "take the ball in basket 1 and put it in basket 4. This is further illustrated 
in Figure
\ref{Fig_9.Permutation_product} that shows how the product permutation is
obtained.
\ \\
{\bf The unit Element:}  As an example, The unit element, $I$, of $S_5$ is
\begin{equation}
    I = \left(\! \! {\begin{array}{*{20}c}
    1 & 2 & 3 & 4 & 5 \\
    1 & 2 & 3 & 4 & 5 \\
  \end{array} } \! \! \right)  ~, \ \Rightarrow I P=PI=P \ \forall P \in S_5,
\end{equation}
where the second part follows the definition and the illustration in Fig. \ref{Fig_9.Permutation_product}. 
\ \\
{\bf The Inverse Element of a Given Permutation} The inverse 
of the element in (\ref{Eq:9.exch.20}) is
\begin{equation}
  \left(\! \! \begin{array}{*{20}c} 1 & 2 &3 \\
  i_1 & i_2& i_3  \\
   \end{array}  \! \! \right)^{-1} =    \left(\!  \!  {\begin{array}{*{20}c} i_1 & i_2 &
     i_3 \\ 1 & 2 & 3  \\
 \end{array} } \! \! \right) ~.
\end{equation}
After this row inversion is carried out, it is useful to juggle columns on the right hand side so that the first row is 
ordered from the left. 
Thus we have established that $S_n$ is a (non-Abelian) group of order $n!$. 
\subsubsection{Equivalence of $S_3$ and $G(\triangle)$: Group Isomorphism} 
It is easy to check that there is a $1 \leftrightarrow 1$ mapping between $S_3$ and $G(\triangle)$ 
written as $P_i \in S_3 \leftrightarrow g_i \in G(\triangle)$ such that $P_i \bullet P_j \in S_3  \leftrightarrow g_i \bullet g_j \in G(\triangle)$. For example, 
\begin{equation} \label{S3Gtriangle}
\binom{123}{231}  \leftrightarrow C, \ \ \binom{123}{312}  \leftrightarrow C2, \ \ \binom{123}{132}  \leftrightarrow D_1,
\end{equation} 
and so on. This equivalence is referred to as {\it group isomorphism}. Alternatively, we write 
$S_3 \sim G(\triangle)$ and say that $S_3$ and $G(\triangle)$ are isomorphic. 
On the other hand, $G(\square)$ is NOT isomorphic to $S_4$.
If, in analogy with Figure \ref{Fig7_3} we inspect the symmetry operations on a perfect square lying on the plane we find the following operations: $C, C^2, C^3, C^4=I,D_{13}, D_{24}, h_{12},h_{23}$ where 
$C$ is a 90$^0$ rotation around an axis perpendicular to the plane passing through the square's center, 
$D_{ij}$ is a 180$^0$ degrees rotation around a diagonal passing through points $i$ and $j$ and $h_{ij}$ 
 is a 180$^0$ degrees rotation around a bisector of a segment joining points $i$ and $j$, all together 8 elements. Hence the order of $G(\square)$ is 8 while the order of $S_4$ is $4!=24$. The reason is that 
in $G(\square)$ there are several constraints forcing some pair of apex to be in certain points, for example, 
apex $a$ and $b$ should occupy adjacent fixed points like 1 and 2 for example. These constraints reduce the number of elements on $G(\square)$ compared with $S_4$. However, 
an important theorem referred to as {\it Cayley's theorem} states that 
{\it every group $G$ of order $n$ is isomorphic to a subgroup of $S_n$};
This means that  $G(\square)$ is isomorphic to a subgroup of $S_8$.   
\subsubsection{Matrix Representation of $S_3$}
Let us now look at the permutation of the three balls between 
three baskets 1,2,3 in a somewhat different way. We form a three dimensional vector 
$(1,2,3)^T$ that represents the initial locations of balls $a,b,c$ respectively. Then if we 
apply a permutation such as $P_1=\binom{123}{213}$ we ask the question: What is the $3 \times 3$ 
matrix $M_1$ that transform the vector ${123}^T$ to the vector ${213}$? The answer is simple,
\begin{spacing}{0.9} 
\begin{equation} \label{Rep1}
\begin{pmatrix} 2\\1\\3 \end{pmatrix}=\begin{pmatrix} 0&1&0\\1&0&0 \\0&0&1 \end{pmatrix} \begin{pmatrix} 1\\2\\3\\ \end{pmatrix}~.
\end{equation}    
\end{spacing}
\ \\
Since the matrix $M_1$ depends on the permutation $P_1$ we write it as $M_1(P_1)$. 
Now we apply a second permutation, say $P_2=\binom{123}{231}$. How do we 
represent the permutation $P_2P_1$ in a matrix form? The answer is now simple. We form the matrix $M_2(P_2)$ that transform $(123)^T$ onto $(231)^T$ and apply it from the left on $M(P_1)$. A simple inspection leads to the form $M(P_2)=\tiny{\begin{pmatrix} 0&1&0\\0&0&1\\1&0&0 \end{pmatrix}}$.
Therefore we find 
\begin{spacing}{1.0}
\begin{equation} \label{Rep2}
P_2P_1=\binom{123}{132}, \ \ M(P_2P_1)\begin{pmatrix} 1\\2\\3\\ \end{pmatrix}=M(P_1)M(P_2)\begin{pmatrix} 1\\2\\3\\ \end{pmatrix}=\begin{pmatrix} 1&0&0 \\0&0&1\\0&1&0 \end{pmatrix}\begin{pmatrix} 1\\2\\3\\ \end{pmatrix}=\begin{pmatrix} 1\\3\\2\\ \end{pmatrix}.
\end{equation}
\end{spacing}
\ \\
It is easy to construct the matrices corresponding to all the six elements of $S_3$. We will list them all 
because we will need to discuss it throughly:
\begin{spacing}{1.0} 
\begin{eqnarray}
&& \binom{123}{123} \to  \begin{pmatrix} 1&0&0 \\0&1&0\\0&0&1 \end{pmatrix} \equiv I ={\bf 1}_3, \ \
 \binom{123}{213} \to  \begin{pmatrix} 0&1&0 \\1&0&0\\0&0&1 \end{pmatrix} \equiv M_2,\ \
 \binom{123}{321} \to  \begin{pmatrix} 0&0&1 \\0&1&0\\1&0&0 \end{pmatrix} \equiv M_3, 
 \nonumber \\
&&  \binom{123}{132} \to  \begin{pmatrix} 1&0&0 \\0&0&1\\0&1&0 \end{pmatrix} \equiv M_4, \ \
  \binom{123}{231} \to  \begin{pmatrix} 0&1&0 \\0&0&1\\1&0&0 \end{pmatrix} \equiv M_5, \ \ 
 \binom{123}{312} \to  \begin{pmatrix} 0&0&1 \\1&0&0\\0&1&0 \end{pmatrix} \equiv M_6.
 \label{Rep3}
\end{eqnarray}
\end{spacing}
\ \\
Thus we have at the following result
\tboxed{To every element $P \in S_3$ there corresponds a matrix $M(P)$ such that to the element $P_2P_1 \in S_3$ the corresponding matrix is $M(P_2P_1)=M(P_2)M(P_1)$. In each column and each row in each matrix there is a single entry 1 and the other two entries are zero.  This is a {\it faithful matrix representation of $S_3$.}  The six matrices form a group ${\cal M}_3$ that is isomorphic to $S_3$. 
} 
\begin{figure}[!ht]
\centering
\includegraphics[scale=0.4]{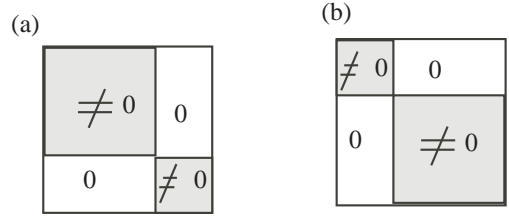}
\caption{Block diagonal structure of the matrices (a) - $M_2$ and (b)-$M_4$ listed in Eq.~.(\ref{Rep3}). }
\label{Fig7_4}
\end{figure}
Let us denote the 6 matrices as $I,M_2,\ldots, M_6$, where $I={\bf 1}_3$ is the unit $3 \times 3$ matrix. We want to relate the structure of these matrices to some important concept in 
group representation by a matrices as we have just established. Consider for example 
the permutations $\binom{123}{123}$ and $\binom{123}{213}$. They form a subgroup  
$S_{12} \subset S_3$. This subgroup is represented by the matrices $I, M_2 \in {\cal M}_3$. 
Therefore $I, M_2$ form a subgroup $M_{12} \subset {\cal M}_3$ that is a matrix representation of the
subgroup   $S_{12} \subset S_3$. The structure of both matrices $I$ and $M_2$ is that of 
{\it block diagonal structure} illustrated in Fig.~\ref{Fig7_4}a. Similar considerations hold for 
 $\binom{123}{123}$ and $\binom{123}{132}$ represented by $I,M_4$ with block diagonal form as 
 in Fig.~\ref{Fig7_4}b. Now we arrive at an important\\
 \tboxed{
 \underline{Definition:} If {\it all} the matrices in a representation of a group $G$ have {\it the same block diagonal form} we say that {\it the matrix representation of $G$ is reducible}. A representation that is NOT reducible is termed as {\it \underline {irreducible representation (irrep) of $G$ }} . 
}
Inspecting the representation of the whole group $S_3$ in terms of the 6 matrices displayed in Eq.~(\ref{Rep3}) from the point of view of reducibility it is immediately clear that not all the six matrices have the same block diagonal form, and therefore the representation is reducible. Therefore we conclude:
\tboxed{ The representation of $S_3$ in terms of the group of 6 matrices $I,M_2,\ldots, M_6$ is irrep. the representation of the the subgroup $S_{12}$ in terms of the group of 2 matrices $I, M_2$ is reducible. }
An important lemma that has been used in proving the theorem in Subsection \ref{Subsec_Qtrits} is:\\
\underline{Schurs lemma:} Let $G$ be a group and let $\{ M(g), \ g \in G \} $ be an irrep of $G$ in terms of $n \times n$ matrices. 
If a matrix $A$ commutes with the matrices $\{ M(g) \}$ then $A$ is a constant times the unit $n \times n $ matrix, namely,
\begin{equation} \label{Rep4} 
[A,M(g)]=0, \ \forall \ g \in G \ \ \Leftrightarrow \ A=\alpha {\bf 1}_n~.
\end{equation}
\subsubsection{Continuous Groups} 
So far we encountered groups of finite order, namely finite groups. This is relevant for classical games 
with finite number of strategies of each player. We can extend this easily to 
infinite albeit discrete groups, for example, the group of integers $\la \{ n \}, n \in  \mathbb{Z}, \bullet=+, I=0 \ra$  is an infinite discrete group. Similarly, the group of rational numbers $\la \{ r \}, \ \ r=p/q, \ \ p,q \in \mathbb{Z}/\{0 \}, \bullet=\times, \ I=1 \ra $ is an infinite discrete group. In the language of set theory we say that the order of an infinite discrete group is $\aleph_0$. The notion of continuous groups requires more than just having an infinite order. It requires some definition of ``distance" between group elements. For example, the group of real numbers $\la \{ x \}, x \in  \mathbb{R}, \bullet=+, I=0 \ra$  is an infinite group satisfying the group axioms, and,  
in addition, we have a natural definition of distance between two group elements $x \in \mathbb{R}$ and $y \in \mathbb {R}$ as $d=|x-y|$. 
Before writing down the formal definition it is useful to consider the group of quantum strategies $U(\phi, \alpha, \theta)$ defined in Eq.~(\ref{11}). Since $U$ is unitary and has unit determinant, the group axioms are satisfied (but will be slightly modified to account for the fact continuous case). The main points to be noticed is as follows: 1) The elements of the group (namely, the matrices  $U(\phi, \alpha, \theta)$  depend on a number (here three) of continuous parameters $0 \le \phi < 2 \pi, \ \ 0 \le \alpha<  2 \pi, \ \ 0 \le \theta < \pi$ denoted here collectively as ${\bf a}=(\phi,\alpha, \theta)$ (instead of ${\bm \gamma}$ as used formerly). The three dimensional box 
containing all points $(0 \le \phi \le 2 \pi, 0 \le \alpha \le 2 \pi, 0 \le \theta \le \pi)$  defines the 
{\it parameter domain} ${\cal A}$.  2) We can define ``distance between two matrices 
$ U({\bf a} )$ and $U({\bfa}' ) $ as, 
\begin{equation} \label{Cont1}
d[U({\bf a} ),(U({\bf a}' )]=\sum_{i,j=1}^2 |U_{i,j}({\bf a} )-U_{i,j}({\bf a}' )|^2~, \ \ \forall \ \ \bfa, \bfa ' \in {\cal A}~. 
\end{equation} 
Each number inside the absolute value is a difference between two complex numbers and hence it is a complex number 
whose absolute value is defined in Eq.~(\ref{55}). 
3) The matrix elements of the matrices $U({\bf a}_i)$ are 
continuous (and differentiable) functions of the three Euler angles.  Therefore, we have, 
\begin{equation} \label{Cont2}
d({\bf a} ,{\bf a}' ) \equiv (\phi -\phi' )^2+(\alpha -\alpha' )^2+(\theta -\theta' )^2 \to 0 \ \Rightarrow \ d[U({\bf a} ),(U({\bf a}' )] \to 0~.
\end{equation} 
\tboxed{
\underline{ Definition:} A continuous group $G$ is a group whose elements $ g({\bm \gamma}) \in G$ depend on a (usually finite) set 
${\bm \gamma}=a_1,a_2,\ldots a_r$ of continuous parameter, endowed with a distance $d(g,g'): G \times G \to \mathbb{R}_+$ such that $d({\bm \gamma},{\bf a}') \to 0$ implies $d(g,g') \to 0$. 
}
In our example above $r=3, \ a_1=\phi, a_2=\alpha, a_3=\theta$. The slightly modified list of axioms for a continuous group are:
\begin{enumerate}
\item{} {\bf Closeness:} Given the (sets of) parameters $\bfa, \bfb \in {\cal A}$ there is a set $\bfc \in {\cal A}$ such that\\
$g(\bfc)=g(\bfb)g(\bfb)$, where $\bfc$ is a {\it real function} of the real parameters $\bfa$ and $\bfb$. 
\begin{equation} \label{Cont3}
c_k=f_k(a_1,a_2,\ldots, a_r; b_1,b_2,\ldots,b_r)=f_k(\bfa;\bfb), \ \ k=1,2,\ldots,r~, \ \ \mbox{or compactly} \ \ \bfc = {\bf f}(\bfa;\bfb)~. 
\end{equation}
\item{}  {\bf Associativity:} $g(\bfc)[g(\bfb)g(\bfa)]=[g(\bfc) g(\bfb)]g(\bfa)], \ \ \forall \ \ \bfa, \bfb, \bfc \in {\cal A}~. $.
\item{} {\bf Unit Element:} There exists a parameter $\bfa=\bfa_0$ (conveniently chosen as $\bfa_0=0$ such that\\
$g(\bfa) g(0)=g(0)g(\bfa)=g(\bfa)$. Occasionally we denote $g(0)=I$. 
\item{} {\bf Existence of Inverse:} For any $\bfa \in {\cal A}, \ \exists \ \bar {\bfa} \in {\cal A} \ \ni g(\bar {\bfa})g(\bfa)=g(\bfa)g(\bar{\bfa})=g(0)=I, \ \ \bar {\bfa}={\bf h}(\bfa) $. 
\item{} {\bf Minimality (or essentiality) of $r$:} The number of parameters $r$ is minimal. If $G$ is a continuous group of $r$ parameters, then it is not possible to obtain all the elements of the group in terms of less than $r$ parameters. 
\end{enumerate} 
The order of a continuous group is equal to the cardinality of the continuum. which in set theory is denoted by $\aleph$.   
\subsubsection{Lie Groups} 
The modified set of axioms listed above is applicable also for finite groups and for discrete groups if infinite ($\aleph_0$) order. Now we require a new condition that is peculiar to continuous groups: \\
{\bf New Condition:} The  
argument $\bfc$ appearing in $g(\bfc)=g(\bfa)g(\bfb)$ of the first axiom is an analytic function of 
$\bfa$ and $\bfb$. By that we mean that the functions ${\bf f}(\bfa;\bfb)$ defined in Eq,~(\ref{Cont3}) are infinitely differentiable with respect to to the $2r$ parameters $\bfa$ and $\bfb$. Similarly, $\bar {\bfa}$ is an analytic function of $\bfa$, namely, the function ${\bf h}(.)$ defined in axiom 4 is infinitely differentiable. 
Under the axioms listed above and the new condition we get an {\it $r$-parameters Lie group}.

\begin{figure}[!ht]
\centering
\includegraphics[scale=0.4]{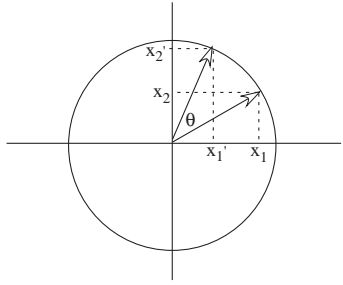}
\caption{\footnotesize{ Demonstration of $O(2)$ Lie group as a rotation of a vector $\bfx =(x_1,x_2) \to \bfx'=(x_1',x_2')$ whose tip lies on a circle of radius $\rho$ through an angle $\theta$. This rotation is such that 
$x_1^2+x_2^2=r'^2=x_1'^2+x_2'^2=r^2$ (Pithagoras theorem). } }
\label{Fig7_5}
\end{figure}

\noindent
{\bf Lie Groups of Transformations:} Lie groups are very often used as {\it groups of transformations}. By this we mean that we have $n$ quantities 
$(x_1,x_2,\ldots x_n)=\bfx$ that undergo a transformation after which we get $(x_1',x_2',\ldots x_n')=\bfx'$. 
The transformation depends on $r$ parameters $\bfa$ and written compactly as 
\begin{equation} \label{Cont4} 
\bfx'={\bf f}(\bfx;\bfa)~.
\end{equation}
For example $\bfx=(x_1,x_2)$ can be the coordinates of a point on the plane and the group operation is to move this point 
as $(x_1,x_2) \to (x_1+a_1,x_2+a_2)$ so that  $r=2$, $\bfa=(a_1,a_2)$ and the transformation group 
operation is $\bfx'=\bff(\bfx,\bfa)=\bfx+\bfa$.  Then clearly, $\bfa_0={\bf 0}$ and $\bar {\bfa}=-\bfa$.  If we apply a second transformation corresponding to set of parameters $\bfb=(b_1,b_2)$ we get 
\begin{equation} \label{Cont5}
\bfx''=\bff(\bfx',\bfb)=\bfx'+\bfb=\bff[\bff(\bfx,\bfa);\bfb]=\bfx+\bfa+\bfb \equiv \bfx+\bfc=\bff(\bfx;\bfc), \ \ \bfc={\bm \phi}(\bfa,\bfb)=\bfa+\bfb~.
\end{equation} 
This almost trivial example illustrates the interpretation of the group axioms for the Lie group of transformations (\ref{Cont4}), that is, 
\begin{enumerate} 
\item{} {\bf Closeness} 
$\bfx'=\bff(\bfx;\bfa), \ \bfx''=f(\bfx';\bfb) \ \Rightarrow \ \exists \ \bfc \in {\cal A} \ni \bfx''=\bff(\bfx'; \bfc), \  \ni \bfc={\bm \phi}(\bfa,\bfb)~.$
\item{} {\bf Associativity:} In the following list of three operations, 
$\bfx'''=\bff(\bfx'';\bfc), \ \ \bfx''=\bff(\bfx';\bfb), \ \ \bfx'=\bff(\bfx;\bfa)$\\
 it does not matter whether the first two are done first or the last two is done last. 
 \item{} {\bf Unit Operation} $\exists \  \bfa_0 \in {\cal A}, \ni \ \bfx'=\bff(\bfx; \bfa_0)=\bfx$. 
 \item{} {\bf Existence of an inverse:} $\bfx'=\bff(\bfx;\bfa) \ \ \Rightarrow \ \exists \ \bar{\bfa} \in {\cal A} \ \ni \ \bff(\bfx';\bar{\bfa})=
  \bff[\bff(\bfx;\bfa);\bar{\bfa}]=\bfx $. 
\end{enumerate} 
{\bf Examples of Lie Groups of Transformations:} \\
1) {\bf Diltation:} $x'=f(x;a)=ax, \ (a \ne 0),  \ \ r=1, \ a_0=1, \ \ \bar{a}=1/a, \ \ c=\phi(a,b)=ab$. The group is Abelian. \\  
2) {\bf Dilatation and translation:} \\ 
$x'=a_1x+a_2, \ (a_1 \ne 0), \ r=2, \ \bfa=(a_1,a_2), \ \bfa_0=(1,0), \ \bar{\bfa}=(1/a_1,-a_2/a_1), \ \bfc=(b_1a_1,b_2+b_1a_2)={\bm \phi}(\bfa,\bfb)$. \\
Note that ${\bm \phi}(\bfa,\bfb) \ne {\bm \phi}(\bfb,\bfa)$, meaning that the group is non-Abelian (non-commutative). \\
3) {\bf GL(2,R)- The linear group in two dimensions:}\\
$ \binom{x_1'}{x_2'}=\binom{a_1a_2}{a_3 a_4}\binom{x}{y} =A\binom{x}{y} , \mbox{Det}[A] \ne 0, \ \ \bfx'=\bff(\bfx;\bfa)$. 
Here $r=4, \bfa=A, \ \bfa_0=\binom{10}{01}, \ \bar {\bfa}=A^{-1}, \ \ \bfc=BA$. \\
Here all the numbers are real. 
$GL(2,R)$ is isomorphic to the group of regular $2 \times 2$ real matrices and it is non-Abelian. \\
4) {\bf SL(2,R) - Special Linear (unimodular) Group in two dimensions:} \\
Consider example 3 but restrict the determinant of  $A$ to be unity Det$[A]=(a_1a_4-a_2a_3)$. This introduces a relation between the four parameters so that $r=3$. The choice of the three parameters is not unique but a convenient 
route is to choose three matrices with the desired property, each one depends on one parameter and then 
construct a general matrix in $SL(2,R)$ using their product. \\
\underline{Iwasawa theorem:} Let $K=\left \{ \binom{~~\cos \theta ~~ \sin \theta}{-\sin \theta ~~ \cos \theta} \right \} 
, L=\left \{ \binom{q ~~ 0}{~0 ~~ 1/q} \right \}, M=\left \{ \binom{1 ~~ x}{~0 ~~ 1} \right \} $. 
Then every matrix $A \in SL(2,R)$ has a unique decomposition $\bfa=k(\theta)l(r)m(x)$, with $k(\theta) \in K, \ l(r) \in L$ and $m(x) \in M$. \\
Here $r=3, \bfa=A, \ \bfa_0=k(0)l(1)m(0), \ \bar {\bfa}=A^{-1}=m(-x) l(1/r) k(-\theta), \ \ 
\bfc=BA$. \\
5) {\bf $O(2)$ - Orthogonal Group in Two Dimensions:} We select from $GL(2,R)$ of example 3) only those transformations that conserve the length of the bilinear form $r^2=x_1^2+x_2^2=r'^2=x_1'^2+x_2'^2$. To illustrate it we note that this transformation is like a rotation of a vector whose tip is on a circle of radius $\rho$, see Fig.~\ref{Fig7_5}. 
Simple manipulations in plan trigonometry lead to the following expressions for $\bfx'$ in terms of $\bfx$ and $\theta$,
\begin{equation} \label{Cont6}
x_1'=x_1 \cos \theta - x_2\sin \theta, \ \ x_2'=x_1 \sin \theta+x_2 \cos \theta, 
\ \ \bfx'=\binom{x_1'}{x_2'}=\binom {\cos \theta ~~ -\sin \theta}{\sin \theta \ \ \cos \theta}\binom{x_1}{x_2} \ \ 0 \le \theta \le 2 \pi~.
\end{equation}
Therefore, in the language of Lie group transformations we have (denoting the matrix as $A(\theta) $) \\
$r=1, \ a=A(\theta), \ \ \bfx'=\bff(\bfx,a)=A(\theta) \bfx, \ \ a_0=A(0), \ \bar{a}=A^{-1}(\theta)=A(-\theta), \ \ , \ \ c(\theta_a, \theta_b)=A(\theta_a+\theta_b)$. \\
By construction $O(2)$ is Abelian, performing two rotations does not depend on order. \\

6) {\bf $Sp(2,R)$- Symplectic Group in Two Dimensions:} 
We consider transformations  $(x_1,x_2) \to \bfx'=(x_1',x_2')$  such that 
$x_1^2-x_2^2=x_1'^2-x_2'^2$. Let us define the $2 \times 2$ matrix $A(u)=\binom{\cosh u ~~ \sinh u}{\sinh u ~~ \cosh(u)}$. Then simple calculations show that 
\begin{equation} \label{Cont7}
\binom{x_1'}{x_2'}=A(u) \binom{x_1}{x_2} \ \ \Rightarrow  x_1^2-x_2^2=x_1'^2-x_2'^2~. \ \ 
\mbox{Det}[A(u)]=1, \ \ A^{-1}(u)=\binom{\cosh u ~~ -\sinh u}{-\sinh u ~~ \cosh(u)}=A(-u)~.
\end{equation} 
Therefore, in the language of Lie group transformations we have \\
$r=1, \ a=A(u), \ \ \bfx'=\bff(\bfx,a)=A(u) \bfx, \ \ a_0=A(0), \ \bar{a}=A^{-1}(u)=A(-u), \ \ , \ \ c(u_a, u_b)=A(u_a+u_b)$. \\
By construction $S_p(2)$ is Abelian. 
\subsubsection{The SU(2) and SU(3) groups}
Finally, we consider Lie groups of complex transformations that are relevant for 
quantum games. Recall that the matrices in the set $U(\phi,\alpha,\theta)$ defined in Eq.~(\ref{11}) operate on qubits and serves as a quantum strategy in a quantum game based on 2-players 2-strategies classical games. The matrices $ \left \{ U(\phi,\alpha,\theta) \right \}$ are unitary 
$2 \times 2$ complex matrices with unit determinant, and the corresponding group of matrices is denoted as $SU(2)$ (Special Unitary matrices of dimension 2).  They depend on three parameters (Euler angles $(\phi,\alpha,\theta)$). They can be generated by exponentials of the Pauli matrices defined in Eq.~(\ref{Paulimat1}) according to the 
prescription of Eq.~(\ref{Herm5}), 
\begin{equation} \label{SU1} 
U(\phi,\alpha,\theta)=e^{i \phi \sigma_3}e^{i \theta \sigma_2} e^{i \alpha \sigma_3}~.
\end{equation}  

The $3 \times 3$ matrices $U$ introduced in our discussion of qutrits (Subsection \ref{Subsec_Qtrits}) operate on qutrits and serves as a quantum strategy in a quantum game based on 2-players 3-strategies classical games. The matrices $ \left \{ U(\alpha_1,\alpha_2,\ldots,\alpha_8) \right \}$ are unitary 
$3 \times 3$ complex matrices with unit determinant, and the corresponding group of matrices is denoted as $SU(3)$ (Special Unitary matrices of dimension 3).  This group plays a crucial role in physics, and at the middle of the sixties it was used by Murray Gellman and Yuval Neeman to classify elementary particles and predict the existence of new particles. 
The matrices $U \in SU(3)$ depend on eight parameters Euler angles $(\alpha_1,\alpha_2,\ldots,\alpha_8)$ (the book written by Gellman and Neeman is called {\it The Eight-fold Way}). Instead of the three Pauli matrices, the $SU(3)$ matrices can be generated by exponentials of the eight $3 \times 3$ Gelleman matrices defined in Eq.~(\ref{Gellmat}),  
\begin{spacing}{0.9}
 \begin{eqnarray}  \label{Gellmat}
&&  \lambda_1 = \left( \! \! \begin{array}{ccc}
	0 & 1 & 0  \\
	1 & 0 & 0  \\
	0 & 0 & 0  \end{array} \! \! \right) , \qquad 
	\lambda_2 = \left( \! \! \begin{array}{ccc}
	0 & {-i} & 0  \\
	{i} & 0 & 0  \\
	0 & 0 & 0  \end{array} \! \! \right) , \qquad 
	\lambda_3 = \left( \! \! \begin{array}{ccc}
	1 & 0 & 0  \\
	0 & {-1} & 0  \\
	0 & 0 & 0  \end{array} \! \! \right) , \nonumber \\
&&  \lambda_4 = \left( \!  \!  \begin{array}{ccc} 
	0 & 0 & 1 \\
	0 & 0 & 0 \\
	1 & 0 & 0  \end{array} \! \! \right) , \qquad 
	\lambda_5 = \left( \! \! \begin{array}{ccc}
	0 & 0 & {-i}  \\
	0 & 0 & 0  \\
	{i} & 0 & 0  \end{array} \! \! \right) , \qquad 
	\lambda_6 = \left( \! \! \begin{array}{ccc}
	0 & 0 & 0  \\
	0 & 0 & 1  \\
	0 & 1 & 0  \end{array} \! \! \right) , \nonumber \\
&&  \lambda_7 = \left( \! \! \begin{array}{ccc}
	0 & 0 & 0  \\
	0 & 0 & {-i}  \\
	0 & {i} & 0  \end{array} \! \! \right) , \qquad 
	\lambda_8 = \frac{1}{\sqrt{3}} \left( \! \! \begin{array}{ccc}
	1 & 0 & 0  \\
	0 & 1 & 0  \\
	0 & 0 & {-2} \end{array} \!  \!  \right) ,
\end{eqnarray} 
\end{spacing}  
\vspace{0.2in}
The prescription of Eq.~(\ref{Herm5}) is applicable also if $A^3=A$ (not necessarily $A^2=1$), 
\begin{equation} \label{SU3}
U \in SU(3)=e^{i \alpha_1 \lambda_3}e^{i \alpha_2 \lambda_2}e^{i \alpha_3 \lambda_3}e^{i \alpha_4 \lambda_5}e^{i \alpha_5 \lambda_3}e^{i \alpha_6 \lambda_2}e^{i \alpha_7 \lambda_3}e^{i \alpha_8 \lambda_8}~.
\end{equation} 
\subsection{Hilbert Space} \label{HS}
A Hilbert space ${\cal H}$ is an inner product space as defined in Subsection \ref{Inner_Prod}. 
Here we give a brief reminder, and introduce the notion of
linear operators that transform one state (vector) $|\psi \ra \in {\cal H}$  to another 
vector $|\phi \ra \in {\cal H}$. 
\subsubsection{Brief Reminder} 
\begin{enumerate}
\item{}  Hilbert space ${\cal H}$  is a complex linear inner product space. In Dirac's ket-bra
notation states (vectors) in ${\cal H}$ are denoted by \emph{ket vectors} $\left|  \psi\right\rangle $
in Hilbert space. Any two state vectors differing only by an overall phase
factor $e^{i\theta}$ ($\theta$ real) represent the same state.
\item{}  Corresponding to a ket vector $\left|  \psi\right\rangle $ there is
another kind of state vector called \emph{bra vector}, which is denoted by
$\left\langle \psi\right|  $. The \emph{inner product} of a bra $\left\langle
\psi\right|  $ and ket $\left|  \phi\right\rangle $ is defined as follows:%
\begin{eqnarray}
&& \la \psi|\phi \ra =\la \phi|\psi \ra^* \in \mathbb{C}, \ \la \psi|\psi \ra > 0 \ \ \mbox{for} \ |\psi \ra \ne 0. \nonumber \\
%
&& \left\langle \psi\right|  \left\{  \left|  \phi_{1}\right\rangle +\left|
\phi_{2}\right\rangle \right\}    =\left\langle \psi\mid\phi_{1}%
\right\rangle +\left\langle \psi\mid\phi_{2}\right\rangle
\left\langle \psi\right|  \left\{  c\left|  \phi_{1}\right\rangle \right\}
=c\left\langle \psi\mid\phi_{1}\right\rangle
\end{eqnarray}
for any $c \in \mathbb{C}$, the set of complex numbers. There is a one-to-one
correspondence between the bras and the kets. 
\item{}  The state vectors in Hilbert space that are relevant for quantum game theory 
(and for quantum physics in general) are normalized which means that the
inner product of a state vector with itself gives unity, i.e.,
$\left\langle \psi\mid\psi\right\rangle =1$. 
\item{} In a finite (N) dimensional Hilbert space ${\cal H}_N$ there is a sequence $(|\phi_1\ra, |\phi_2 \ra, \ldots, |\phi_N \ra)$ of linearly independent vectors (states) that can be chosen to be orthonormal, $\la \phi_i|\phi_j \ra=\delta_{ij}$ such that every state $|\psi \ra \in {\cal H}_N$ can be expanded as 
\begin{equation} \label{HS1} 
|\psi \ra=\sum_{i=1}^N c_i |\phi_i \ra, \ \ c_i=\la \phi_i|\psi \ra \in \mathbb{C}~. \ \ \la \psi|\psi \ra=\sum_{i=1}^N|c_i|^2~.
\end{equation}
\end{enumerate}

Anticipating the developments of quantum games 
based on classical games with infinite number of strategies, 
it is worth while to augment these definitions by 
 two additional properties denoted
here as I and II. If $N$ is finite, these properties can be proved
from the previous definitions of an inner product space, but for an
infinite dimensional space they should be considered as axioms.  \\
(I) ${\cal H}$ is complete.  If an infinite sequence $|\psi_1 \ra,
|\psi_2 \ra, \ldots \in {\cal H}$ satisfies the Cauchy convergence
criterion [for each $\varepsilon >0$ there exists a positive integer
$M(\varepsilon)$, such that for $m,n > M(\varepsilon), \ || \, |\psi_n
\ra-|\psi_m \ra \, || < \varepsilon$], then the sequence is
convergent, i.e., it possesses a limit $|\psi \ra$ such that 
$\lim_{n \to \infty} || |\psi_n \ra-|\psi \ra ||=0$.  \\
(II) ${\cal H}$ is separable: that is, there is a sequence $|\psi_1
\ra, |\psi_2 \ra, \ldots \in {\cal H}$ which is everywhere dense in
${\cal H}$.\footnote{An example of a dense sequence is the rational
numbers which are dense in $\mathbb{R}$.} Roughly speaking, elements
of a dense set come arbitrary close to any element in ${\cal H}$.  The
property of separability is equivalent to the statement that there is
a countably infinite complete orthonormal set $\{ |\phi_n \ra \}$ such
that every vector $|\psi \ra \in {\cal H}$ can be expanded as,
\begin{equation} \label{Eq_LA_Hilbert1}
    |\psi \ra=\sum_{n=1}^\infty \la \phi_n|\psi \ra |\phi_n \ra \ \
    \Leftrightarrow \lim_{N \to \infty} || \, \sum_{n=1}^N \la \phi_n|\psi
    \ra |\phi_n \ra-|\psi \ra \, || = 0~.
\end{equation}
A necessary and sufficient condition for convergence is
$\sum_{n=1}^\infty |\la \phi_n|\psi \ra|^2<\infty$.  Hence, this
sum equals $\la \psi|\psi \ra$.
\subsubsection{Operators in Hilbert Space}
For finite dimensional Hilbert space ${\cal H}_N$  the notion of operators and matrices 
are the same, and operators are represented by $N \times N$ matrices. For an infinite 
dimensional Hilbert space the situation is different, since the treatment of 
matrices with infinite dimensions requires some care. We give a short list of the main definitions 
using the nomenclature of operators, but for ${\cal H}_N$, replacement ``operator $\to$ matrix is justified. 

\noindent
Operations can be performed on a ket $\left|  \psi\right\rangle $ and
transform it to another ket $\left|  \chi\right\rangle $. There are operations
on kets which are called \emph{linear operators}, which have the following
properties. For a linear operator ${\cal O}$ we have
\begin{align}
{\cal O}\left\{  \left|  \psi\right\rangle +\left|  \chi\right\rangle
\right\}   &  ={\cal O}\left|  \psi\right\rangle +{\cal O}\left|
\chi\right\rangle \nonumber\\
{\cal O}\left\{  c\left|  \psi\right\rangle \right\}   &  =c{\cal O} \left|  \psi\right\rangle
\end{align}
for any $c\in\mathbf{C}$.
\begin{itemize}
\item  The sum and product of two linear operators ${\cal O}$ and
${\cal P}$ are defined as:%
\begin{equation}
\{  {\cal O}+{\cal P} \}  |\psi \rangle  
={\cal O} | \psi \rangle +{\cal P} |  \psi \rangle
\end{equation}
\begin{equation}
\left \{  {\cal O}{\cal P} \right \}  |  \psi \rangle  
={\cal O} \{ {\cal P} |  \psi \rangle \}
\end{equation}
Generally speaking ${\cal O}{\cal P}$ is not necessarily equal to
${\cal P}{\cal O}$, i.e. $\left[  {\cal O},{\cal P} \right]  \neq0$
\item  The \emph{adjoint} ${\cal O}^{\dagger}$ of an operator ${\cal P}$ 
is defined by the requirement:%
\begin{equation}
\left\langle \psi\mid{\cal O}\chi\right\rangle =\left\langle{\cal O}^{\dagger}\psi\mid\chi\right\rangle
\end{equation}
for all kets $\left|  \psi\right\rangle $, $\left|  \chi\right\rangle $ in the
Hilbert space.
\item  An operator ${\cal O}$ is said to be \emph{self-adjoint} or
\emph{Hermitian} if:%
\begin{equation}
{\cal O}^{\dagger}={\cal O}%
\end{equation}
\end{itemize}
Hermitian operators are the counterparts of real numbers in operators. In
quantum mechanics, the dynamical variables of physical systems are represented
by Hermitian operators. More specifically, every experimental arrangement in
quantum mechanics is associated with a set of operators describing the
dynamical variables that can be observed. These operators are usually called
\emph{observables}.

\subsection{Basic Concepts of Quantum Mechanics} \label{QM}
Quantum theory\cite{BA} is the theoretical basis of modern physics that explains the
nature and behavior of matter and energy on the atomic and subatomic level.
The physical systems at these levels are known as \emph{quantum systems}. Thus
quantum mechanics is a mathematical model of the physical world that describes
the behavior of quantum systems. A physical model is characterized by how it
represents \emph{physical states}, \emph{observables}, \emph{measurements},
and \emph{dynamics} of the system under consideration.

\subsubsection{Postulates of quantum mechanics}
We formulate below the postulates of quantum mechanics 
that are most relevant for quantum games. 

\underline{\bf {Postulates}}
\begin{enumerate}
\item{}
    At each instant of time, $t$, the state of a physical system is
    represented by a vector (sometimes called a ket) $|\psi(t)
    \rangle$ in the vector space of states.
    Every observable attribute of a physical system is described by an
    operator that acts on the ket that describes the system.

    \item{}  
    The only possible result of the measurement of an observable
    $A$ (for example energy of a system) is one of the eigenvalues of 
    an Hermitian operator $\hat {\cal
    A}$ representing the observable $A$.  An observable must be
    represented a Hermitian operator.

    Since measurement results are real numbers, the eigenvalues of
    operators corresponding to observables are real.  The operator
    representing an observable is often called an observable operator,
    or observable for short.  All eigenvalues of Hermitian operators
    are real, and their eigenvectors are orthogonal, $\langle \phi_i |
    \phi_j \rangle = \delta_{ij}$.
    
    \item{} 
    When a measurement of an observable $A$ is made on a
    generic state $|\psi \rangle$, the probability of obtaining an
    eigenvalue $a_i$ is given by $\left| \langle \phi_i|\psi
    \rangle \right|^2$, where $|\phi_i \rangle$ is the eigenstate of
    the observable operator $\hat {\cal A}$ with eigenvalue $a_i$, 
    that is, ${\cal A}|\phi_i \ra=a_i |\phi_i \ra$. 
    
    The complex number $\langle \phi_i|\psi \rangle$ is known as the
    ``probability amplitude'' to measure $a_i$ as the value for
    ${\cal A}$ in the state $|\psi \rangle$.

    \item{} 
    Immediately after the measurement of an observable ${\cal A}$ that
    has yielded a value $a_i$, the state of the system is the
    normalized eigenstate $| \phi_i \rangle$.
    
    With the system initially in state $|\psi \rangle$, measurement of
    an observable ${\cal A}$ collapses the wave function.  If the
    result of the measurement is $a_i$, the wave function
    collapses to state $|\phi_i \rangle$.  
    
    
%
    \item{} 
    For each physical system there is a unique operator $\hat {H}(t)$ 
    known as the Hamiltonian, that determines the evolution of the system with time. 
    Physically, $H(t)$ represents the energy of the system. 
    The time evolution of the state of a quantum system is specified
    by the state vector $|\psi(t) \rangle = {\hat {\cal
    U}}(t,t_0)|\psi(t_0) \rangle$, where the operator ${\hat {\cal
    U}}$ is unitary (${\hat {\cal U}} {\hat {\cal U}}^\dag = 1$), and
    therefore preserves the normalization of the associated ket, and
    is called the {\em evolution operator}: \tbox{
    \begin{equation} \label{Eq:3.post_1}
      \left| {\psi \left( t \right)} \right\rangle = {\hat {\cal U}}
      \left( {t,t_0} \right)\left| {\psi \left(t_0 \right)}
      \right\rangle ~.
    \end{equation} }
    For a time independent Hamiltonian, ${\hat {\cal U}}(t,t_0) = \exp
    ( -i{\hat H}(t-t_0) /\hbar)$.  In general (i.e., even for
    time-dependent Hamiltonians) \tbox{
    \begin{equation} \label{Eq:3.post_1'}
      i \hbar \frac{{\partial {\hat {\cal U}}(t,t_0)}} {{\partial t}}
      = {\hat H}(t) {\hat {\cal U}}(t,t_0) ~.
    \end{equation} }
    
    This is equivalent to saying that $|\psi(t) \rangle$ satisfies the
    Schr\"{o}dinger equation, $i\hbar \frac{\partial}{\partial t}
    |\psi (t) \rangle = \hat{H} |\psi (t) \rangle$.
\end{enumerate}
\end{spacing}

\end{document}